\documentclass{jfm}
\usepackage{graphicx}
\usepackage{epstopdf, epsfig}
\usepackage{hyperref}
\usepackage{amsmath}
\usepackage{amssymb, color}
\usepackage{subfig}
\usepackage{float}
\usepackage{multirow}
\usepackage[normalem]{ulem}
\usepackage[export]{adjustbox}
\usepackage{longtable}
\usepackage{afterpage}

\shorttitle{Inverse cascade and flow speeds}
\shortauthor{ S. Maffei, M. J. Krouss, K. Julien and M. A. Calkins}

\title{On the inverse cascade and flow speed scaling behavior in rapidly rotating Rayleigh-B\'{e}nard convection}

\author{  S. Maffei\aff{1}$^{,}$\aff{3}
    \corresp{\email{S.Maffei@leeds.ac.uk}},
    M. J. Krouss\aff{1},
    K. Julien\aff{2}
 \and M. A. Calkins\aff{1}}

\affiliation{\aff{1}Department of Physics, University of Colorado, Boulder, USA
\aff{2}Department of Applied Mathematics, University of Colorado, Boulder, USA
\aff{3}School of Earth and Environment, University of Leeds,  Leeds, UK
}

\def\be{\begin{equation}}
\def\ee{\end{equation}}

\def\begineqn{\begin{equation*}}
\def\endeqn{\end{equation*}}

\def\beginar{\begin{eqnarray}}
\def\endar{\end{eqnarray}}
\def\beginarn{\begin{eqnarray*}}
\def\endarn{\end{eqnarray*}}

\def\lb{\left ( }
\def\rb{\right ) }

\def\Rat{\widetilde{Ra}}
\def\Ret{\widetilde{Re}}
\def\Re{Re}
\def\Ra{Ra}

\def\Pr{Pr}
\def\Ek{Ek}
\def\Ro{Ro}

\def\Nu{Nu}

\def\Num{Nu}
\def\Retm{\Ret}
\def\Kbtm{\Kbt}
\def\Kbcm{\Kbc}
\def\Km{K}
\def\Kzm{K_z}

\def\Kbt{K_{bt}}
\def\Kbc{K_{bc}}

\def\ub{\mathbf{u}}

\def\mth{\overline{\Theta}}
\def\pth{\vartheta}

\def\dsx{{\partial_x}}
\def\dsy{{\partial_y}}

\def\dst{{\partial_t}}

\def\dz{{\partial_Z}}
\def\dzt{{\partial^2_Z}}

\def\hz{{\bf\widehat z}}

\def\lp{{\nabla_\perp^2}}

\def\alphaRaNum{0.1883}
\def\betaRaNum{1.1512}
\def\gammaRaNum{-1.2172}

\def\alphaNuNum{0.2169}
\def\betaNuNum{1.1791}
\def\gammaNuNum{0.0481}

\begin{document}

\maketitle

\begin{abstract}
Rotating Rayleigh-B\'enard convection is investigated numerically with the use of an asymptotic model that captures the rapidly rotating, small Ekman number limit, $\Ek \rightarrow 0$. The Prandtl number ($Pr$) and the asymptotically scaled Rayleigh number ($\Rat = Ra \Ek^{4/3}$, where $Ra$ is the typical Rayleigh number) are varied systematically. For sufficiently vigorous convection, an inverse kinetic energy cascade leads to the formation of a depth-invariant large-scale vortex (LSV). With respect to the kinetic energy, we find a transition from convection dominated states to LSV dominated states at an asymptotically reduced (small-scale) Reynolds number of $\Ret \approx 6$ for all investigated values of $Pr$. The ratio of the depth-averaged kinetic energy to the kinetic energy of the convection reaches a maximum at $\Ret \approx 24$, then decreases as $\Rat$ is increased. This decrease in the relative kinetic energy of the LSV is associated with a decrease in the convective correlations with increasing Rayleigh number. The scaling behavior of the convective flow speeds is studied; although a linear scaling of the form $\Ret \sim \Rat/Pr$ is observed over a limited range in Rayleigh number and Prandtl number, a clear departure from this scaling is observed at the highest accessible values of $\Rat$. Calculation of the forces present in the governing equations shows that the ratio of the viscous force to the buoyancy force is an increasing function of $\Rat$, that approaches unity over the investigated range of parameters. 

\end{abstract}


\section{Introduction}

Convection is common in the fluid regions of planet and stars. In particular, convection is the primary energy source for the generation of large-scale planetary and stellar magnetic fields \citep{cJ11b,tG14,jmA15}, and it is thought to be a source of energy for the observed zonal flows in the atmospheres of the giant planets \citep[e.g.][]{mH16}. The flows in many of these natural systems are considered turbulent and strongly influenced by rotation; previous studies have shown that the combination of these physical ingredients can lead to an inverse kinetic energy cascade \citep{smith1999transfer}. The inverse cascade transfers kinetic energy from small-scale convection up to domain-scale flows, and, in a planar geometry of square cross section, results in the formation of depth-invariant, large-scale vortices (LSVs) \citep{kJ12,rubio_upscale_2014,cG14,bF14,sS14}. Such vortices tend to be characterized by flow speeds that are significantly larger than the underlying convection, and can have an influence on heat transport and magnetic field generation \citep{cG14,cG15}. However, the convective vigor required to excite LSVs, how fluid properties (via the thermal Prandtl number) influence their formation, and the scaling of their saturated amplitude with buoyancy forcing still remain poorly understood. 

In planetary and astrophysical fluid systems, rapid rotation is thought to play an essential role in shaping the dynamics of convection. The importance of rotation on the dynamics of such systems is quantified by the Ekman and Rossby numbers, respectively defined as 
\be
\Ek = \frac{\nu}{2\Omega H^2}, \quad \Ro = \frac{U}{2 \Omega H} ,
\ee
where $\nu$ is the kinematic viscosity, $\Omega$ is the rate of rotation, $H$ is the spatial scale of the system (i.e. the depth of the fluid region) and $U$ is a characteristic flow speed. $\Ek$ and $\Ro$ quantify, respectively, the ratio of viscous forces to the Coriolis force, and the ratio of inertia to the Coriolis force. 
Systems in which $(\Ek,\Ro) \ll 1$ are said to be rapidly rotating and rotationally-constrained. For the Earth's outer core, for instance, estimates suggest that $\Ek\simeq10^{-15}$ and $\Ro\simeq10^{-5}$  \citep{cF11,de1998viscosity,rutter2002towards}. It is currently impossible to use such extreme values of the governing parameters with direct numerical simulation (DNS). Quasi-geostrophic (QG) models have helped to overcome this deficiency, and have been critical for elucidating several convective phenomena that are thought to be of significant interest for planets, including identification of the primary flow regimes of rotating convection, heat transport behavior, and the convection-driven inverse cascade \citep{kJ12,rubio_upscale_2014}. QG models accurately capture the leading order dynamics in systems characterized by small values of $\Ek$ and $\Ro$.

Previous work has shown that the structure of the LSVs is dependent on the relative importance of rotation: for sufficiently small values of  $\Ek$ and $\Ro$ the vortices are dipolar in structure with zero net (spatially-averaged) vorticity \citep{kJ12,rubio_upscale_2014,sS14}; whereas for larger values of $\Ro$ the vortices tend to be cyclonic with a net vorticity that is parallel to the rotation axis \citep{kC13,cG14,bF14}. QG models find only dipolar LSVs since they capture asymptotically-small values of  $\Ek$ and $\Ro$ only, in which there is no preferred sign for the vorticity. DNS studies have found dipolar LSVs when $\Ek \approx 10^{-7}$ \citep{sS14}, and cyclonic LSVs when $\Ek \gtrsim 10^{-6}$ \citep{kC13,cG14,bF14}. This distinction in LSV structure may have consequences for the influence of LSVs on heat transport and flow speeds. Moreover, for $\Ro = O(1)$, the presence of LSVs appears to be dependent upon the initial condition used in the simulations \citep{bF19}. In contrast, for the rapidly rotating regime studied here, we find that LSV formation is not dependent upon initial conditions.

Natural systems are characterized by a broad range of fluid properties and, as a result, the Prandtl number $\Pr = \nu / \kappa$ (where $\kappa$ is the thermal diffusivity) can take on a wide range of values, ranging from $\Pr =O(10^{-6})$ in stellar interiors \citep{mO03} to  $\Pr=O(10^{-2})$ for the liquid metals characteristic of planetary interiors \citep{mP13}. More generally, the density heterogeneities that lead to buoyancy-driven convection can also result from compositional differences, as is expected to be the case within terrestrial planetary interiors, for instance; under such circumstances the thermal Prandtl number in the governing equations is replaced by the compositional Schmidt number $Sc = \nu/D$ where $D$ is a chemical diffusivity. For the Earth's outer core, studies suggest $Sc = O(100)$ for representative chemical species \citep[e.g.][]{mP13}. This wide range of diffusivities that characterize geophysical and astrophysical fluids motivates the need for additional investigations that explore the influence of the Prandtl number on the dynamics, since all previous numerical calculations investigating the inverse cascade have focussed on fluids with $\Pr=1$. QG simulations \citep{kJ12} have characterized the flow regimes that occur when $\Pr \ge 1$. In general, it is found that low $\Pr$ fluids reach turbulent regimes for a lower value of the thermal forcing (as measured by the Rayleigh number $\Ra$) than higher $\Pr$ fluids. More turbulent flows can be characterized by an increase in the Reynolds number
\be
 \Re = U H /\nu,
\ee 
which is a fundamental output parameter for convection studies. Many of the results found in QG studies have been confirmed by DNS calculations \citep{sS14}. In both approaches, the formation of LSVs has been tied to the geostrophic turbulence regime, which, prior to the present work, has not been observed for $\Pr>3$.

Laboratory experiments are an important tool for exploring the dynamics of rapidly rotating convection \citep[e.g.][]{pV02,eK09,rK10,sS14,rE14} and, much like natural systems, can access a wide range of $\Pr$ values. Liquid metals ($\Pr\approx0.025$) \citep{jmA01,jA01,eK13,dZ14,mA15,jmA18,tV18}, water ($\Pr\approx7$) \citep{iS02,pV02,eK09,jC15}, and gases ($\Pr\approx1$) \citep{jN10b,rE14}, have all been utilized. As for numerical calculations, fluids with smaller values of $\Pr$ reach higher values of $\Re$ than higher $\Pr$ fluids for equivalent $\Ra$. From this perspective, such fluids are of significant interest for exploring the geostrophic turbulence regime of rapidly rotating convection \citep[e.g.][]{kJ12b}. However, the use of such low $\Pr$ fluids is not always practical and can reduce or eliminate flow visualization opportunities. Furthermore, whereas small $\Pr$ fluids can lead to more turbulent flows, lower Ekman numbers must be used to provide a sufficiently large parameter regime over which the fluid remains rotationally-constrained \citep[e.g.][]{eK13}. It would therefore be of use to identify the general dynamical requirements for observing inverse-cascade-generated LSVs for a variety of fluid properties.

One of the basic goals in convection studies is to determine the functional dependence of $\Re$ on the input parameters, namely, determination of the functional form $\Re = f(\Ra, \Pr)$. 
Power-law scalings of the form $\Re = c_1 \Ra^{c_2} \Pr^{c_3}$ are often sought, where each $c_i$ is a constant. A well-known example is the so-called `free-fall' scaling of the form $\Re \sim (\Ra/\Pr)^{1/2}$ observed in non-rotating convection \citep[e.g.][]{gA09,rO18}. The free-fall scaling arises from a balance between nonlinear advection and the buoyancy force in the momentum equation, and represents a `diffusion-free' scaling in the sense that the flow speeds are independent of both the thermal and viscous diffusion coefficients. Motivated by this free-fall form of the scaling, and the assumption that natural systems are expected to be highly turbulent in the sense that $\Re \gg 1$, rotating convection studies have also sought to find diffusion-free scalings for the flow speeds. For instance, the recent work of \cite{cG19} observed $\Re \sim \Ra \Ek/\Pr$ in spherical convection simulations, which is also a diffusion-free scaling for the flow speeds. In the present work we show that this scaling is equivalent to $\Ret \sim \Rat/\Pr$, where $\Ret = \Ek^{1/3} \Re$ and $\Rat= \Ek^{4/3} \Ra$ are, respectively, a Reynolds number and a Rayleigh based on the small convective scale $\ell$. The $\Re \sim \Ra \Ek/\Pr$ scaling appears to be present for our larger $Pr$ cases over a finite range in $\Rat$; as $\Rat$ is increased a significant departure in this scaling is observed. 

In the present work we investigate the properties of the inverse cascade for varying $\Rat$ and $\Pr$ in the rapidly rotating asymptotic (QG) limit for thermal convection in a plane layer geometry (rotating Rayleigh-B\'enard convection). This choice allows for comparison with previous results from QG \citep{mS06,kJ12,rubio_upscale_2014} and DNS \citep{sS14, cG14} plane-layer calculations.
The paper is organised as follows: in section \ref{S:methods} we describe the governing equations and diagnostic quantities; in section \ref{S:Results} the results of the simulations are presented and analysed; and a discussion is given in section \ref{S:Discussion}.

\section{Methodology}
\label{S:methods}

\subsection{Governing equations}

We consider rotating Rayleigh-B\'enard convection in a plane layer Cartesian geometry of depth $H$, with constant gravity vector $\boldsymbol{g} = - g \hz$ 
pointing vertically downward, perpendicular to the planar boundaries. The fluid is Boussinesq with thermal expansion coefficient $\alpha$. The top boundary is held at constant temperature $T_1$ and the bottom boundary is held at constant temperature $T_2$ such that $\Delta T = T_2 - T_1 > 0$. The system rotates about the vertical with rotation vector $\mathbf{\Omega} = \Omega \hz$. 
In the limit of strong rotational constraint (i.e. small Rossby and Ekman numbers), the governing equations, can be reduced to the following set of equations \citep{kJ06,mS06}
\begin{equation}
\partial_t \zeta + J[\psi,\zeta] - \dz w = \lp \zeta, \label{E:vort0}
\end{equation}
\begin{equation}
\partial_t w + J[\psi,w] +  \dz \psi = \frac{\Rat}{\Pr}  \pth  + \lp w, \label{E:mom0}
\end{equation}
\begin{equation}
\partial_t \pth + J[\psi,\pth] +  w \dz \mth = \frac{1}{\Pr} \lp \pth , \label{E:theat0}
 \end{equation}
\begin{equation}
\dz (\overline{w\pth}) = \frac{1}{\Pr} \dzt \mth , \label{E:Theat0}
 \end{equation}
where $\zeta$ is the vertical component of the vorticity, $\psi$ is the dynamic pressure and also the geostrophic streamfunction, and $w$ is the vertical component of the velocity. The vertical component of vorticity and the streamfunction are related via $\zeta= \nabla^2_\perp \psi$, where $\nabla^2_\perp = \dsx^2 + \dsy^2$ . The non-dimensional temperature $\theta$ is decomposed into mean and fluctuating components $\mth$ and $\pth$, respectively, such that $\theta = \mth +\Ek^{1/3} \pth$. Here the mean is defined as an average over the small  spatial ($x,y,z$) and the fast temporal $(t)$ scales. The Jacobian operator $J[\psi,f] = \partial_x\psi\partial_y f - \partial_y\psi\partial_x f = \textbf{u}_\perp\cdot\nabla_\perp f$ describes advection of the generic scalar field $f$ by the horizontal velocity field $\textbf{u}_\perp = (u, v, 0) = ( -\dsy \psi, \dsx \psi, 0)$. The reduced Rayleigh number $\Rat$ is defined by
\be
\Rat = \Ek^{4/3} \Ra,
\label{eqn:RatRaH}
\ee
where the standard Rayleigh number is
\be
\Ra = \frac{g \alpha \Delta T H^3}{\nu \kappa} .
\ee
Equations \eqref{E:vort0}-\eqref{E:Theat0} have been nondimensionalized by the small-scale viscous diffusion time $\ell^2/\nu$, where the small horizontal convective length scale is $\ell = H \Ek^{1/3}$. The derivation of the reduced system relies on the assumption that the Coriolis force and pressure gradient force are balanced at leading order, i.e.~geostrophic with $\hz\times\textbf{u}=-\nabla_\perp \psi$. 
This force balance implies the Taylor-Proudman constaint is satisfied on small vertical scales such that $\partial_{z} (\textbf{u},\psi)=0$ \citep[e.g.][]{kS79}. Therefore, along the vertical direction all fluid variables vary on an $O(H)$ dimensional scale associated with the coordinate $Z = \Ek^{1/3} z$. As a result, fast inertial waves with dimensional frequency $O(\Omega)$ are filtered from the above equations, allowing for substantial computational savings. However, slow, geostrophically-balanced inertial waves are retained \citep{kJ12}. 

The mean temperature $\mth$ evolves on the slow time-scale $\tau = \Ek^{2/3} t$ associated with the vertical diffusion time $H^2/\nu$. However, \citet{kJ98a,mP18} found that spatial averaging over a sufficient number of convective elements on the small scales is sufficiently accurate to (i) omit fast-time averaging and (ii) assume a statistically stationary state where the slow evolution term   $\partial_\tau \mth$ that would appear in (\ref{E:Theat0})  is omitted.

Finally, we note that three-dimensional incompressibility is invoked through the solenoidal condition for the ageostrophic, sub-dominant horizontal velocity, 
 \be
 \nabla_\perp \cdot\textbf{u}^{ag}_{\perp} + \dz w =0,
 \ee
 where $ \textbf{u}^{ag}_{\perp} = O( \Ek^{1/3} \textbf{u}_{\perp})$.

The equations are solved using impenetrable, stress-free mechanical boundary conditions, and constant temperature boundary conditions. However, it should be noted that the specific form of the thermal boundary conditions are unimportant in the limit of rapid rotation \citep{mC15c}, and the present model can be generalized to no-slip mechanical boundary conditions \citep{kJ16}. Each variable is represented with a spectral expansion consisting of Chebyshev polynomials in the vertical $(Z)$ dimension, and Fourier series in the horizontal $(x, y)$ dimensions. The resulting set of equations are truncated and solved numerically with a pseudo-spectral algorithm that uses a 3rd-order implicit/explicit Runge-Kutta time-stepping scheme \citep{pS91}. The code has been benchmarked successfully and used in many previous investigations \citep{pM16,sM19,mY19}. 

Spatial and temporal resolutions are given in table 1. The horizontal dimensions of the domain are periodic and scaled by the critical horizontal wavelength $\lambda_c = 2 \pi/k_c \approx 4.8154$, measured in small-scale units $\ell$. Most of the simulations use horizontal dimensions of $10\lambda_c \times 10\lambda_c$, though some additional simulations with different domain size were also carried out to quantify the influence of the geometry. We find that a domain size of $10\lambda_c \times 10\lambda_c$ is sufficient for accurate computation of statistical quantities, though the role of LSVs appears to become increasingly important with increasing domain size; we discuss this effect in our results.

\subsection{Depth-averaged dynamics and energetics}
\label{sed:dept-averaged}
For the purpose of investigating the inverse energy cascade, we decompose the vertical vorticity into a depth-averaged (barotropic) component, $\langle \zeta \rangle$, and a fluctuating (baroclinic) component, $\zeta'$, such that 
\be
\zeta = \langle \zeta \rangle + \zeta', 
\ee
where, by definition, $ \langle \zeta' \rangle=0$. 
The depth-averaged (barotropic) vorticity equation is then found by vertically-averaging equation \eqref{E:vort0}, and is given by
\be
\partial_t \langle \zeta\rangle + J[\langle \psi \rangle, \langle \zeta\rangle ] = - \langle J[\psi', \zeta' ] \rangle + \nabla_\perp^2 \langle \zeta\rangle .
\label{E:baro}
\ee
Thus, the barotropic dynamics are governed by a two-dimensional vorticity equation in which the sole forcing comes from convective dynamics, represented by the first term on the righthand side of the above equation. 

The barotropic, time-dependent kinetic energy density is defined as follows:
\be
K_{bt}(t) =\frac{1}{2} \left ( \overline{\langle u \rangle^2 +\langle v \rangle^2}^\mathcal{V} \right)  =\frac{1}{2}  \overline{ \vert \nabla_\perp \langle\psi\rangle \vert^2}^\mathcal{V}.
\label{eqn:Kbt}
\ee
where $ \overline{ \ \cdot \ } ^\mathcal{V}$ indicates an average over the small, horizontal spatial scales, consistent with the notation employed in \citet{mP18}. 
In Fourier space, the barotropic kinetic energy equation is derived by multiplying the Fourier representation of equation \eqref{E:baro} by the complex conjugate of $-\langle \psi \rangle_{\mathbf{k}} \exp{(i \mathbf{k} \cdot \mathbf{x})}$, the spectral representation of $\langle \psi \rangle$, and integrating over physical space to obtain
\begin{equation}
\dst\Kbt(k) = T_k + F_k + D_k,
\label{eqn:dt_Kbt}
\end{equation}
where the box-normalized horizontal wavenumber vector is $\mathbf{k} = (k_x, k_y, 0)$, and $k = |\mathbf{k}|$. This equations describes the evolution of the kinetic energy contained in the barotropic mode of wavenumber $k$ that is due to (1) the interaction with the other barotropic modes, 
\be
T_k = \sum_{|\textbf{k}|=k}\textrm{Re}\left\{\left<\psi\right>^*_{\textbf{k}} \circ \mathcal{F}_\textbf{k}\left[ J[\left<\psi\right>,\left<\zeta\right>] \right]\right\};
\ee
(2) the interaction with the baroclinic, convective modes, 
\be
F_k = \sum_{|\textbf{k}|=k}\textrm{Re}\left\{\left<\psi\right>^*_{\textbf{k}} \circ\mathcal{F}_\textbf{k}\left[ \left<J[\psi',\zeta']\right> \right] \right\};
\ee
and (3) the viscous dissipation of the barotropic mode, 
\be
D_k = \sum_{|\textbf{k}|=k}\textrm{Re}\left\{|\textbf{k}|^2 \left<\psi\right>^*_{\textbf{k}}\circ\left<\zeta\right>_{\textbf{k}} \right\} .
\ee
In the above definitions, the superscript $*$ denotes a complex conjugate, $\mathcal{F}_\textbf{k}[\cdot]$ indicates the horizontal Fourier transform of the argument in square brackets, the symbol $\circ$ indicates a Hadamard (element-wise) product, $\textrm{Re}\left\{\cdot\right\}$ is the real part of the argument in curly brackets and the sum is taken over all horizontal wavenumbers. The barotropic-to-barotropic and baroclinic-to-barotropic transfer functions $T_k$ and $F_k$ can be explicitly expressed in terms of a triadic interaction due to the Jacobian (i.e.~non-linear) terms \citep{rubio_upscale_2014}. The formation of LSVs is due to a positive contribution from $T_k$ and $F_k$ in equation \eqref{eqn:dt_Kbt} at the domain-scale wavenumber $k=1$. As the LSV forms, the kinetic energy grows in time until dissipation balances the positive transfer at $k=1$. Eventually a statistically stationary state is reached where $\overline{D}_k \approx - (\overline{T}_k + \overline{F}_k)$, where we notice that, for these quantities, $\overline{ \ \cdot \ }$ is equivalent to an average over the fast temporal scale only; 
in contrast with previous work, all of the simulations presented here have reached this stationary state. Hereafter, in order to simplify notation we omit the averaging operator and refer only to the time-averaged values of $\Kbt$, $T_k$, $F_k$, and $D_k$, unless otherwise stated.

\subsection{Diagnostic quantities}

Here we define several diagnostic quantities that will be used to characterize the dynamical state of the convective system. The heat transfer across the fluid layer is quantified by the non-dimensional Nusselt number
\be
 \Nu = 1 + \Pr \overline{ \langle w\theta \rangle } .
 \label{eqn:Nu_def}
\ee
In the present study the small-scale, or convective, Reynolds number is defined as 
\be
\Ret = \frac{\left<W_{rms}\right> \ell}{\nu} = \left<w_{rms}\right>, 
\ee
where $W_{rms} = (\overline{W^2})^{1/2}$ and $w_{rms}= (\overline{w^2})^{1/2}$ are the rms values of the dimensional and non-dimensional vertical velocity component, respectively. The above definition is particularly useful for characterizing the amplitude of the convective motions, rather than the large-amplitude horizontal motions that occur in the presence of a strong inverse cascade. We also find it useful to refer to instantaneous values of the Nusselt and Reynolds number, and denote these by $Nu(t)$ and $\Ret(t)$, respectively.


Together with the barotropic kinetic energy \eqref{eqn:Kbt} we will also consider the time-averaged baroclinic, vertical and total kinetic energy densities, respectively defined as: 

\be
\Kbc =\frac{1}{2}\langle  \overline{ (u')^2 + (v)'^2}\rangle  =\frac{1}{2}  \langle \overline{ \vert \nabla_\perp \psi' \vert^2}\rangle;
\ee
\be
K_z = \frac{1}{2}\langle \overline{ w^2} \rangle ,
\ee
\be
K =\frac{1}{2} \langle \overline{u^2 + v^2 +w^2 }\rangle  =\frac{1}{2}  \langle \overline{ \vert \nabla_\perp \psi\vert^2} \rangle + K_z.
\ee
With the above definitions, the Reynolds number can be expressed as $ \Ret = \sqrt{2 K_z}$. As for $Nu$ and $\Ret$, we find it useful to refer to the instantaneous values of the total kinetic energy density as $K(t)$.



\section{Results}
\label{S:Results}

\subsection{Flow morphology: two-scale flows}

\begin{figure*}
\centering
\subfloat[][]
{\includegraphics[width=0.45\textwidth]{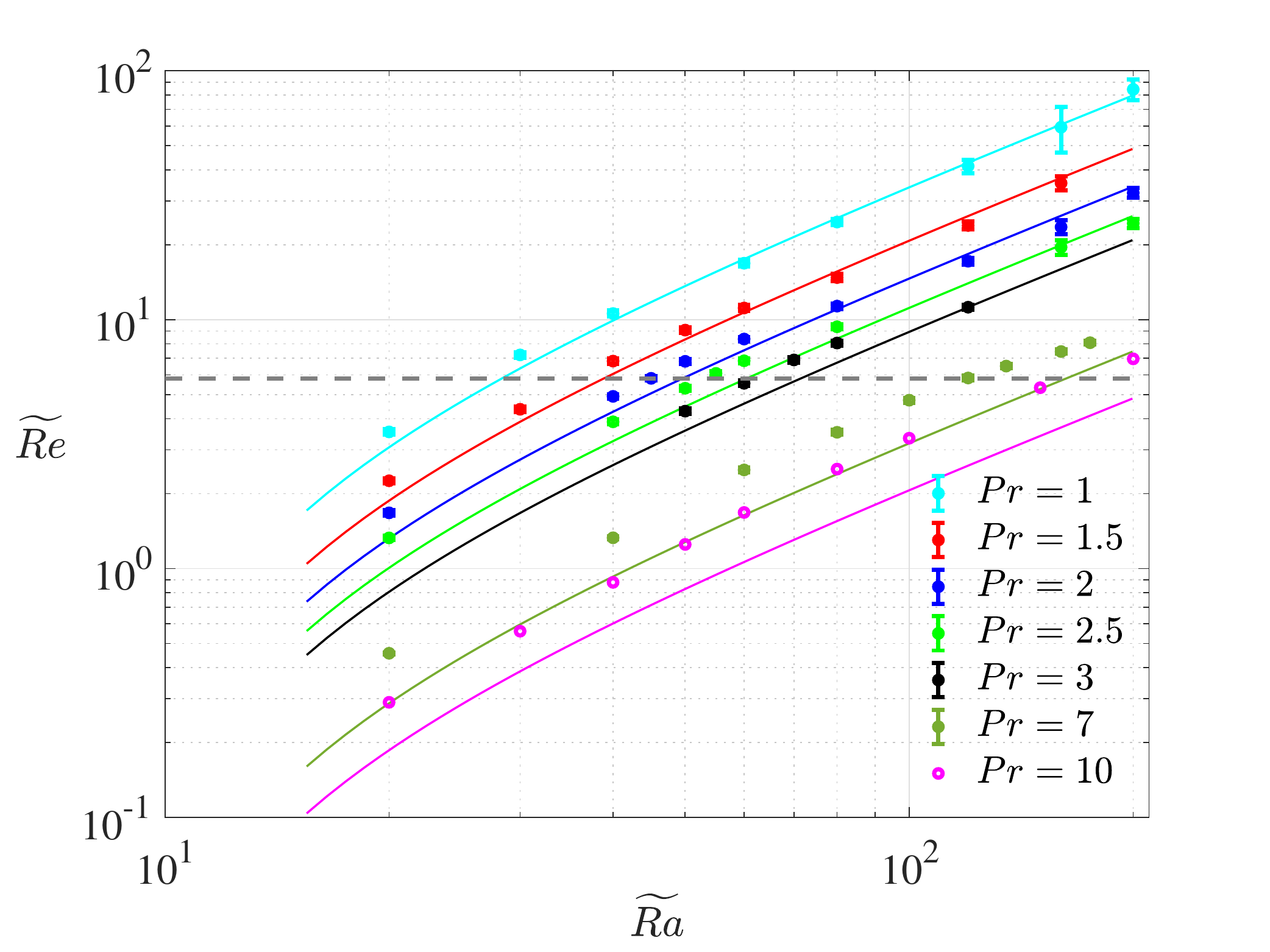}\label{fig:ReRa}}\quad
\subfloat[][]
{\includegraphics[width=0.45\textwidth]{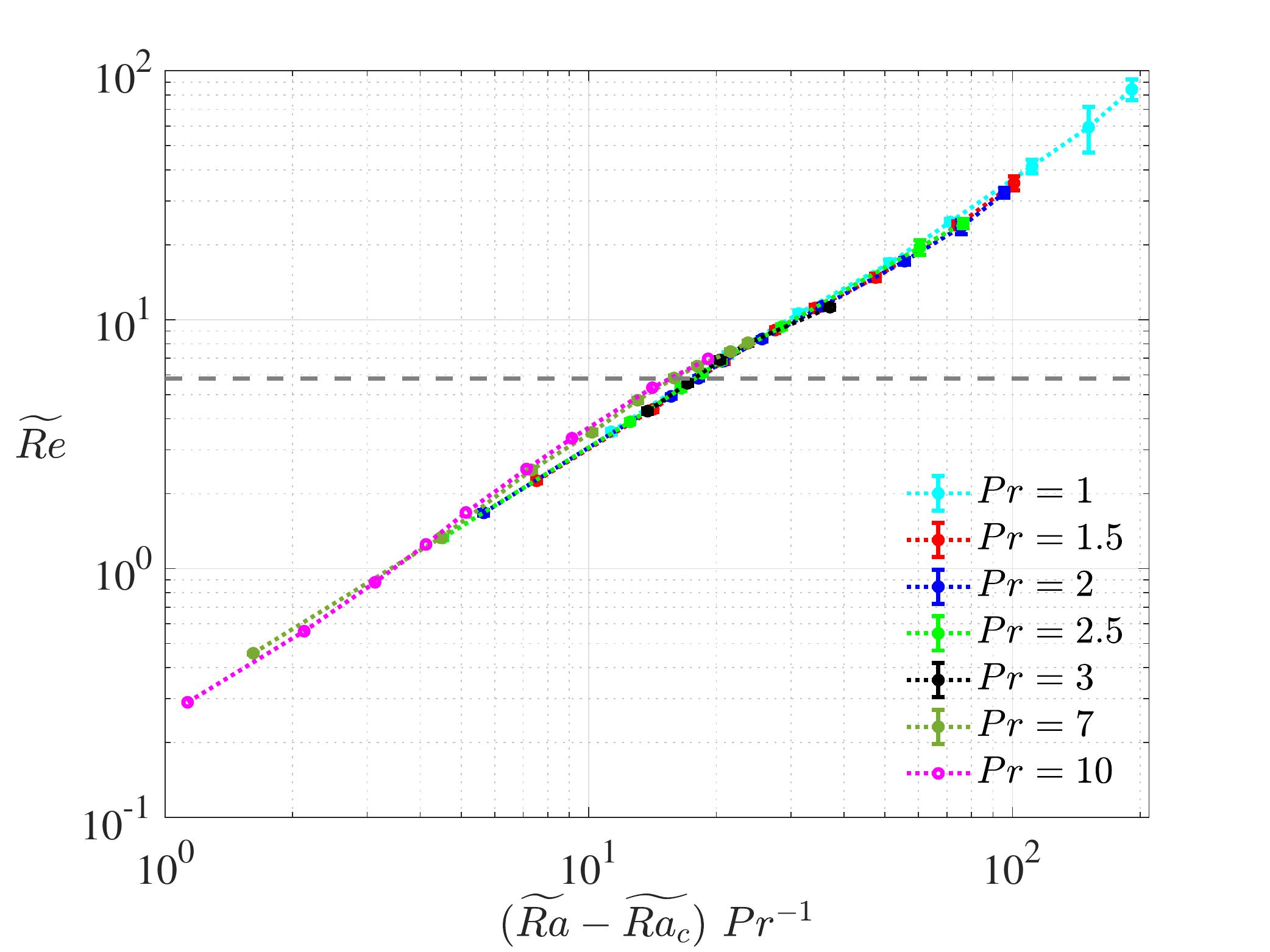}\label{fig:ReRa_over_Pr}}\\
\subfloat[][]
{\includegraphics[width=0.45\textwidth]{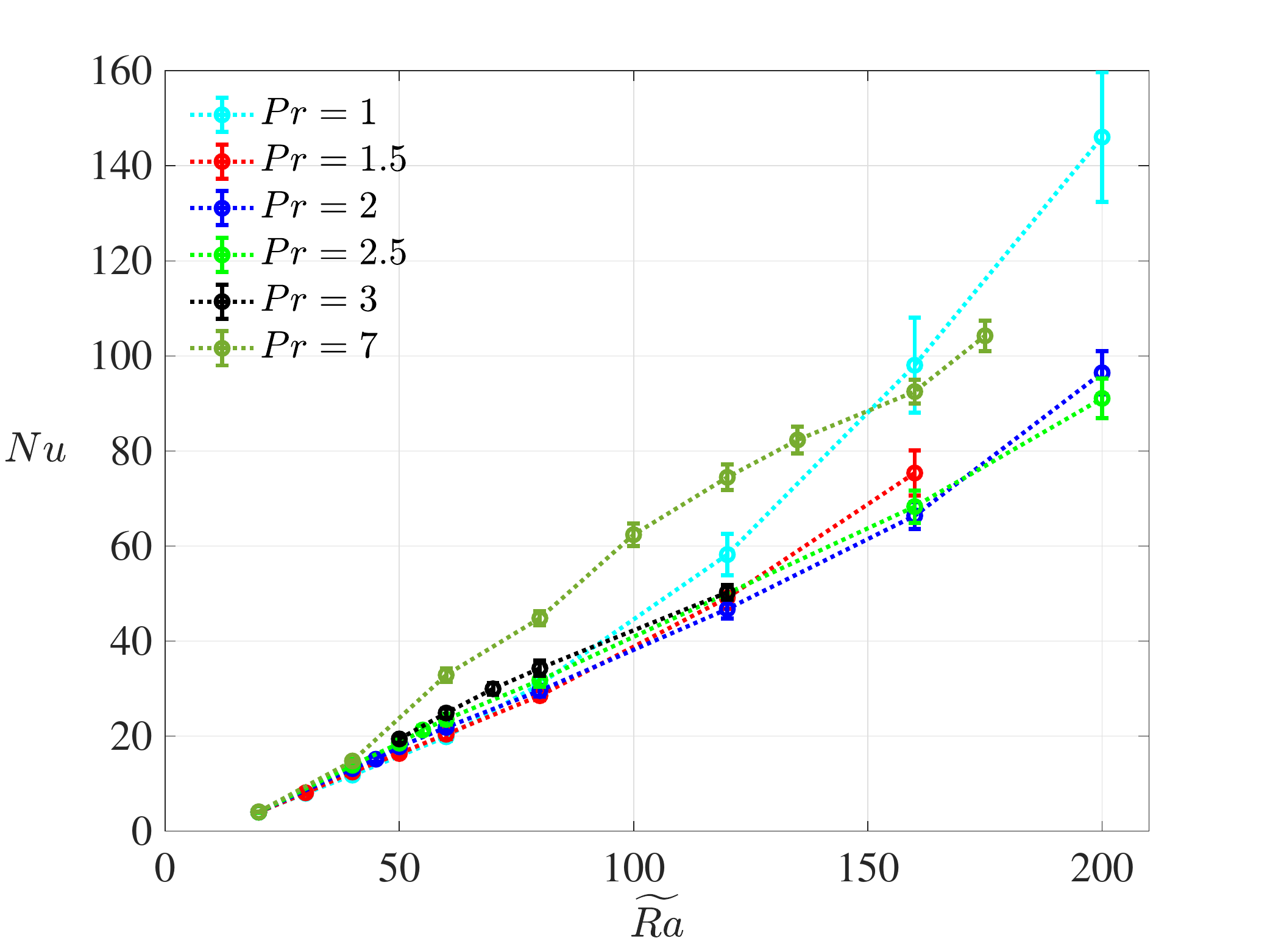}\label{fig:NuRa}}\quad
\subfloat[][]
{\includegraphics[width=0.45\textwidth]{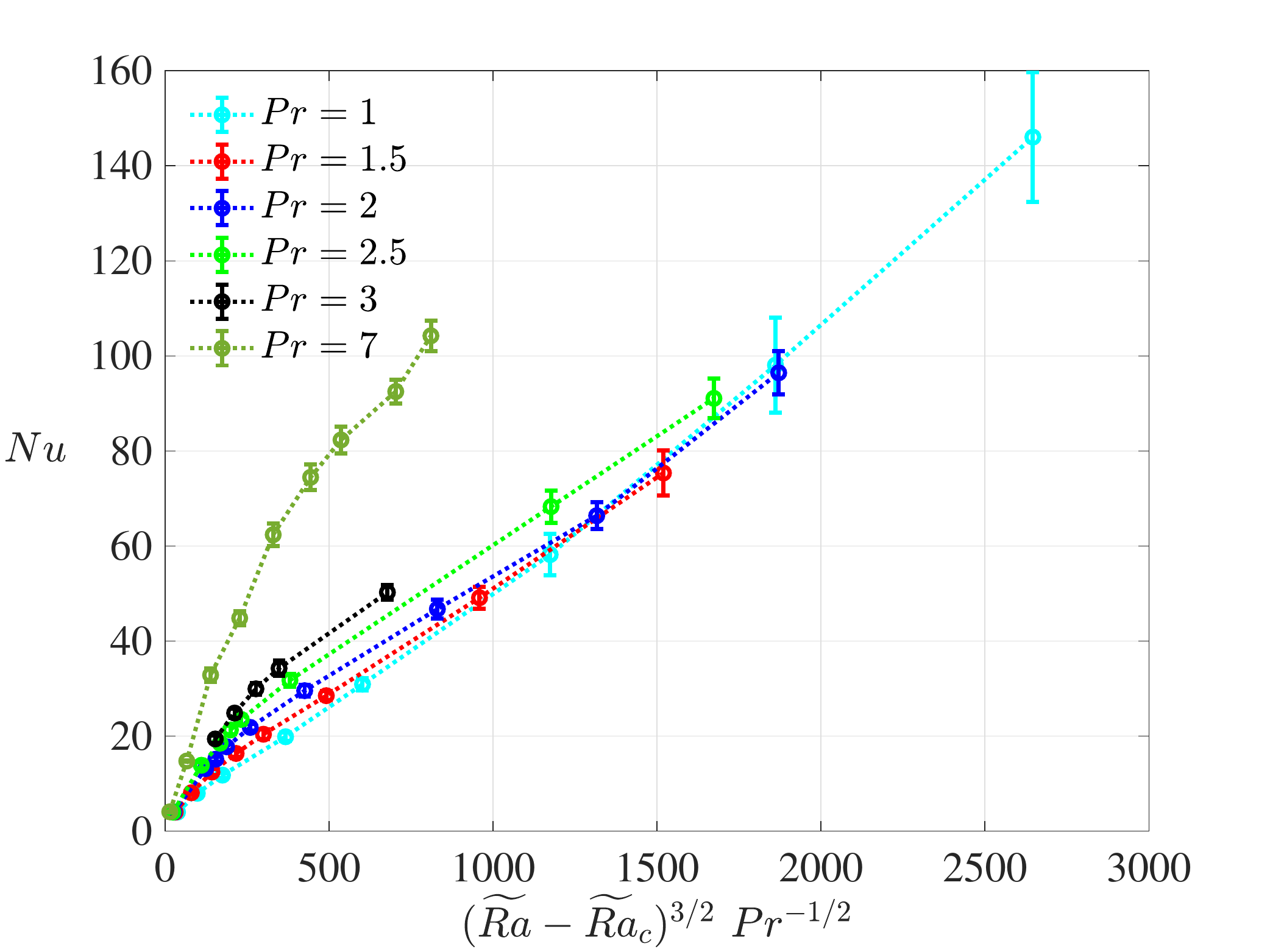}
\label{fig:NuRePr}}
\caption{(a,b) Temporally averaged Reynolds number $\Retm $ and (c,d) Nusselt number $\Num$ for all of the simulations. (a) $\Retm$ versus the reduced Rayleigh number $\Rat$; (b) $\Retm$ versus the rescaled quantity $(\Rat-\Rat_c)/Pr$. (c) $\Num$ versus $\Rat$; (d) $\Num$ versus the rescaled coordinate 
$(\Rat-\Rat_c)^{3/2} Pr^{-1/2}$. Continuous lines in (a) show the best-fit, three parameter power-law scaling $\Retm = \alpha_r (\Rat-\Rat_c)^{\beta_r} \Pr^{\gamma_r}$, where $\alpha_r =  \alphaRaNum $, $\beta_r =  \betaRaNum $ and $\gamma_r =  \gammaRaNum$ (see Section \ref{s:Scaling}). Data for $\Pr = 10$ in (a,b) are from \citet{mC16}. The dashed horizontal line in (a,b) show the Reynolds number at which the box-scale depth averaged kinetic energy becomes dominant (see Section \ref{subsec:LSV}).}       
\label{fig:Re_Nu}
\end{figure*}

The details of the simulations performed for this study are given in table \ref{tab:results}. The choice of parameters allows us to refine the results of previous QG calculations \citep{kJ12} in the range $1\le\Pr\le3$ and for $\Pr=7$, of particular relevance for laboratory experiments. The temporally averaged values of $\Ret$ and $\Nu$  displayed in table \ref{tab:results} are calculated over a temporal window in which the system reached a statistically stationary state.
\afterpage{
\clearpage
\setlength{\tabcolsep}{0.2cm} 


\begingroup
\renewcommand\arraystretch{1.1}

\begin{longtable}{lccccc}

\hline\\[-1.4ex]
	$\Pr $ & $ \Rat $ & $ N_x \times N_y \times N_z $ &  $ \Delta t $ & $\Retm \pm \sigma_{\Retm} $ & $\Num \pm \sigma_{\Num}$ \\[1ex] 
  	\hline
	\hline
	$1$ & $20$ & $128 \times 128 \times 64$ & $5 \times 10^{-3}$ & $3.5408 \pm 0.0860$ & $4.010 \pm 0.130$\\
	$1$ & $30$ & $128 \times 128 \times 64$ & $5 \times 10^{-4}$ & $7.2190 \pm 0.1805$ & $7.9603 \pm 0.3688$\\
	$1$ & $40$ & $256 \times 256 \times 64$ & $5 \times 10^{-4}$ & $10.586 \pm 0.307$ & $11.788 \pm 0.569$\\
	$1$ & $60$ & $256 \times 256 \times 96$ & $1 \times 10^{-4}$ & $16.822 \pm 0.558$ & $19.961 \pm 0.953$\\
	$1$ & $80$ & $384 \times 384 \times 128$ & $5 \times 10^{-5}$ & $24.685 \pm 0.711$ & $30.92 \pm 1.25$\\
	$1$ & $120$ & $384 \times 384 \times 192$ & $5 \times 10^{-5}$ & $41.40 \pm 2.59$ & $58.20 \pm 4.32$\\
	$1$ & $160$ & $256 \times 256 \times 256$ & $1 \times 10^{-4}$ & $59.4 \pm 12.4$ & $98.06 \pm 9.95$\\
	$1$ & $200$ & $384 \times 384 \times 384$ & $5 \times 10^{-5}$ & $84.21 \pm 6.76$ & $146.24 \pm 12.38$\\
	\hline
	$1.5$ & $20$ & $128 \times 128 \times 96$ & $1 \times 10^{-3}$ & $2.2527 \pm 0.0482$ & $3.990 \pm 0.108$\\
	$1.5$ & $30$ & $128 \times 128 \times 96$ & $1 \times 10^{-3}$ & $4.366 \pm 0.103$ & $8.075 \pm 0.315$\\
	$1.5$ & $40$ & $192 \times 192 \times 122  $ & $5 \times 10^{-4}$ & $6.7529 \pm 0.227$ & $12.439 \pm 0.584$\\
	$1.5^*$ & $50^*$ & $164 \times 164 \times 108 $ & $5 \times 10^{-4}$ & $9.082 \pm 0.239$ & $16.348 \pm 0.758$\\
	$1.5$ & $60$ & $192 \times 192 \times 136 $ & $5 \times 10^{-4}$ & $11.0592 \pm 0.434$ & $20.39 \pm 1.01$\\
	$1.5$ & $80$ & $244 \times 244 \times 136 $ & $1 \times 10^{-4}$ & $14.772 \pm 0.506$ & $28.51 \pm 0.999$\\
	$1.5$ & $120$ & $256 \times 256 \times 256$ & $1 \times 10^{-4}$ & $23.899 \pm 0.881$ & $49.11 \pm 2.32$\\
	$1.5$ & $160$ & $256 \times 256 \times 256$ & $1 \times 10^{-4}$ & $35.39 \pm 2.30$ & $75.37 \pm 4.76$\\
	\hline
	$2$ & $20$ & $128 \times 128 \times 72 $ & $1 \times 10^{-3}$ & $1.6744 \pm 0.0442$ & $4.0143 \pm 0.0852$\\
	$2$ & 40 & $192 \times 192 \times 122 $ & $5 \times 10^{-4}$ & $4.9194 \pm 0.0965$ & $13.149 \pm 0.427$\\
	$2^*$ & $45^*$ & $192 \times 192 \times 128$ & $1 \times 10^{-4}$ & $5.812 \pm 0.124$ &  $15.142 \pm 0.525$\\
	$2$ & $50$ & $192 \times 192 \times 128$ & $1 \times 10^{-4}$ & $6.803 \pm 0.164$ & $17.749 \pm  0.674$\\
	$2$ & $60$ & $192 \times 192 \times 136 $ & $1 \times 10^{-4}$ & $8.361 \pm 0.171$ & $21.39 \pm 1.05$\\
	$2$ & $80$ & $192 \times 192 \times 136 $ & $1 \times 10^{-4}$ & $11.342 \pm 0.413$ & $29.56 \pm 1.28$\\
	$2$ & $120$ & $404 \times 404 \times 256 $ & $1 \times 10^{-4}$ & $17.186 \pm 0.531$ & $46.74 \pm 1.96$\\
	$2$ & $160$ & $224 \times 224 \times 256 $ & $1 \times 10^{-4}$ & $23.57 \pm 1.52$ & $66.38 \pm 2.81$\\
	$2$ & $200$ & $256 \times 256 \times 256$ & $1 \times 10^{-4}$ & $30.4862 \pm 1.66$ & $93.80 \pm 4.14$\\
	\hline
	$2.5$ & $20$ & $128 \times 128 \times 72 $ & $1 \times 10^{-3}$ & $1.3277 \pm 0.0276$ & $3.994 \pm 0.108$\\
	$2.5$ & $40$ & $192 \times 192 \times 122 $ & $5 \times 10^{-4}$ & $3.8869 \pm 0.0903$ & $13.863 \pm 0.665$\\
	$2.5^*$ & $55^*$ & $216 \times 216 \times 128$ & $1 \times 10^{-4}$ & $6.067 \pm 0.134$ & $21.286 \pm 0.850$\\
	$2.5^*$ & $60^*$ & $216 \times 216 \times 128$ & $5 \times 10^{-4}$ & $ 6.837 \pm 0.168$ & $23.50 \pm 1.01$\\
	$2.5$ & $80$ & $216 \times 216 \times 136 $ & $1 \times 10^{-4}$ & $9.364 \pm 0.236$ & $31.75 \pm 1.31$\\
	$2.5$ & $160$ & $216 \times 216 \times 256$ & $1 \times 10^{-4}$ & $19.55 \pm 1.35$ & $68.3 \pm 3.39$\\
	$2.5$ & $200$ & $256 \times 256 \times 256$ & $1 \times 10^{-4}$ & $24.7205 \pm 1.55$ & $91.00 \pm 5.42$\\
	\hline
	$3$ & $50$ & $288 \times 288 \times 128$ & $5 \times 10^{-4}$ & $4.290 \pm  0.106$ & $19.402 \pm 0.770$\\
	$3^*$ & $60^*$ & $192 \times 192 \times 128$ & $1 \times 10^{-4}$ & $5.551 \pm 0.139$ & $24.853 \pm 0.943$\\
	$3$ & $70$ & $512 \times 512 \times 128$ & $1 \times 10^{-4}$ & $ 6.919\pm0.155 $ & $ 30.14 \pm 1.26$\\
	$3^*$ & $80^*$ & $256 \times 256 \times 128$ & $1 \times 10^{-4}$ & $8.060 \pm 0.184$ & $34.27 \pm 1.58$\\
	$3$ & $120$ & $216 \times 216 \times 186 $ & $1 \times 10^{-4}$ & $11.226 \pm 0.246$ & $50.24 \pm 1.55$\\
	\hline
	$7$ & $20$ & $(512 \times 512 \times 32)^\dagger$ & $1 \times 10^{-2}$ & $0.45662 \pm 0.00387$ & $4.0696 \pm 0.0369$\\
	$7$ & $40$ & $256 \times 256 \times 32$ & $1 \times 10^{-3}$ & $1.3293 \pm 0.0179$ & $14.761 \pm 0.119$\\
	$7$ & $60$ & $256 \times 256 \times 64$ & $1 \times 10^{-3}$ & $2.4907 \pm 0.0522$ & $32.86 \pm 1.42$\\
	$7$ & $80$ & $256 \times 256 \times 64$ & $2 \times 10^{-4}$ & $3.5423 \pm 0.0792$ & $44.73 \pm 1.55$\\
	$7$ & $100$ & $384 \times 384 \times 92$ & $2 \times 10^{-4}$ & $4.748 \pm 0.108$ & $62.39 \pm 2.37$\\
	$7$ & $120$ & $384 \times 384 \times 92$  & $2 \times 10^{-4}$ & $5.831 \pm 0.122$ & $74.46 \pm 2.67$\\
	$7$ & $135$ & $384 \times 384 \times 92$ & $1 \times 10^{-4}$ & $6.150 \pm 0.133$ & $82.32 \pm 2.80$\\
	$7$ & $160$ & $512 \times 512 \times 92$ & $1 \times 10^{-4}$ & $7.448 \pm 0.138$ & $92.51 \pm 2.51$\\
	\hline
	\hline

\caption{Details of the numerical simulations. $\Pr$ is the Prandtl number; $\Rat$ is the reduced Rayleigh number; $N_x$, $N_y$ and $N_Z$ are, respectively, the number of Fourier modes in the $x$ and $y$ directions and the number of Chebyshev modes in the $Z$ direction; $\Delta t$ is the timestep size used during the simulation; $\Retm = \overline{\left<w_{rms}\right>}$ is the time-averaged, reduced Reynolds number based on the vertical component of the velocity; $\Num$ is the time-averaged Nusselt number;  $\sigma_{\Retm}$ and $\sigma_{\Num}$ are the standard deviations of $\Retm$ and $\Num$, respectively. The superscript $\dagger$ indicates that the horizontal box size for the simulation is taken to be $20 \lambda_c \times 20 \lambda_c$, where $\lambda_c = 2\pi/k_c$ is the critical wavelength for the onset of thermal convection; for all other cases the  box size is $10 \lambda_c \times 10 \lambda_c$. The superscript $*$ indicated cases for which the influence of the lack or presence of an LSV in the initial condition on the saturated state has been checked (see section \ref{s:IC} for details).
} 
\label{tab:results}
\end{longtable}

\endgroup

} 
If an LSV is present in the domain, $\Ret$ and $\Nu$ might reach stationary values only when the barotropic kinetic energy has saturated, in accordance with previous studies \citep{kJ12, bF14, cG14, rubio_upscale_2014} where $\Nu$ has been shown to evolve over the time needed for the total kinetic energy to saturate. The interested reader is directed to supplemetary figure 1 for an example of this behavior. Figures \ref{fig:ReRa} and \ref{fig:NuRa} shows $\Retm$ and $\Num$ as functions of $\Rat$ and $\Pr$. The continuous lines in figure \ref{fig:ReRa} indicate the least-square fit to a power law of the kind $\Retm = \alpha_r (\Rat - \Rat_c)^{\beta_r} \Pr^{\gamma_r}$ with $\alpha_r = \alphaRaNum $, $\beta_r = \betaRaNum$ and $\gamma_r = \gammaRaNum$. In figure \ref{fig:ReRa_over_Pr} we illustrate the collapse of the $\Retm$ data points to the law $\Retm\sim (\Rat-\Rat_c) \Pr^{-1}$, empirically found and consistent with the the coefficients $\beta_r$ and $\gamma_r$. Figure \ref{fig:ReRa_over_Pr} suggests that the reduced Grashof number, $\Rat \Pr^{-1}$ plays a key role in controlling the dynamics. Figure \ref{fig:NuRePr} shows the collapse of the $\Num$ data points to a power law of the kind $\Num\sim (\Rat-\Rat_c)^{3/2} \Pr^{-1/2}$, distinctive of the ultimate regime of thermal convection.  Further details concerning these fits are given in section \ref{s:Scaling}, but are shown here to summarize the cases that were computed.

\begin{figure*}
\centering

\subfloat[][$\Rat = 40, \Pr = 2 $ (CTC/P)]
{\includegraphics[width=0.3\textwidth,trim={0 0 10cm 0},clip]{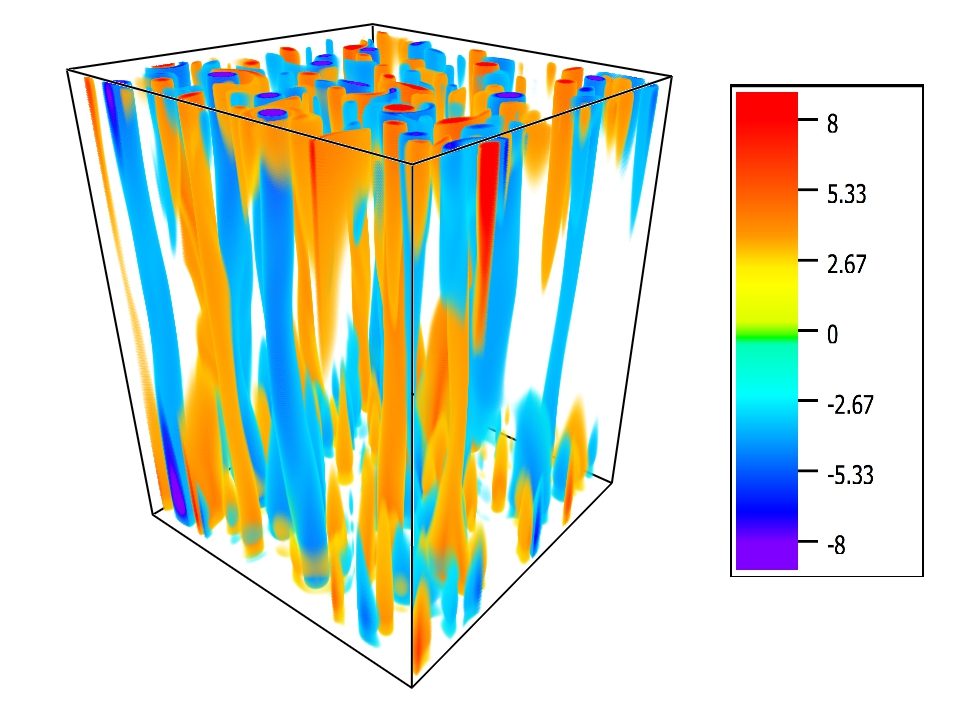}}\quad
\subfloat[][$\Rat = 60, \Pr = 2 $ (P)]
{\includegraphics[width=0.3\textwidth,trim={0 0 10cm 0},clip]{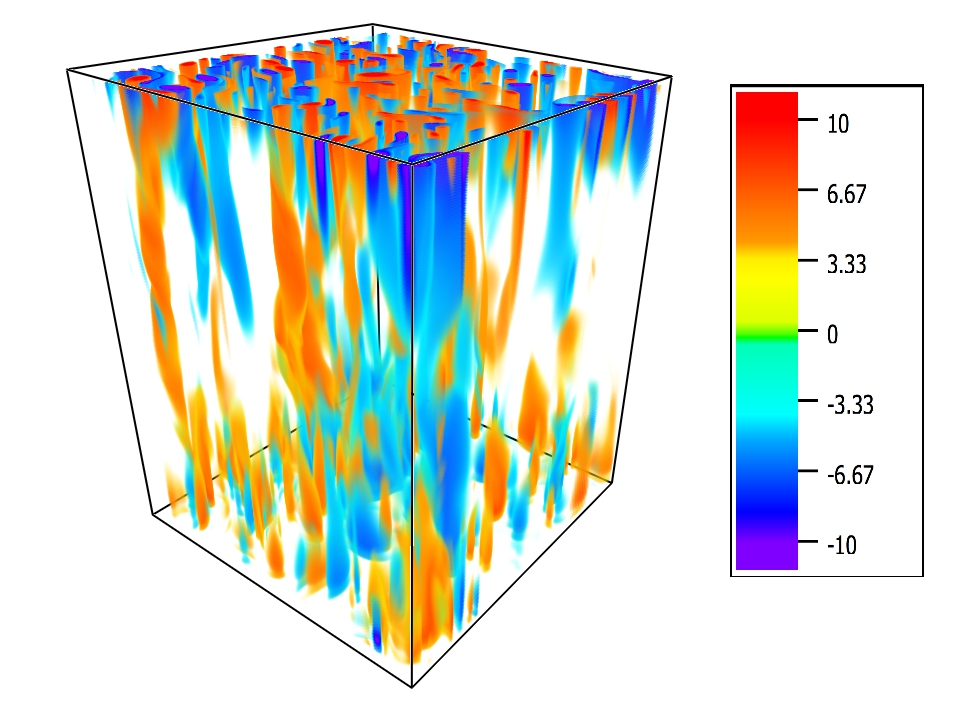}}\quad
\subfloat[][$\Rat = 200, \Pr = 2 $ (G)]
{\includegraphics[width=0.3\textwidth,trim={0 0 10cm 0},clip]{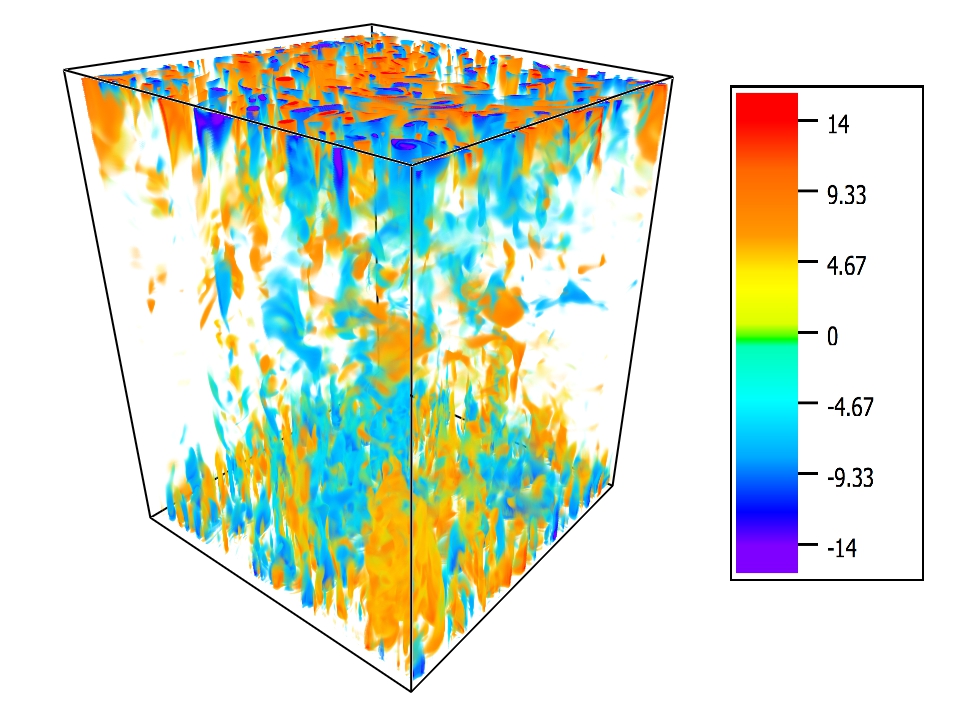}}

\subfloat[][$\Rat = 60, \Pr = 3 $ (CTC/P)]
{\includegraphics[width=0.3\textwidth,trim={0 0 10cm 0},clip]{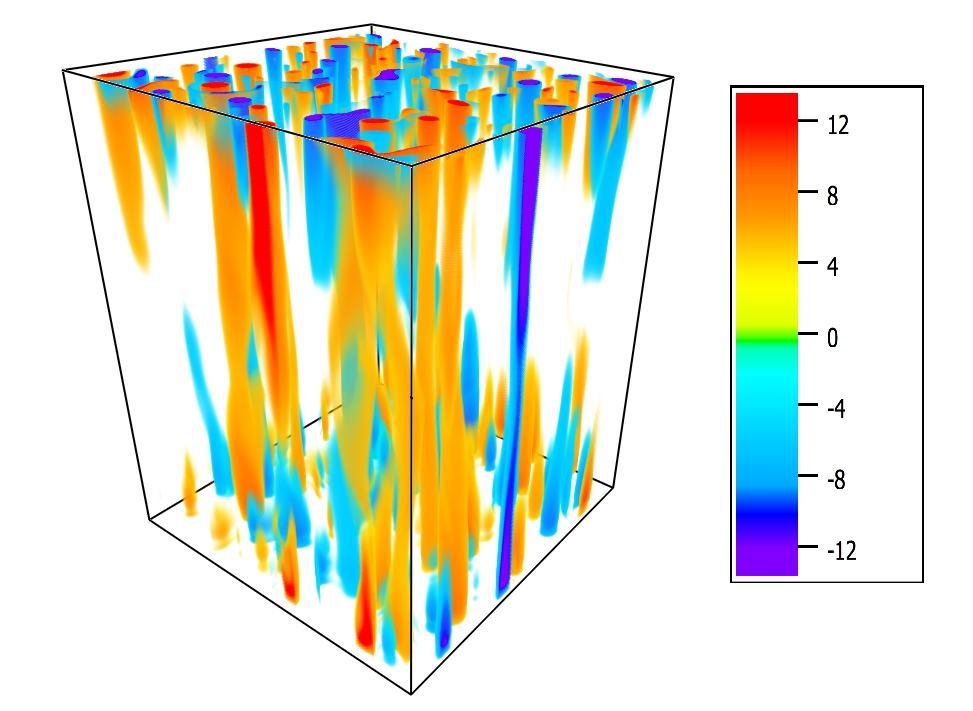}}\quad
\subfloat[][$\Rat = 80, \Pr = 3 $ (P)]
{\includegraphics[width=0.3\textwidth,trim={0 0 10cm 0},clip]{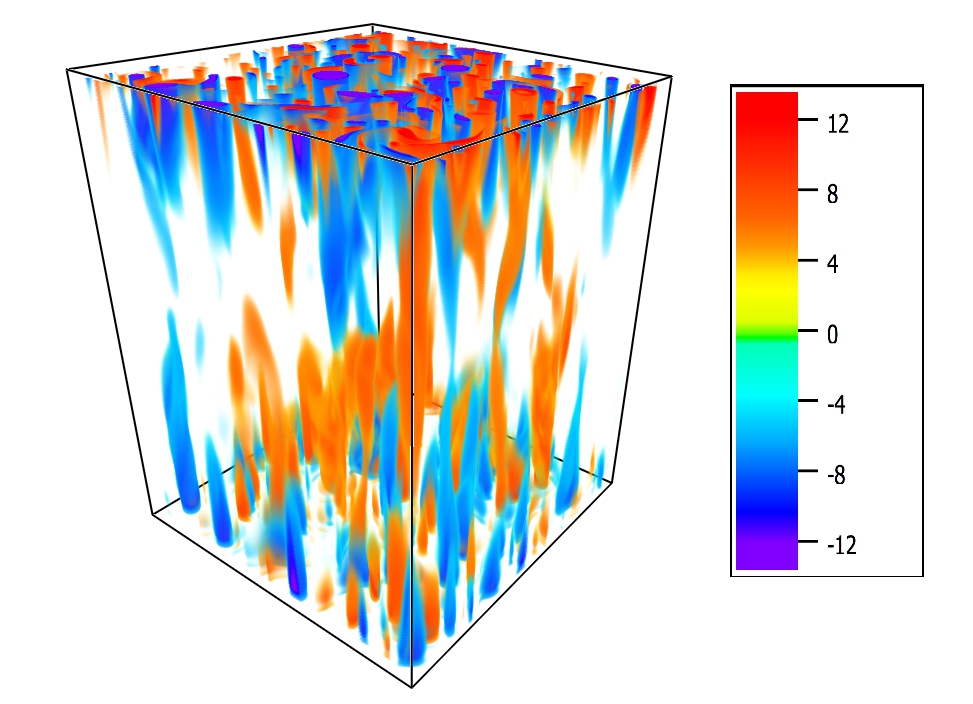}}\quad
\subfloat[][$\Rat = 120, \Pr = 3 $ (P)]
{\includegraphics[width=0.3\textwidth,trim={0 0 10cm 0},clip]{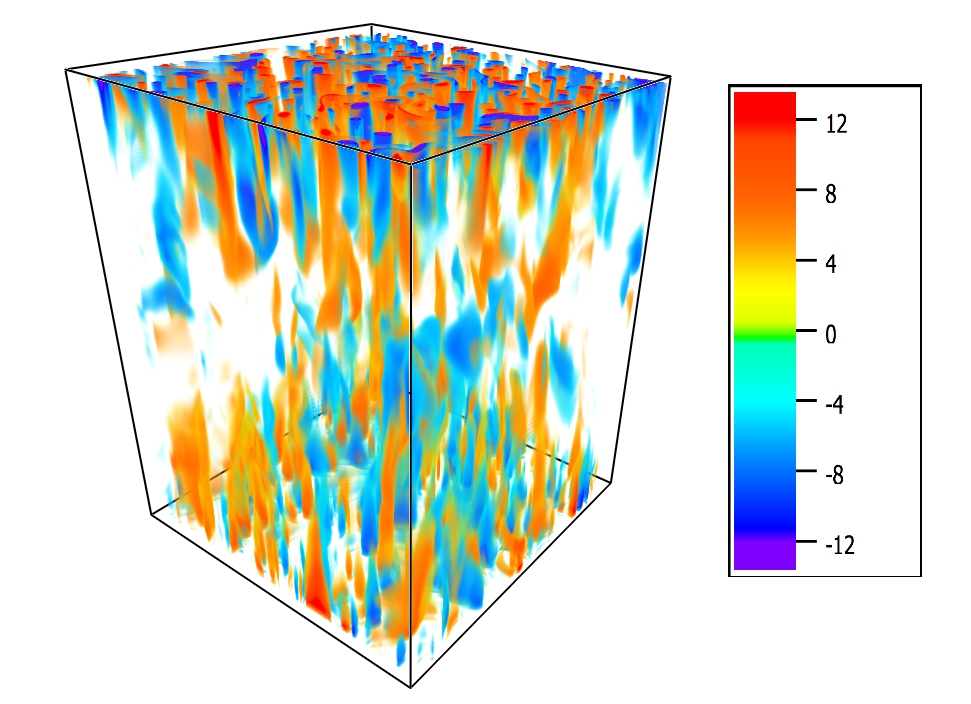}}\\

\subfloat[][$\Rat = 80, \Pr = 7 $ (CTC)]
{\includegraphics[width=0.3\textwidth,trim={0 0 10cm 0},clip]{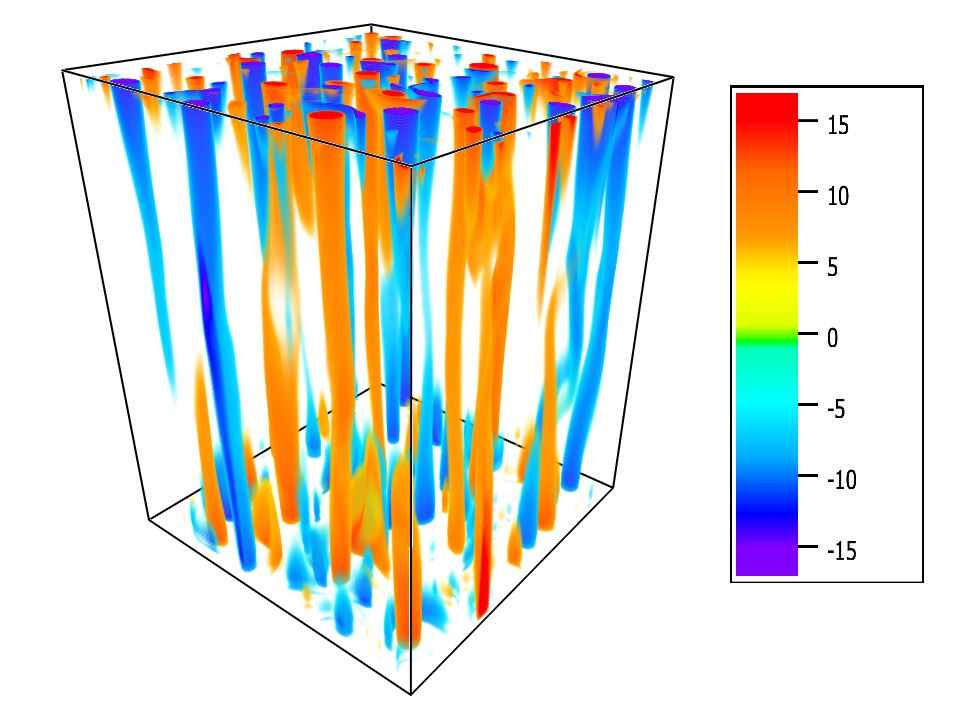}}\quad
\subfloat[][$\Rat = 135, \Pr = 7 $ (P)]
{\includegraphics[width=0.3\textwidth,trim={0 0 10cm 0},clip]{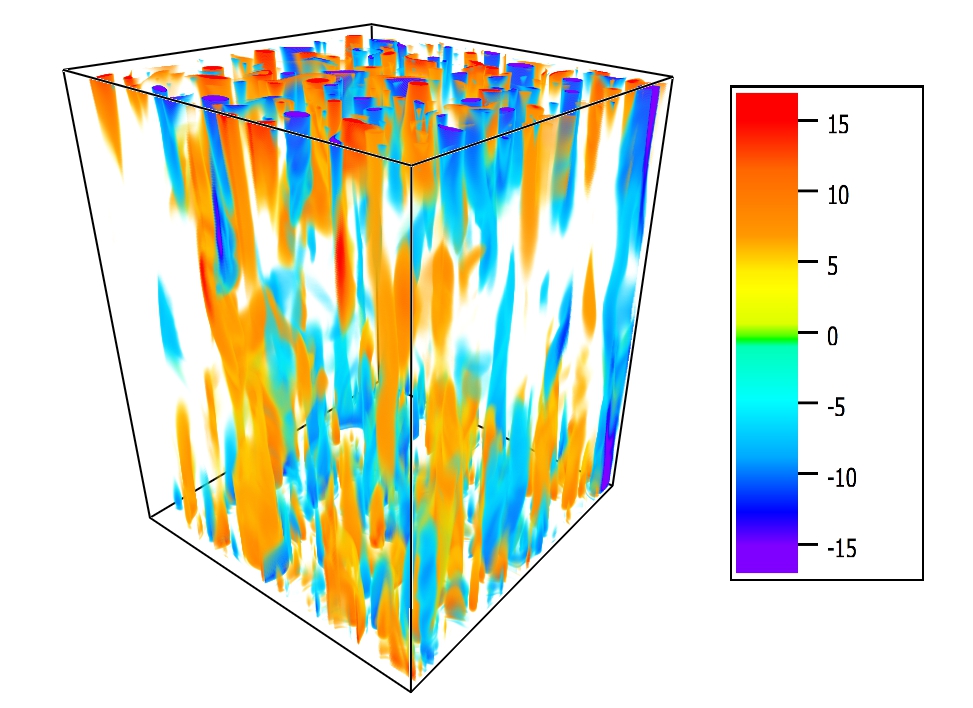}}\quad
\subfloat[][$\Rat = 160, \Pr = 7 $ (P)]
{\includegraphics[width=0.3\textwidth,trim={0 0 10cm 0},clip]{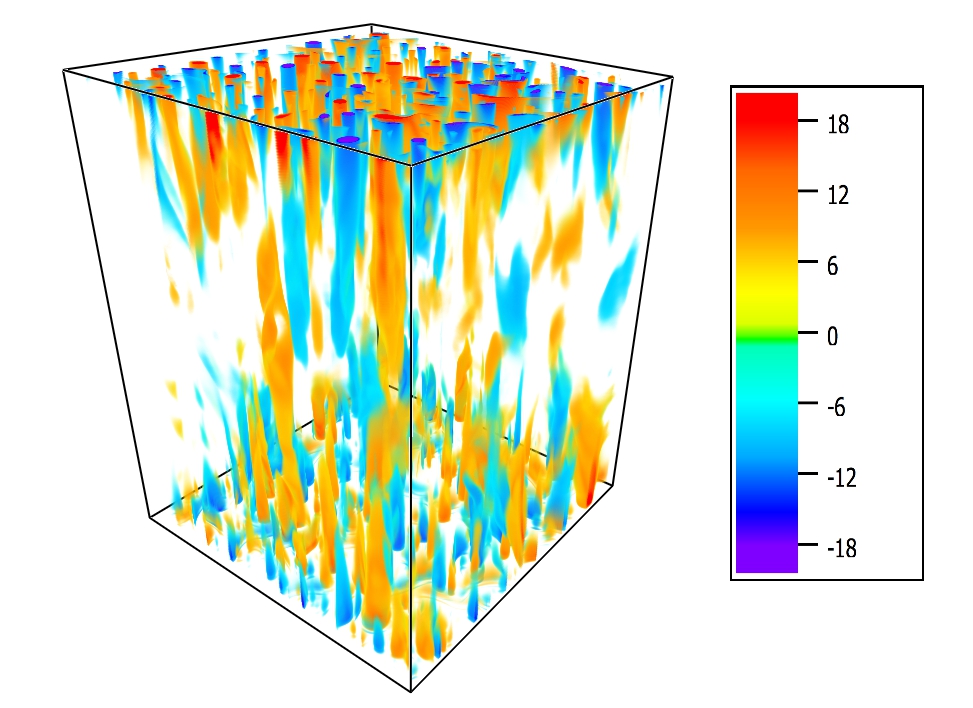}}\quad

\caption{Volumetric renderings of fluctuating temperature, $\pth$, showing the different convective regimes for increasing Rayleigh number (left to right) and increasing Prandtl number (top to bottom). The abbreviations correspond to: convective Taylor column (CTC); plume (P); geostrophic turbulence (G). See online supplementary material for movies illustrating the temporal evolution of the fluctuating temperature for selected values of $\Rat$ and $\Pr$.}
\label{fig:temp_visualization}
\end{figure*}

Following \citet[][]{kJ12}, inspection of volumetric renderings of the fluctuating temperature (figure \ref{fig:temp_visualization}) suggests that we can qualitatively classify the flows into: the cellular regime (C); the convective Taylor column regime (CTC); the plume regime (P); and the geostrophic turbulence regime (G) . Regime C is only obtained close to the onset of thermal convection, i.e.~for Rayleigh numbers not much larger than the critical value of $\Rat_c \simeq8.7$; the CTC regime is characterized by columns that stretch across the fluid layer, surrounded by ``sleeves'' of oppositely signed vorticity (also visible in the fluctuating temperature) that prevent columns from interacting with each other; in the P regime the insulation mechanism weakens and column-column interaction shortens these structures, transforming them into plumes; finally, geostrophic turbulence prevails at sufficiently large Rayleigh numbers where no obvious coherence in the fluctuating temperature field is observed. Although distinct transitions in the flow statistics can sometimes be used to separate these flow regimes \citep{dN14}, an obvious distinction cannot always be made, e.g.~cases $(\Rat=40, \Pr=2)$ and ($\Rat = 60, \Pr=3$) shown in figure \ref{fig:temp_visualization}, where plumes generated at each horizontal boundary seem to coexist with columns spanning the whole vertical extension of the computational domain.

\begin{figure*}
\centering

\subfloat[][$\Rat = 40, \Pr = 2 $]
{\includegraphics[width=0.3\textwidth,trim={0 0 10cm 0},clip]{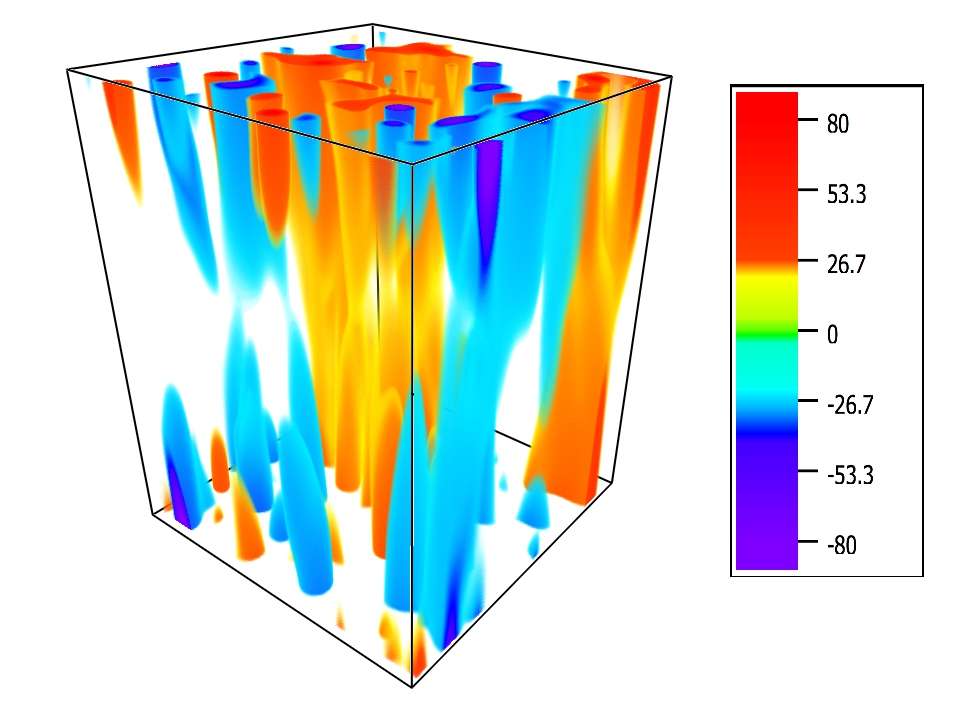}}\quad
\subfloat[][$\Rat = 60, \Pr = 2 $]
{\includegraphics[width=0.3\textwidth,trim={0 0 10cm 0},clip]{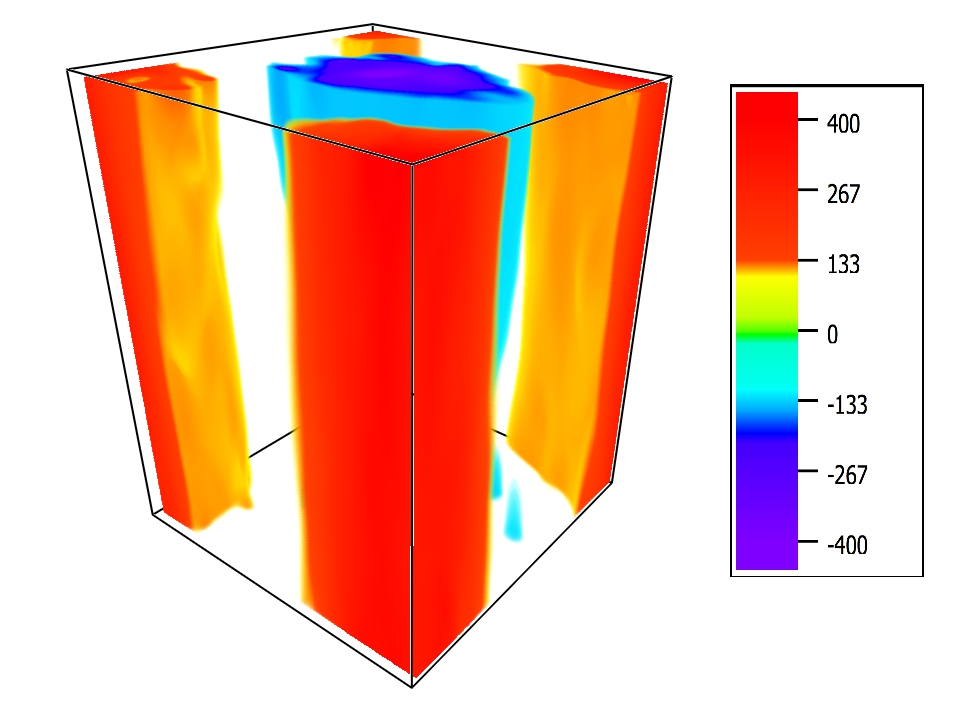}}\quad
\subfloat[][$\Rat = 200, \Pr = 2 $]
{\includegraphics[width=0.3\textwidth,trim={0 0 10cm 0},clip]{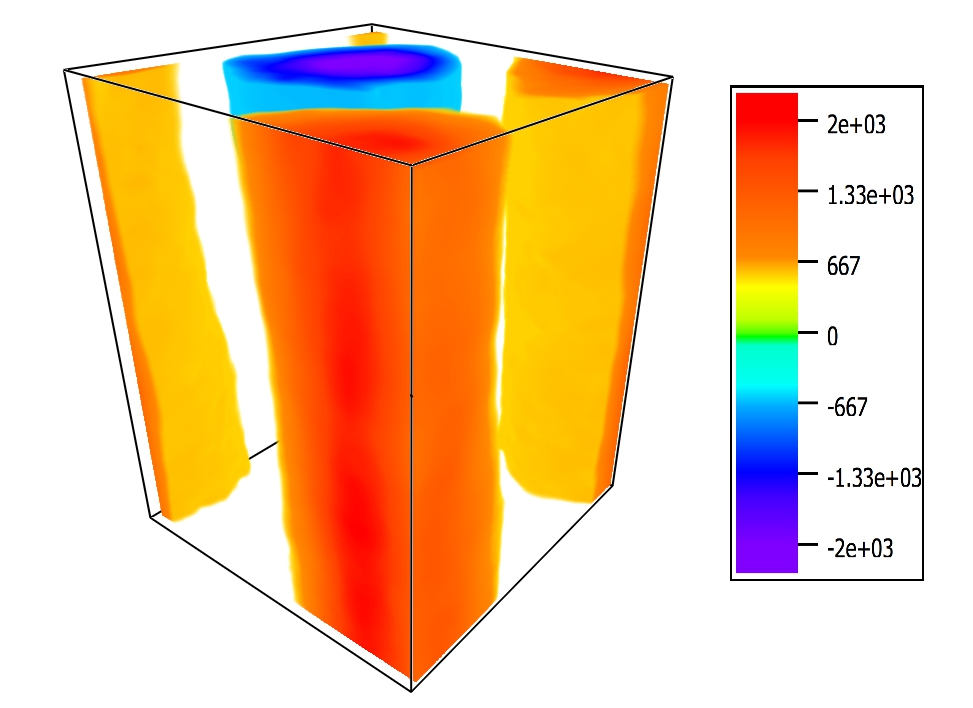}}

\subfloat[][$\Rat = 60, \Pr = 3 $]
{\includegraphics[width=0.3\textwidth,trim={0 0 10cm 0},clip]{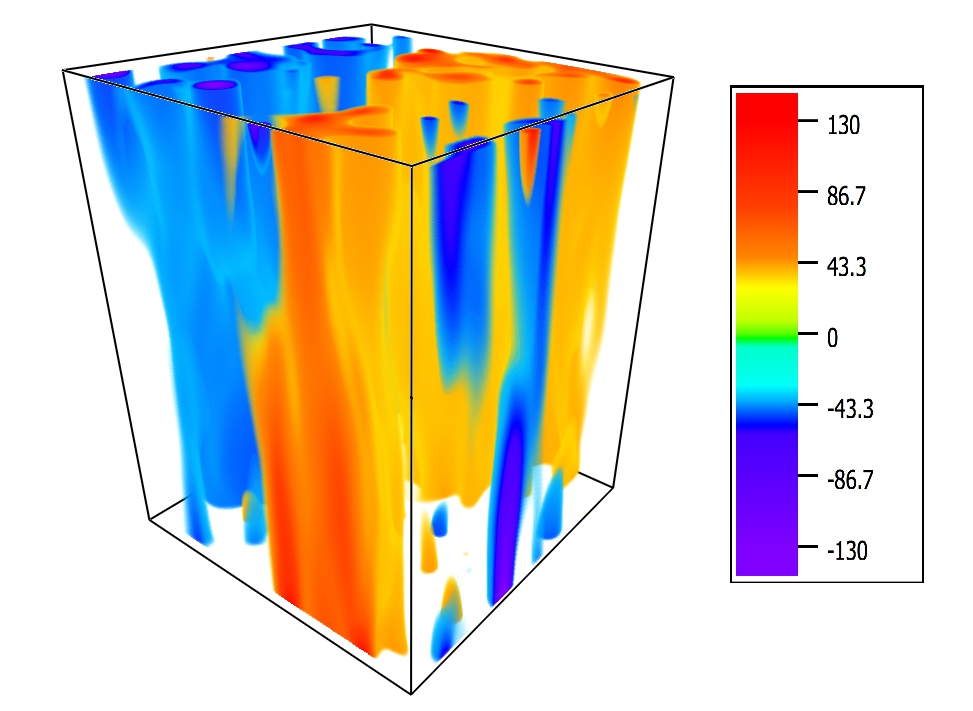}}\quad
\subfloat[][$\Rat = 80, \Pr = 3 $]
{\includegraphics[width=0.3\textwidth,trim={0 0 10cm 0},clip]{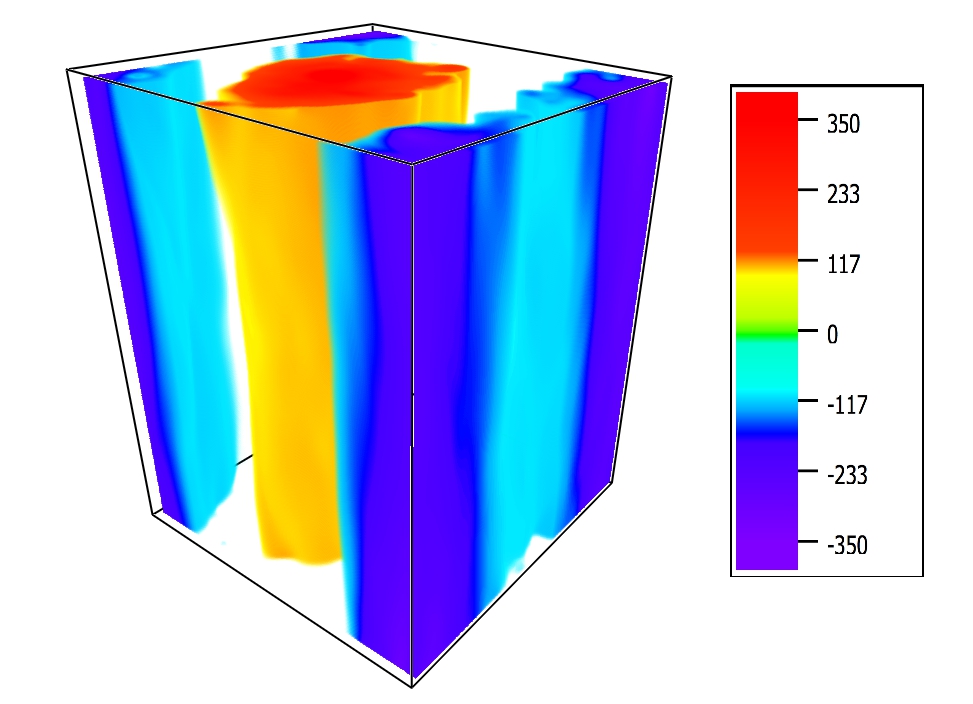}}\quad
\subfloat[][$\Rat = 120, \Pr = 3 $]
{\includegraphics[width=0.3\textwidth,trim={0 0 10cm 0},clip]{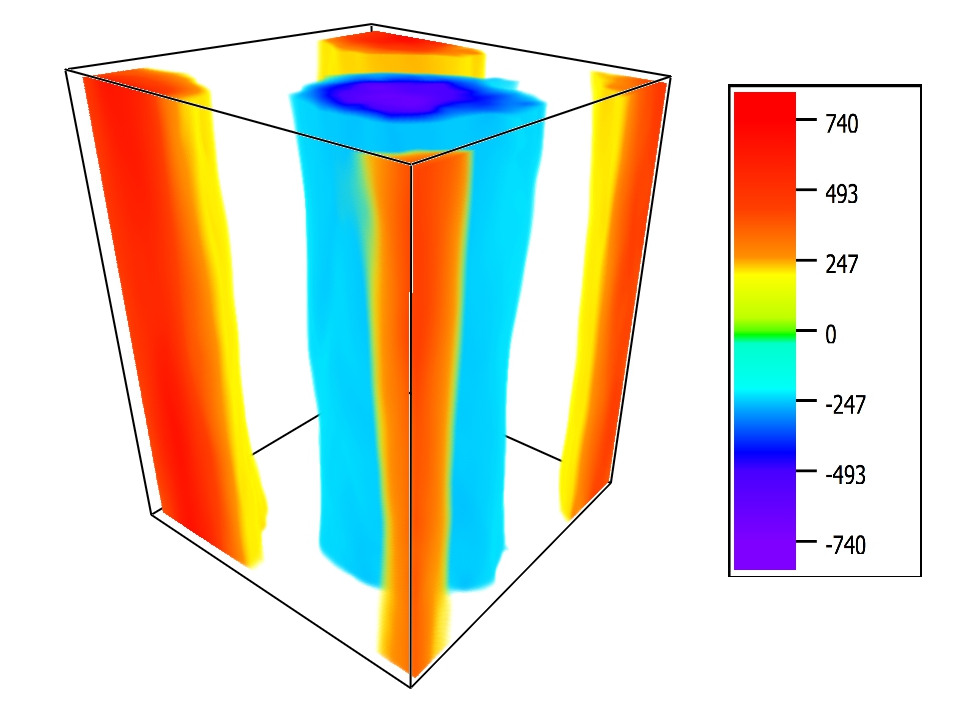}}\\

\subfloat[][$\Rat = 80, \Pr = 7 $]
{\includegraphics[width=0.3\textwidth,trim={0 0 10cm 0},clip]{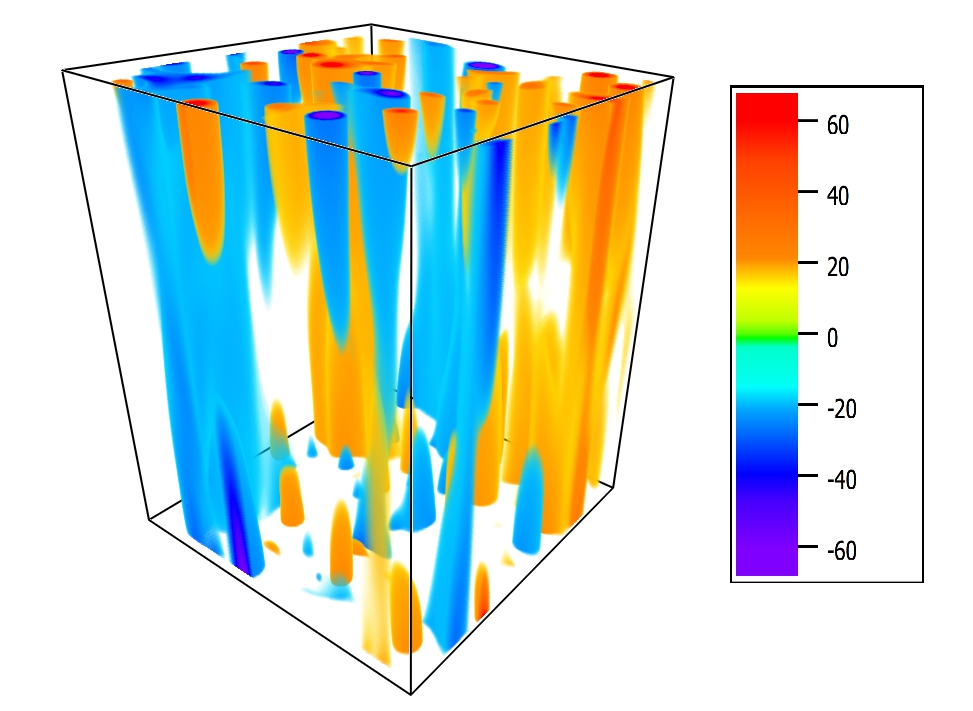}}\quad
\subfloat[][$\Rat = 135, \Pr = 7 $]
{\includegraphics[width=0.3\textwidth,trim={0 0 10cm 0},clip]{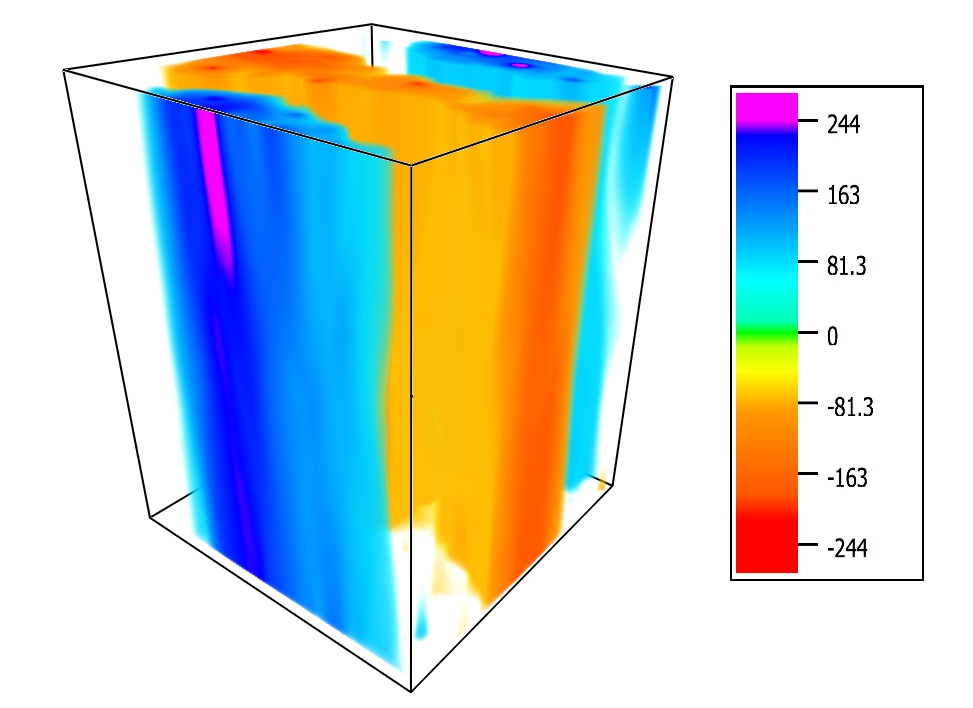}}\quad
\subfloat[][$\Rat = 160, \Pr = 7 $]
{\includegraphics[width=0.3\textwidth,trim={0 0 10cm 0},clip]{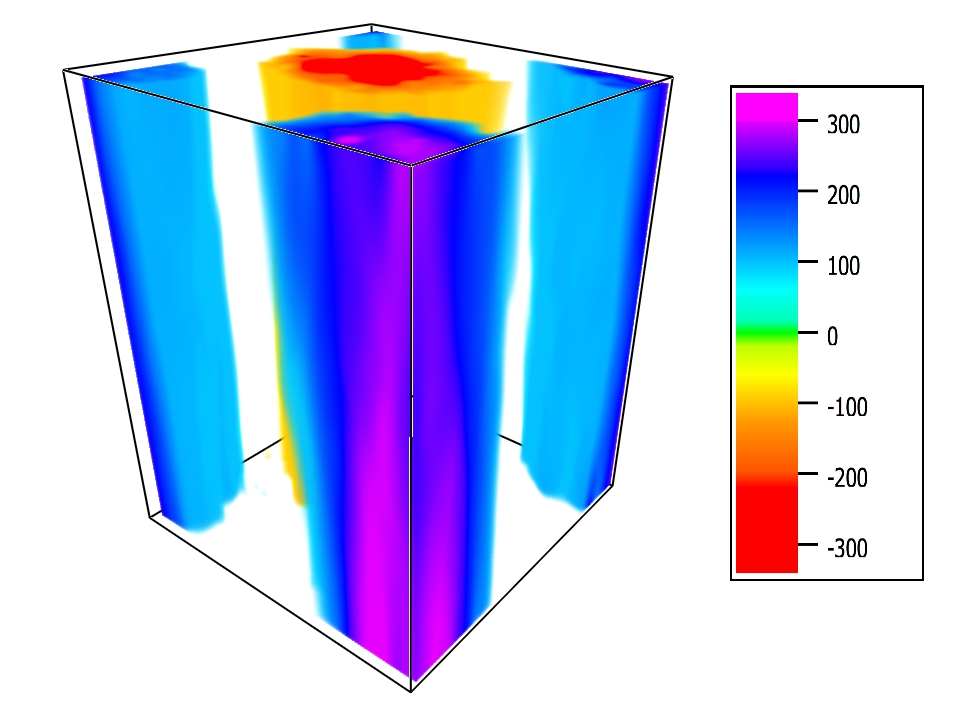}}\quad

\caption{Volumetric renderings of the geostrophic streamfunction (pressure), $\psi$, showing the development of large-scale vortices for increasing Rayleigh number (left to right) and increasing Prandtl number (top to bottom).}
\label{fig:psi_visualization}
\end{figure*}

For a given value of $\Pr$, we observe the formation of LSVs as $\Rat$ is increased. These depth-invariant, dipolar vortices are readily identified from visual inspection of the geostrophic streamfunction $\psi$. Some representative cases are shown in figure \ref{fig:psi_visualization}. Crucially, we observe LSV formation for all $\Pr$ values reported in table \ref{tab:results}, including, for the first time to our knowledge, $\Pr>3$. We find that for an LSV to be present in the domain, convection does not need to be in the geostrophic turbulent regime, as was previously suggested \citep{kJ12,sS14}.

\subsection{LSV characterization}\label{subsec:LSV}

\begin{figure*}
\centering
\subfloat[][]
{\includegraphics[width=0.4\textwidth]{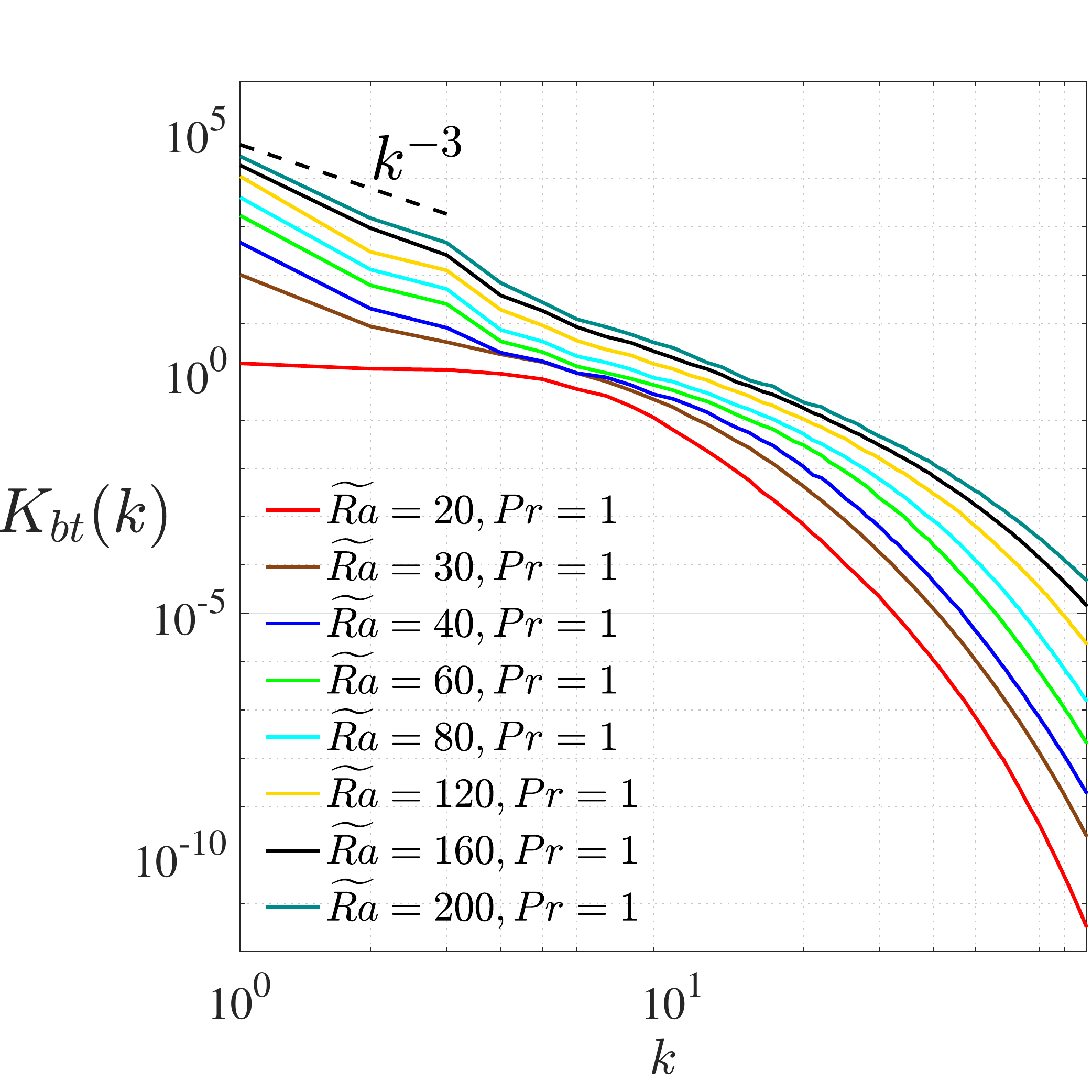}\label{subfig:Kbt_spectra_Pr1}}\quad
\subfloat[][]
{\includegraphics[width=0.4\textwidth]{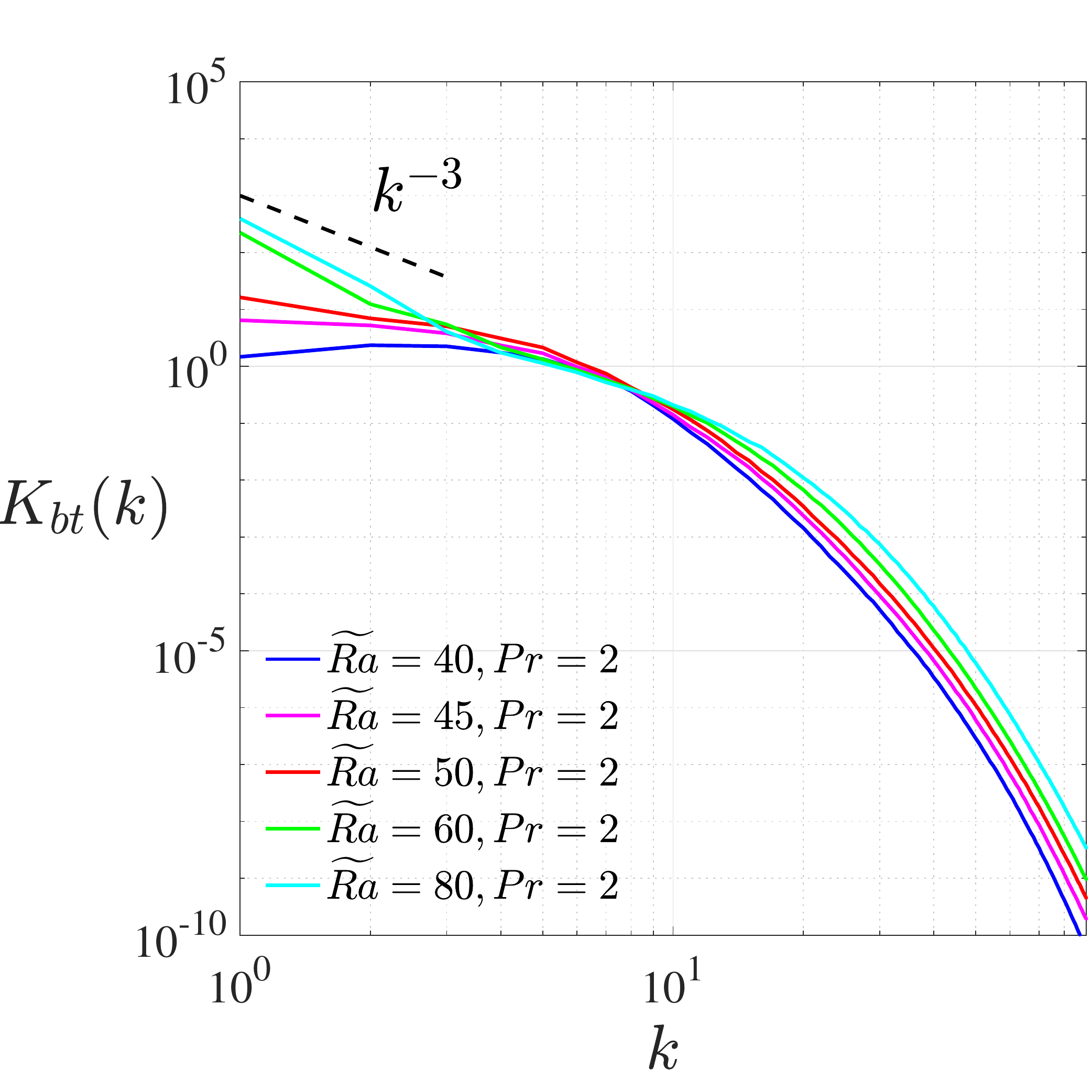}\label{subfig:Kbt_spectra_Pr2}}\\
\subfloat[][]{\includegraphics[width=0.36\textwidth]{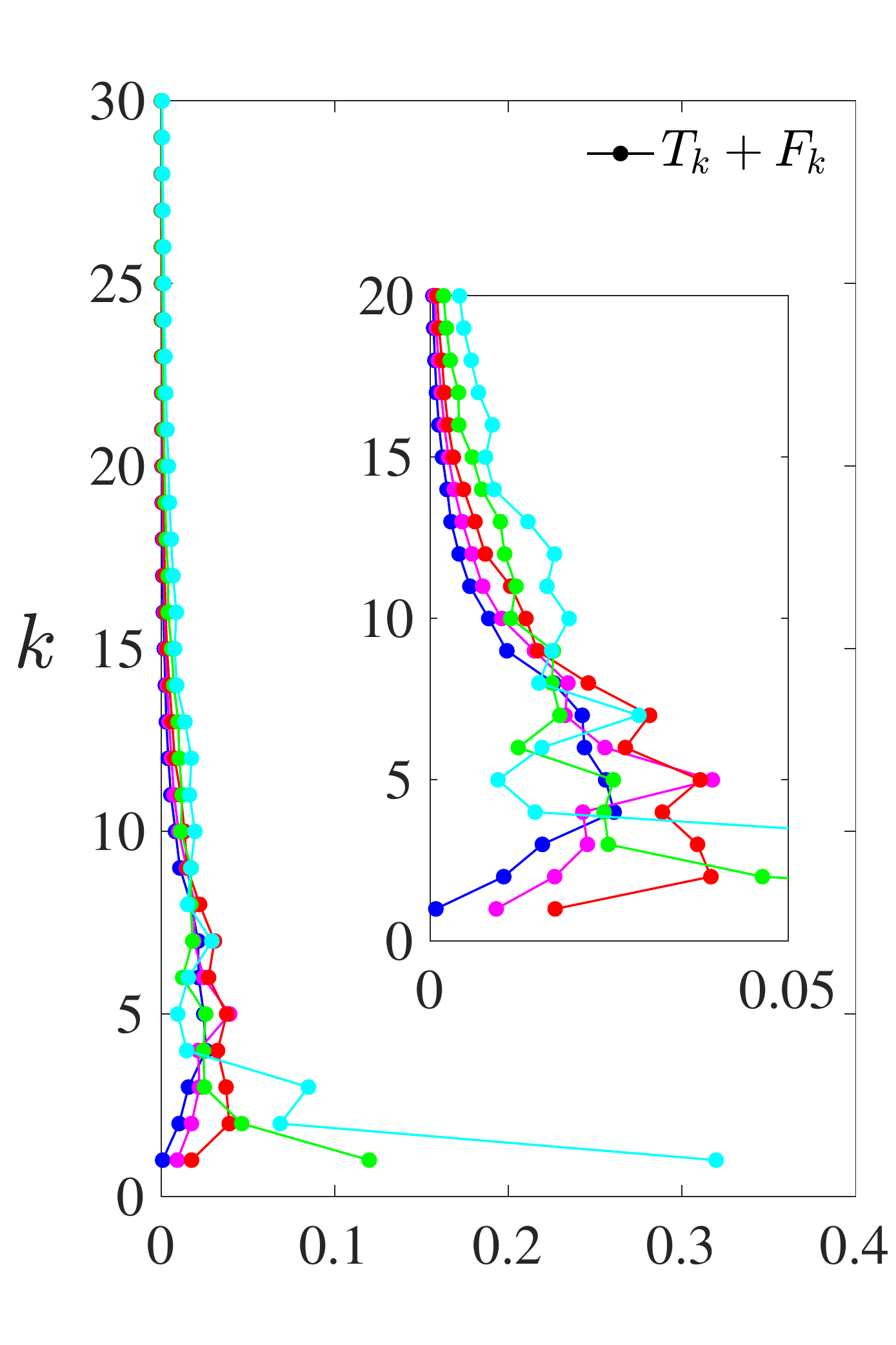}\label{subfig:TF}}\quad
\subfloat[][]
{\includegraphics[width=0.36\textwidth]{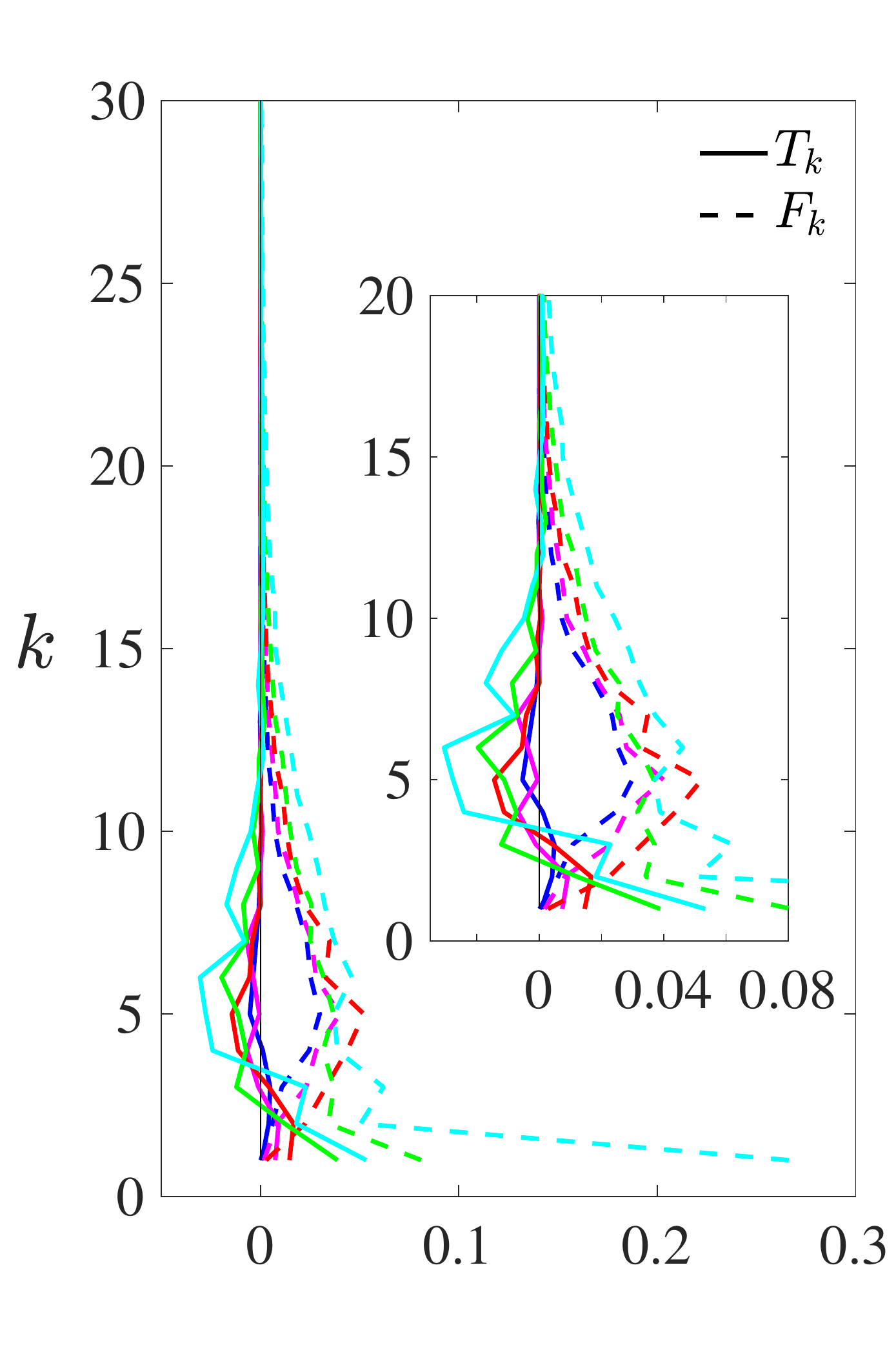}\label{subfig:TF_detail}}\quad

\caption{Spectra of the barotropic kinetic energy $\Kbt(k)$ for select values of $\Rat$ and for (a) $\Pr=1$ and (b) $\Pr=2$. The scaling $K_{bt}(k)\sim k^{-3}$ is shown for comparison. (c): barotropic transfer $T_k+F_k$ for the $\Pr=2$ cases shown in (b). (d): barotropic-to-barotropic ($T_k$) and baroclinic-to-barotropic ($F_k$) for the $\Pr=2$ case separately illustrated in continuous and dashed lines, respectively. Color legend for figures (c) and (d) is the same as figure (b). A black vertical line is drawn in correspondence of $T_k, F_k=0$. The insets highlight the behaviour for $\Rat\le50$. All quantities have been time-averaged over a statistically stationary state for which $T_k + F_k \approx -D_k$.}
\label{fig:Kbt_transfer}
\end{figure*}

To quantify the presence of an LSV in the domain we analysed the time-averaged barotropic kinetic energy spectra $\Kbt(k)$. We define flows in which the LSV is energetically  dominant by the two conditions: $ \Kbtm> \Kbcm$; and $ \Kbtm(k=1) \ge  \Kbtm(k>1)$. As examples, figures \ref{subfig:Kbt_spectra_Pr1} and \ref{subfig:Kbt_spectra_Pr2} show the barotropic kinetic energy spectra for $\Pr=1$ and $\Pr=2$ over a range in $\Rat$. The transition to LSV-dominant states occurs within the ranges $20<\Rat<30$ and $40<\Rat<45$ for the $\Pr=1$ and the $\Pr=2$ cases, respectively. As $\Rat$ is further increased beyond the transition, the LSV becomes increasingly dominant, as shown by the spectra. Note that for the $(\Rat=20, \Pr=1)$ case, the barotropic spectra has a maximum at $k=1$. However, for this case, the barotropic kinetic energy is not dominant, rather, we find that $ \Kbcm\simeq 3  \Kbtm$ for this case. Therefore, there is no energetically dominant LSV in the domain for this particular case. A similar process is observed for all $\Pr$ considered in the present study, although the threshold Rayleigh number for an LSV-dominant state increases with $\Pr$. However, we observe that the LSV becomes energetically dominant provided $\Retm\ge 5.812$, independent of $\Pr$. This value of $\Retm$ (indicated by the horizontal dashed line in figures \ref{fig:ReRa} and \ref{fig:ReRa_over_Pr}) corresponds to the case $(\Rat=45, \Pr=2)$ and it is the lowest $\Ret$ value for which LSV formation has been observed. The only exception is the $(\Rat=55, \Pr=2.5)$ case for which no energetically dominant LSV is observed, although $\Retm = 6.067 \pm 0.134$. This discrepancy can be explained by noting that these two values of $\Ret$ are (considering their temporal fluctuations) consistent with each other and by admitting that the transition to the LSV-dominated regime is not abrupt. A more detailed exploration of the parameter space around the transition could reveal other exceptions to the threshold we identified and possibly a subtle $\Pr$ dependence.

\begin{figure*}
\centering

\subfloat[][$\Rat = 60, \Pr = 7 $]
{\includegraphics[width=0.3\textwidth,valign=t]{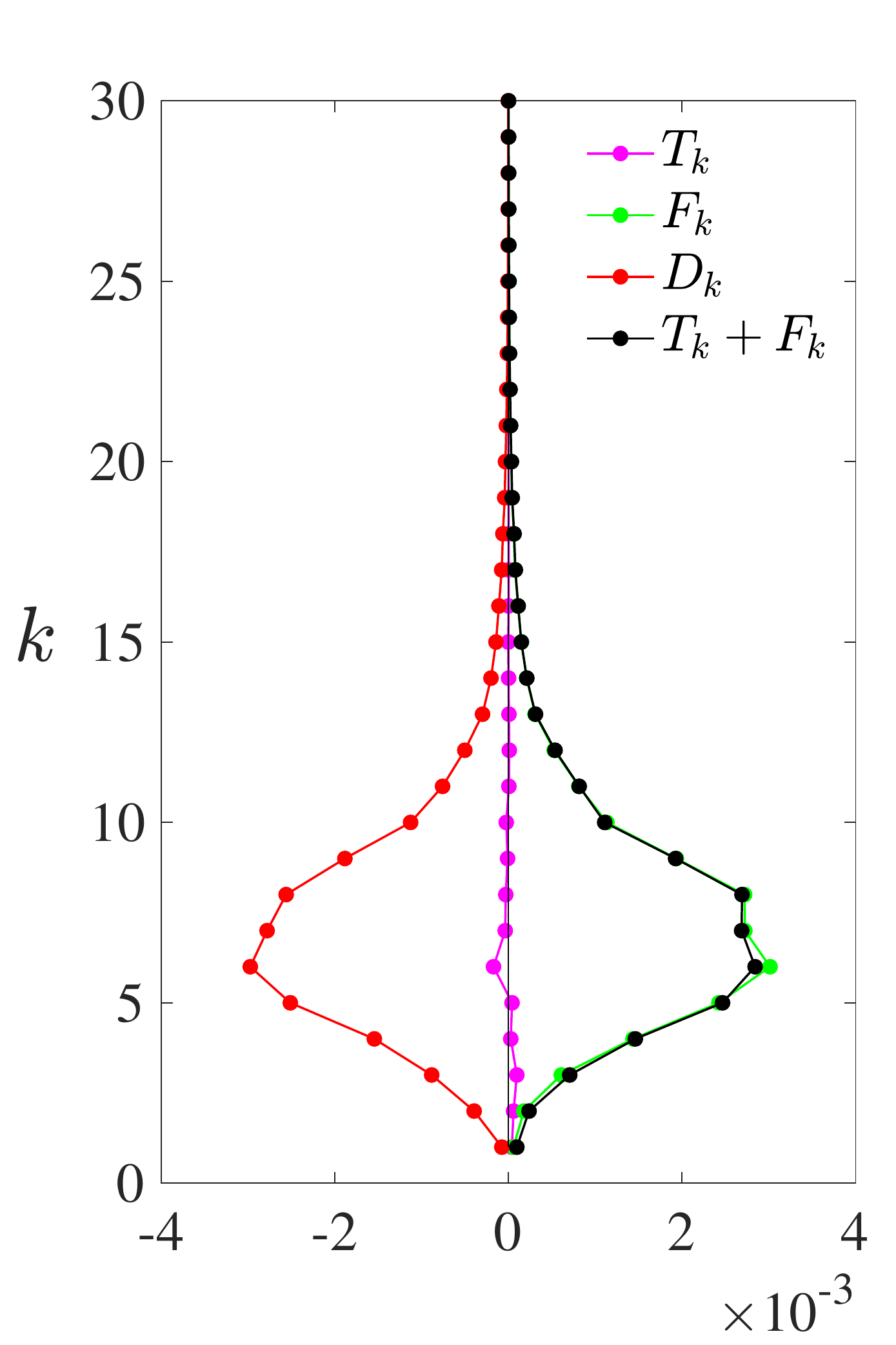}\label{subfig:bt_transfers_R60P7}
}\quad
\subfloat[][$\Rat = 60, \Pr = 2.5 $]
{\includegraphics[width=0.3\textwidth,valign=t]{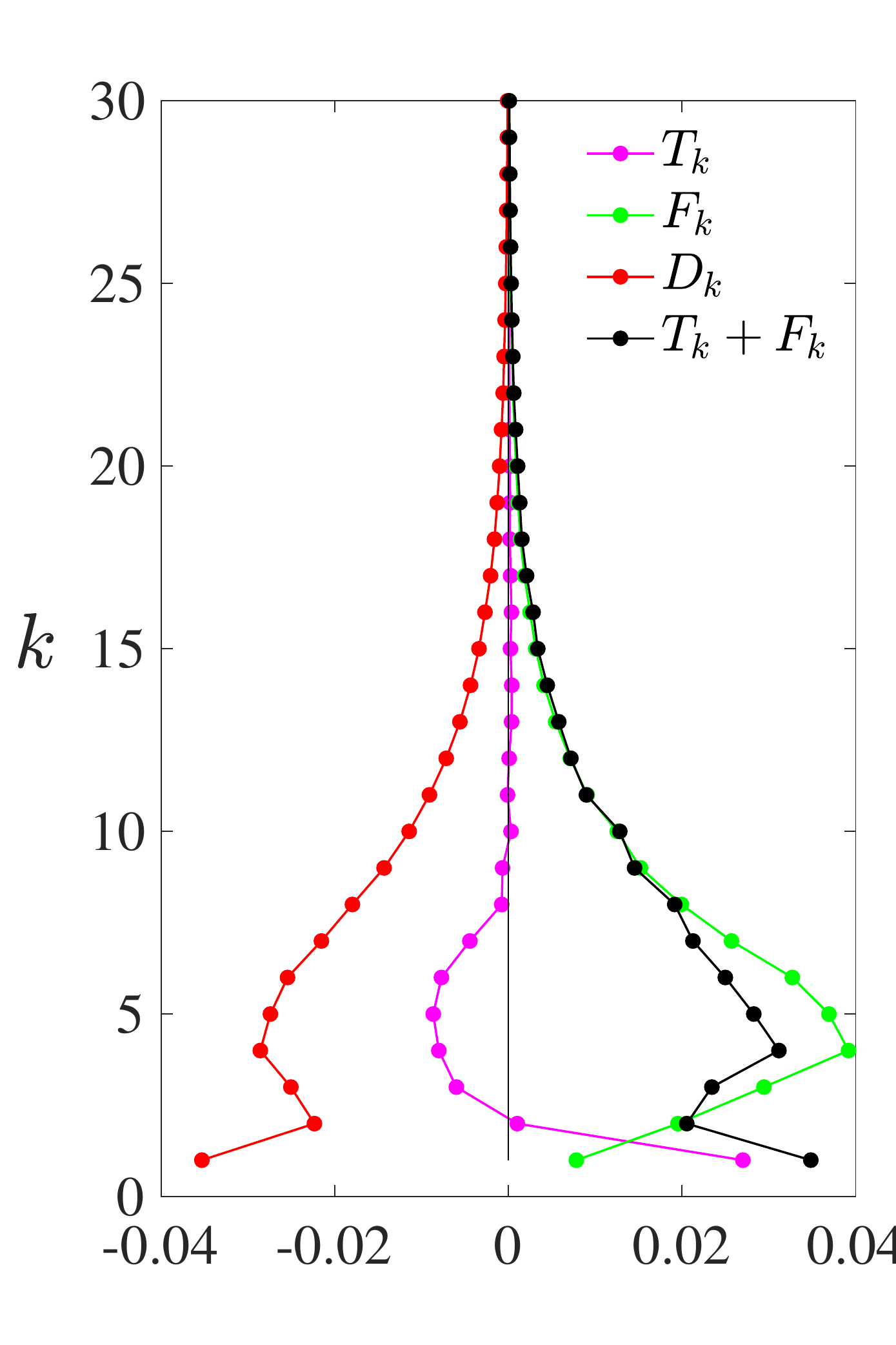}\label{subfig:bt_transfers_R60P2point5}
}\quad
\subfloat[][$\Rat = 60, \Pr = 1 $]
{\includegraphics[width=0.3\textwidth,valign=t]{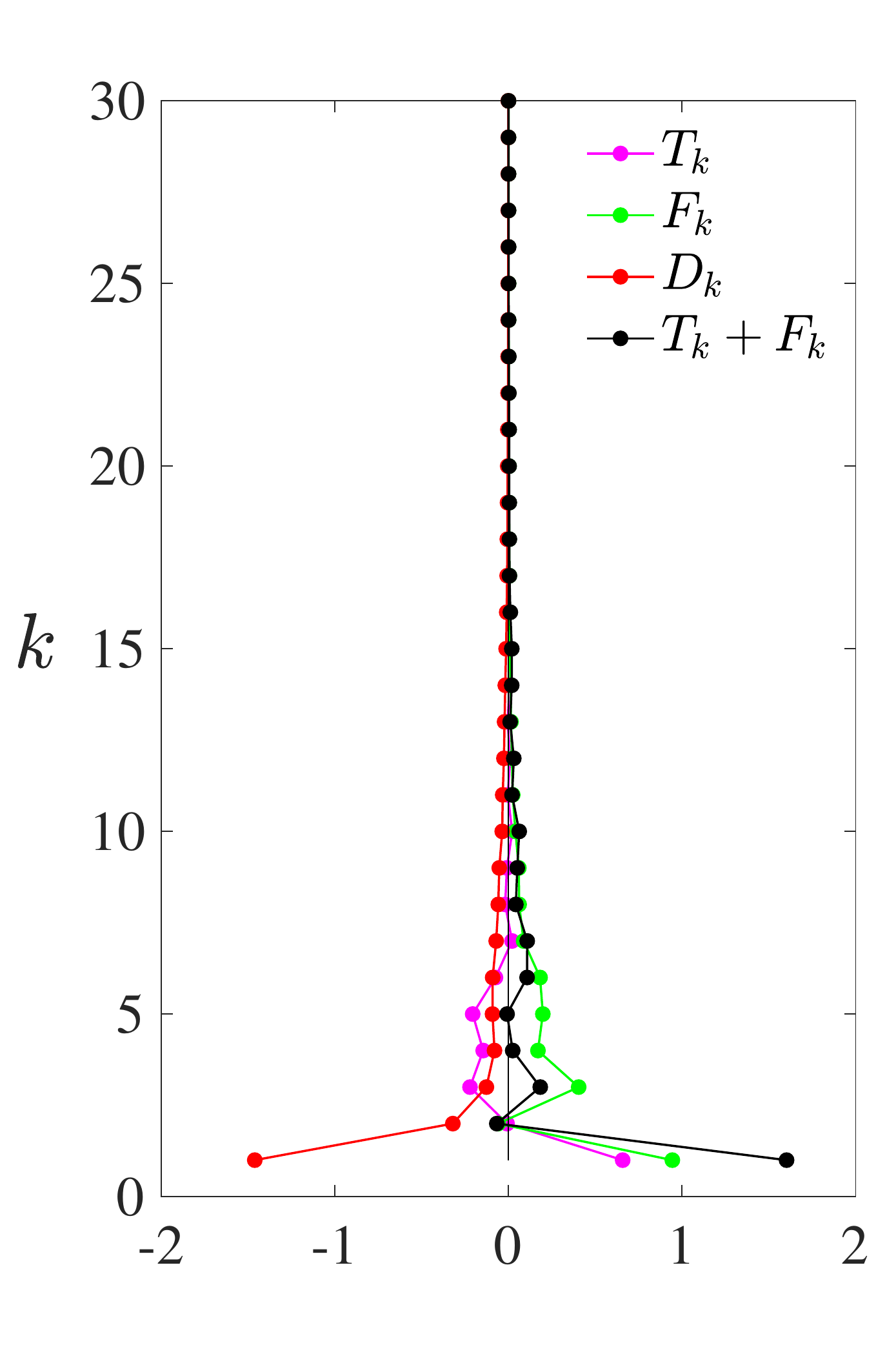}\label{subfig:bt_transfers_R40P7}
}

\caption{Time-averaged, barotropic transfer functions showing the change in energy transfer behavior as the inverse cascade becomes more prominent. All cases use $\Rat = 60$ and Prandtl numbers (a) $\Pr=7$ ($\Ret \approx 2.5 $), (b) $\Pr=2.5$ ($\Ret \approx 6.8 $) and (c) $\Pr=1$ ($\Ret \approx 16.8 $). These three cases are representative for, respectively, the CTC/P regime, showing no LSV in the domain and $\Ret\ll 5.812$; the P regime, showing an LSV in the domain and $\Ret\gtrsim 5.812$; the G regime, with a robust LSV and $\Ret\gg 5.812$.}
\label{fig:bt_transfers_examples}
\end{figure*}


In addition to the transition shown in the barotropic kinetic energy spectra, we also find (with increasing $\Rat$) a distinct transition in the character of the three terms present in the spectral kinetic energy equation \eqref{eqn:dt_Kbt}. In figure \ref{subfig:TF} we illustrate how the time-averaged, barotropic energy transfer $T_k+F_k$ evolves with $\Rat$ for the specific case of $\Pr=2$. We note that $[T_k+F_k]_{k=1} > 0$ for all of the cases investigated, indicating that energy is always being transferred to the $k=1$ mode, regardless of the value of $\Rat$. However, we find that $T_k+F_k$ changes from possessing a peak at $k > 1$, to then peaking at $k=1$ for a sufficiently large value of $\Rat$;  for the $Pr=2$ data shown this transition occurs when $\Rat>50$. Analyzing all of our simulations shows that this transition occurs when $\Retm\ge 6.491$, for any value of $\Pr$. This threshold value of corresponds to the simulation $(\Rat=135, \Pr=7)$ and it is noted that no simulation with $\Ret < 6.491$ satisfies $[T_k+F_k]_{k=1} > [T_k+F_k]_{k>1}$. The only exception is the case $(\Rat=40, \Pr=1.5)$, for which  $\Retm = 6.7529 \pm 0.227$ and the energy transfer at $k=1$ is subdominant. Again, given the finite fluctuations in the $\Retm$ values, we argue that a transition region may exists for which a simple threshold rule may not always work; a more detailed exploration of the parameter space may reveal subtle $\Pr$ dependencies in the transition into the regime for which the barotropic energy transfer at $k=1$ is dominant.

 Closer inspection of $F_k$ and $T_k$ separately (figure \ref{subfig:TF_detail}) indicates that the baroclinic, convective dynamics primarily transfers energy to the barotropic dynamics around $k\simeq 5$ (notice the positive peak in $F_k$ for $k\simeq 5$). Energy transferred to the barotropic dynamics is then transferred upscale by the barotropic non-linear interactions, as indicated by negative values of $T_k$ for wavenumbers $k > 4$, and positive values at the largest scales. The inverse cascade that leads to LSV formation is therefore directly driven by the barotropic non-linear interactions in \eqref{E:baro}, whereas the energy is provided by the interaction of the baroclinic dynamics with the barotropic flows. The fact that $\sum_{k\ge0}T_k = 0$  confirms that the non-linear barotropic interactions do not inject or extract barotropic energy and that the saturation of $\Kbt$ is controlled by the balance between energy injected from the baroclinic dynamics and energy dissipated through viscosity:
\be
0 \approx \sum_{k\ge 0 } F_k + \sum_{k\ge 0 } D_k.
\label{eqn:BT_balance}
\ee
This mechanism is in agreement with the formation of large-scale condensates in 2D calculations \citep{chertkov2007dynamics,laurie2014universal} where the transfer of energy is due to the non-linear interactions between different scales of the 2D flow (equivalent to the barotropic-to-barotropic energy transfer described by $T_k$). 
This view is also confirmed by recent 3D studies \citep{buzzicotti2018energy}.

As mentioned in Section \ref{sed:dept-averaged}, in the saturated state $D_k \approx - (T_k + F_k)$. Examples are shown in figure \ref{fig:bt_transfers_examples} for $\Rat=60$ and $\Pr=7$, $\Pr=2.5$ and $\Pr=1$. These cases are representative of, respectively, a low-turbulence case at the edge of the CTC and P regimes where $\Retm\ll5.812$ and, consequently, an inverse cascade that is not strong enough to drive LSV-formation; a case in the P regime with $\Retm\gtrsim5.812$, slightly above the critical value for LSV formation and an inverse cascade; and a case in the G regime, with a robust inverse cascade possessing a strong peak at $k=1$ driving an energetically dominant LSV. This figure illustrate how the pattern observed in figure \ref{fig:Kbt_transfer} for fixed $\Pr$ and increasing $\Rat$ can be discerned for fixed $\Rat$ and decreasing $\Pr$.

\begin{figure*}
\centering
\subfloat[][]
{\includegraphics[width=0.45\textwidth]{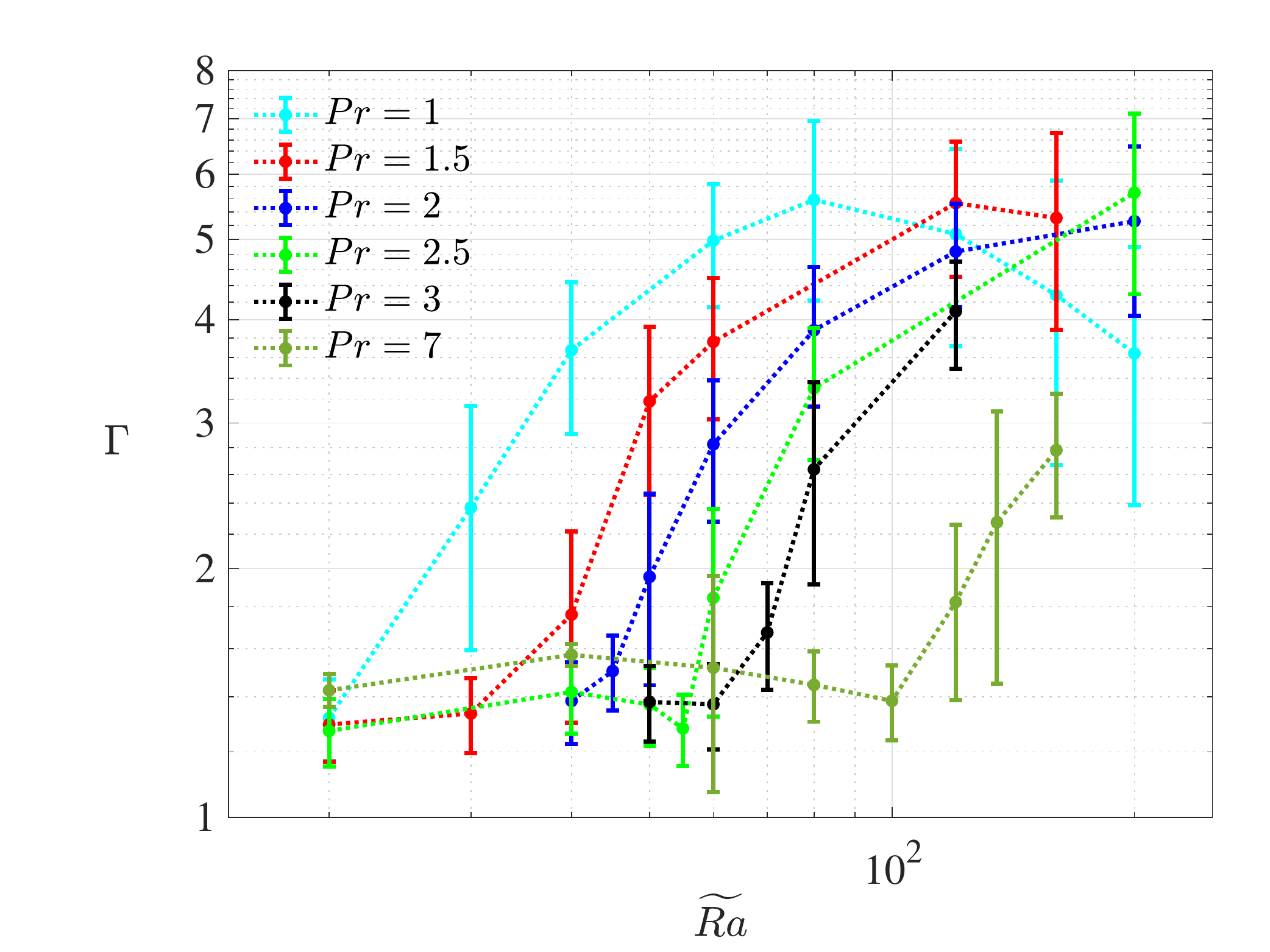}\label{fig:Kz_over_K_Ra}}\quad
\subfloat[][]
{\includegraphics[width=0.45\textwidth]{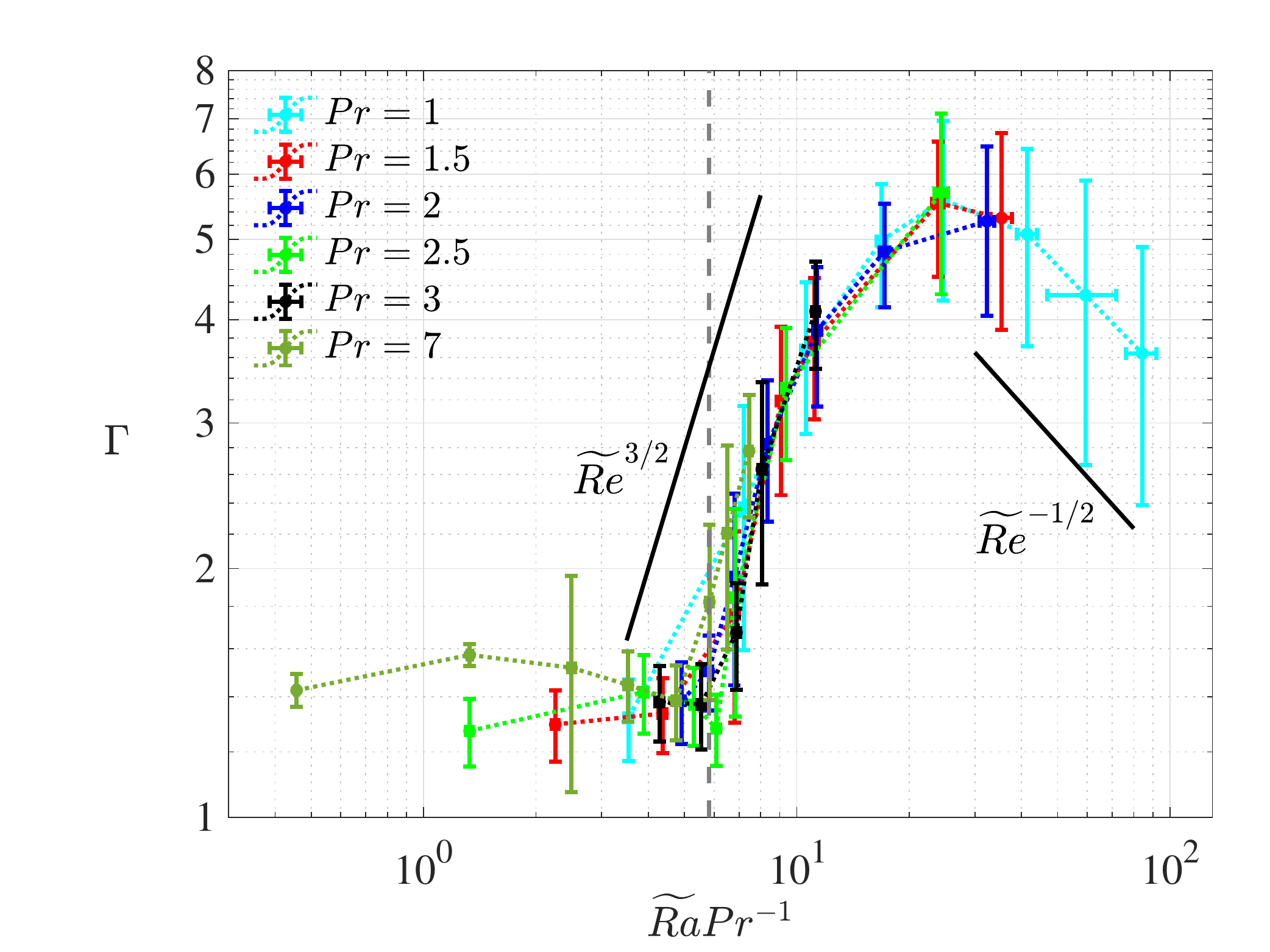}\label{fig:Kz_over_K_Re}}\\
\subfloat[][]
{\includegraphics[width=0.45\textwidth]{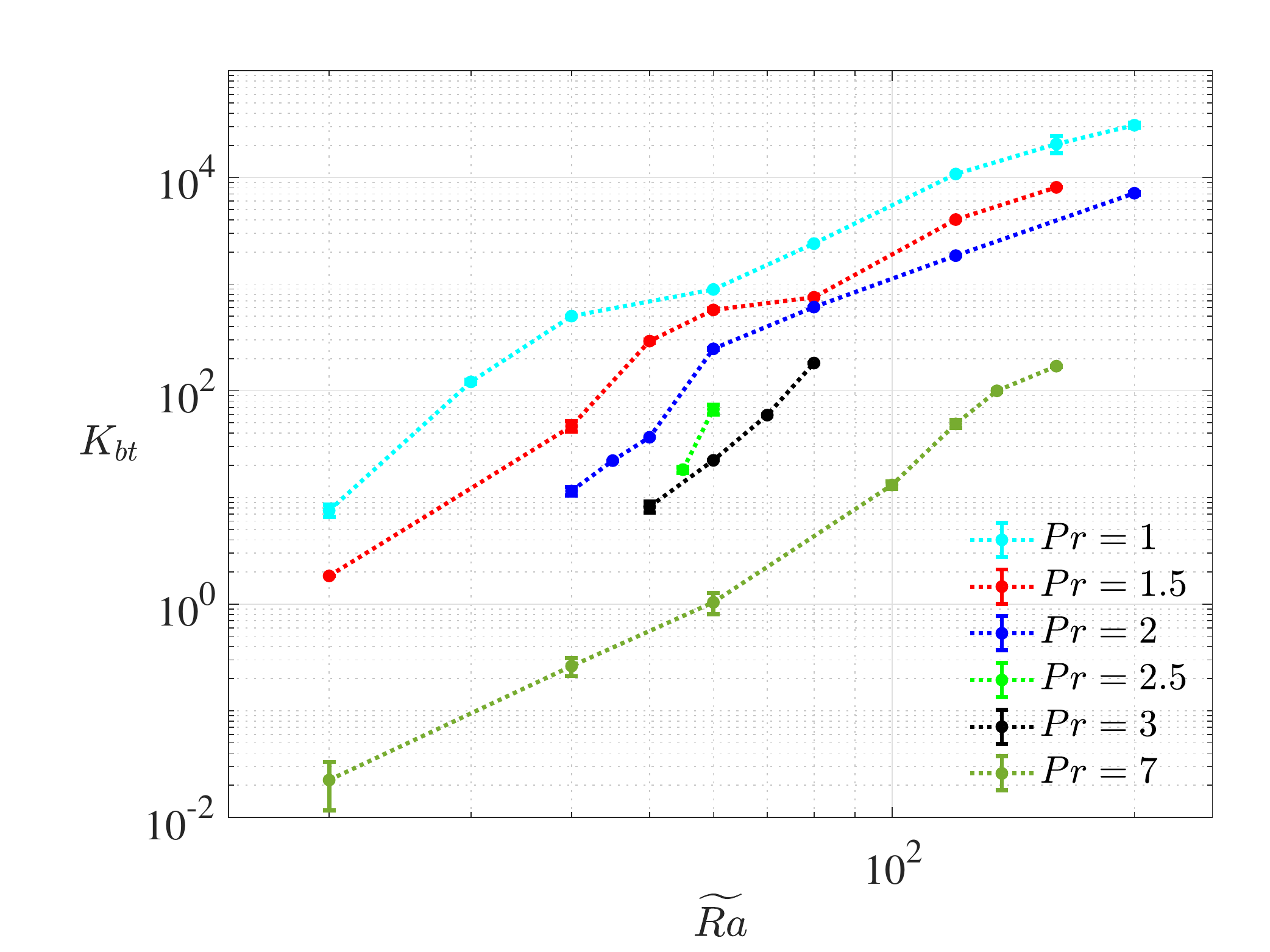}\label{fig:Kbt_Ra}}\quad
\subfloat[][]
{\includegraphics[width=0.45\textwidth]{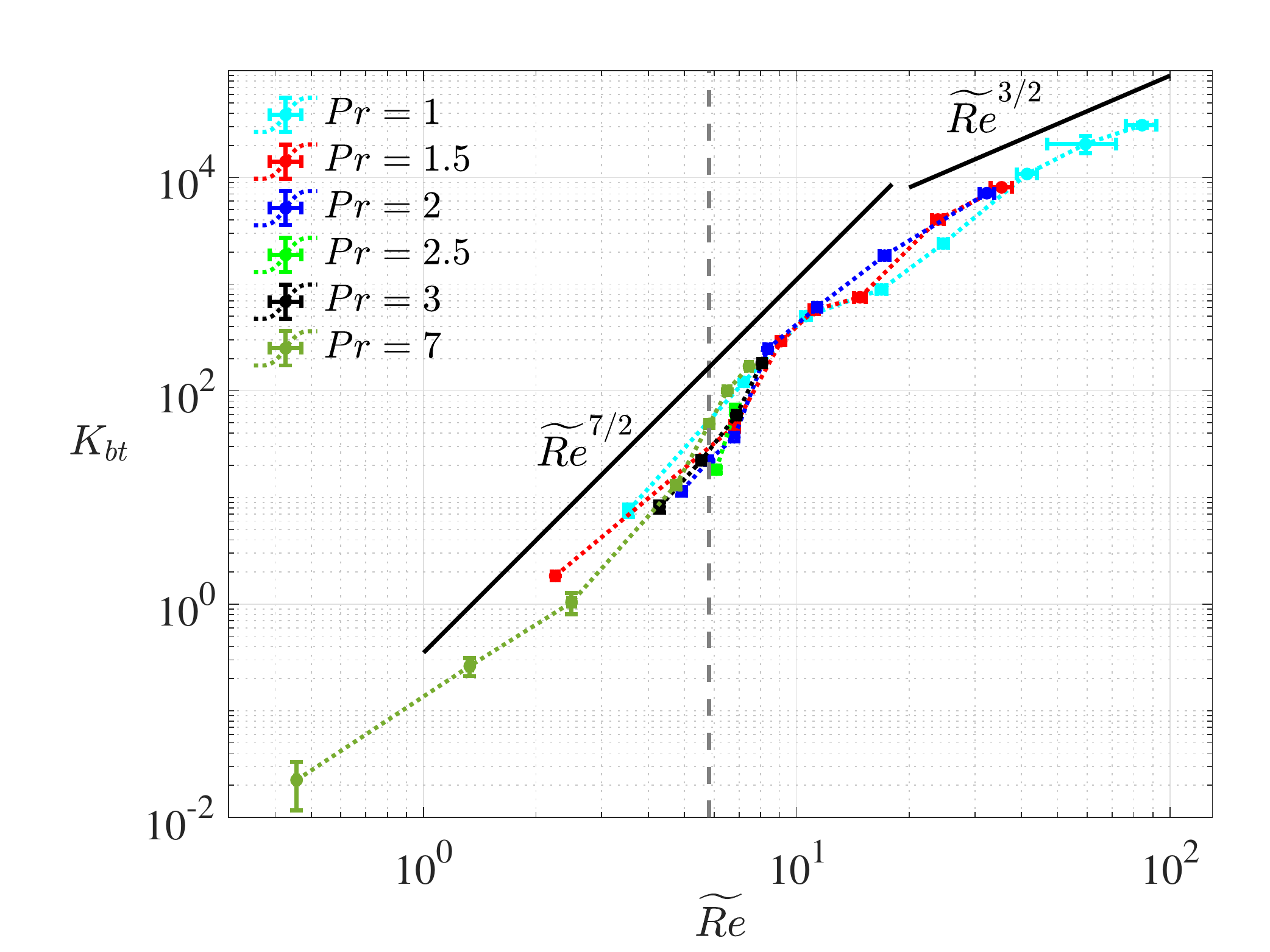}\label{fig:Kbt_Re}}\\
\caption{Scaling behavior of the kinetic energy for all of the simulations. (a) Ratio of the total kinetic energy to the vertical kinetic energy ($\Gamma$) versus $\Rat$; (b) $\Gamma$ versus $\Retm$; (c) barotropic kinetic energy $K_{bt}$ versus $\Rat$; (d) $K_{bt}$ versus $\Retm$. The vertical dashed line at $\Retm =5.812$ in (b) and (d) demarcates the onset of LSV-dominant behavior. Slopes in (b) and (d) are shown for reference.}
\label{fig:Kz_over_K}
\end{figure*}

Following \citet{cG14}, we can characterize the kinetic energy of the barotropic flow using the ratio of total kinetic energy to vertical kinetic energy,  
\be
\Gamma =  \Km / (3  \Kzm) .
\ee
The factor of 3 in the denominator ensures that $\Gamma \rightarrow 1$ if the kinetic energy is equipartitioned between the horizontal and vertical components of the velocity. Conversely, when the barotropic kinetic energy dominates, we expect this ratio to become significantly larger than unity. Figures \ref{fig:Kz_over_K_Ra} and \ref{fig:Kz_over_K_Re} show $\Gamma$ for all of the simulations as a function of $\Rat$ and $\Retm$, respectively. In agreement with the DNS calculations of \cite{cG14}, $\Gamma \approx 1$ for small values of $\Rat$, then increases rapidly once LSVs begin to form. We find that $\Gamma$ reaches a maximum of $\Gamma_{max} \approx 5.5$, that appears to be independent of the particular value of  $\Pr$, though only the  $\Pr=1$ and  $\Pr=1.5$ simulations show a maximum value. For  $\Pr=1$, $\Gamma$ reaches a maximum value at $\Rat = 80$, whereas for $\Pr=1.5$, $\Gamma_{max}$ occurs at $\Rat=120$, suggesting that the value of $\Rat$ at which $\Gamma_{max}$ is observed increases rapidly with Prandtl number.

Since the value of $\Rat$ at which LSVs begin to form is $\Pr$-dependent, $\Gamma$ is also plotted as a function of $\Retm$ in figure \ref{fig:Kz_over_K_Re}. The data suggests that the evolution of $\Gamma$ is uniquely determined by $\Retm$ (or $(\Rat-\Rat_c) \Pr^{-1}$ according to figure \ref{fig:ReRa_over_Pr}) since all curves show self-similar behaviour, independent of the particular value of $\Pr$. The dashed vertical line denotes $\Retm = 5.812$, the approximate value at which the ($k=1$) LSV becomes dominant. For cases in which $\Retm$ is below this threshold value, $\Gamma$ is close to 1 for all values of $\Pr$, and the convective pattern, or flow regime, for all of these cases can be qualitatively classified as cellular or convective Taylor columns. Above this threshold value of the Reynolds number, we find both plumes and eventually geostrophic turbulence as $\Retm$ grows. For both the $\Pr=1$ and $\Pr=1.5$ cases, $\Gamma_{max}$ is reached for $\Retm\simeq24$.

We emphasize that since all of the calculations in the present work were carried out with the QG model, the observed decrease of $\Gamma$ for large values $\Rat$ is not due to a loss of rotational constraint. Although the DNS study of \cite{cG14} also report a decrease in $\Gamma$ for sufficiently large forcing, their observed decrease might be caused by an increase in the Rossby number with increasing forcing.  In the present study, $\Ro$ remains asymptotically small, regardless of the thermal forcing. Also, as pointed out previously, the LSV observed in \cite{cG14} is cyclonic, whereas the LSV observed in the present simulations is dipolar.

Figures \ref{fig:Kbt_Ra} and \ref{fig:Kbt_Re} show the barotropic kinetic energy versus $\Rat$ and $\Retm$, respectively. In figure \ref{fig:Kbt_Re}, slopes of $\Retm^{7/2}$ and $\Ret^{3/2}$ are shown as reference, along with the vertical dashed line denoting the threshold Reynolds number $\Retm=5.812$. The `s-shaped' behavior of the data, along with $\Gamma$, suggests that the barotropic mode is growing at an ever-decreasing rate as $\Rat$ is increased. Taking the barotropic kinetic energy $ \Kbtm$ scaling with $\Retm$ as illustrated in figure \ref{fig:Kbt_Re} and with $\Kbc\sim\Retm^2$ (see supplementary figure 2), we can derive the expected evolution of $\Gamma$ in the growing ($5.812\lesssim\Retm\lesssim24$) and decaying ($\Retm>24$) regimes. In the former, under the assumption that  $ \Kbtm\gg \Kbcm,  \Kzm$, we obtain $\Gamma\sim\Retm^{3/2}$; in the latter, taking $\Retm\to\infty$, we obtain $\Gamma\sim\Retm^{-1/2}$. These slopes are illustrated in figure \ref{fig:Kz_over_K_Re}, for reference.

\begin{figure*}
\centering
\includegraphics[width=0.65\textwidth]{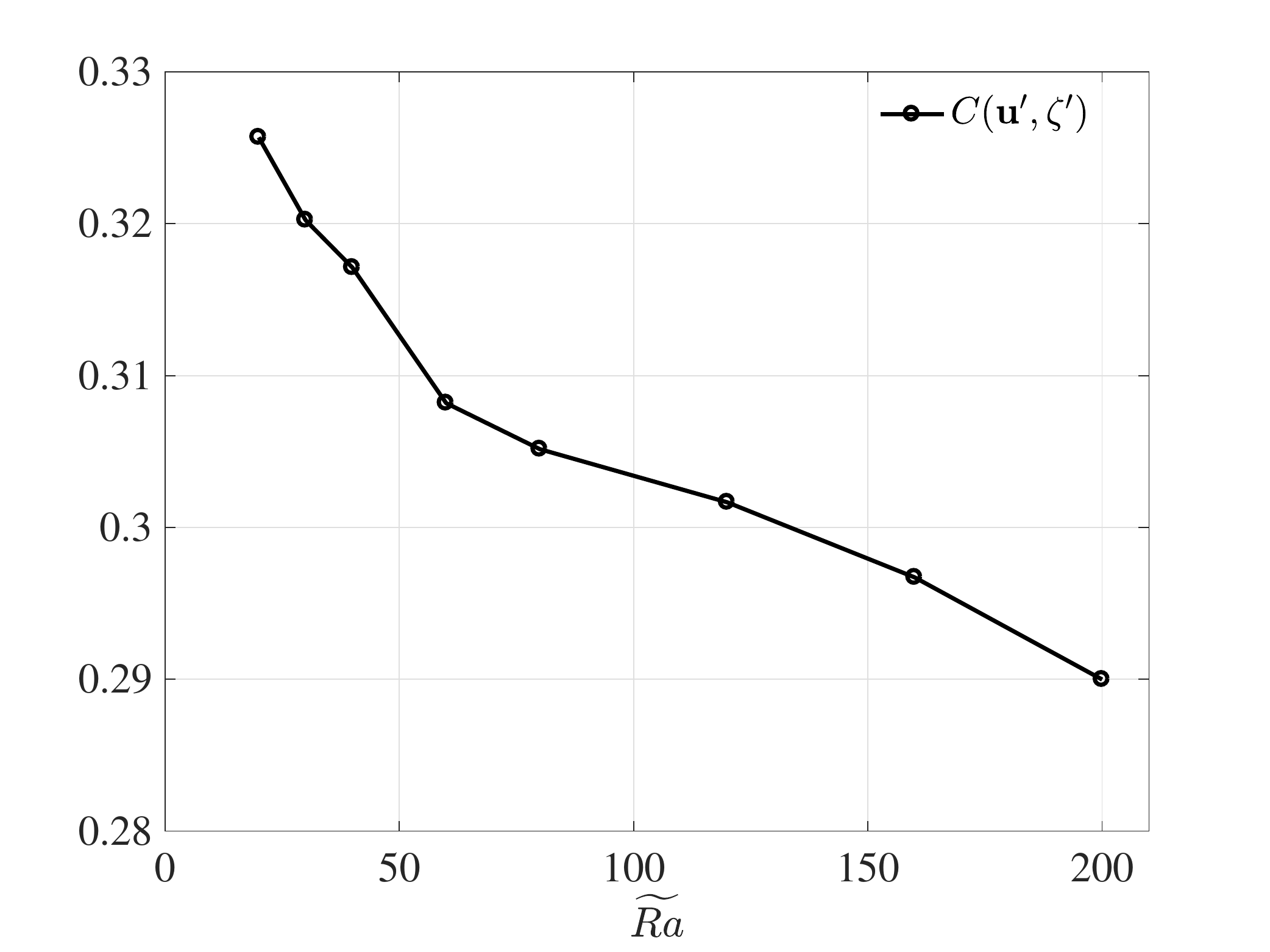}
\caption{Correlation coefficient between the baroclinic vorticity and the $x$-component of the baroclinic velocity as a function of the Rayleigh number $\Rat$. The Prandtl number is fixed at $\Pr=1$. A value of $C = 0.5$ is perfect correlation for one component of the velocity vector.}
\label{fig:corr}
\end{figure*}

To better understand the change in scaling behavior of the barotropic kinetic energy with increasing Rayleigh number, we examine the nonlinear convective forcing term in the barotropic vorticity equation \eqref{E:baro}. In particular, the nonlinear baroclinic term can be written as 
\be
\langle J[\psi', \zeta' ] \rangle = \nabla_\perp \cdot \langle \ub' \zeta' \rangle ,
\ee
which suggests that the decreased efficiency of the barotropic mode is due to a drop in correlations between the baroclinic velocity and baroclinic vorticity. We calculated the cross correlation coefficient for the $x$-component of the baroclinic velocity vector and baroclinic vorticity, defined as 
\be
C(u',\zeta') = \frac{\overline{\langle \lb \zeta' u' \rb^2 \rangle}}{\sqrt{\overline{\langle \lb \zeta' \zeta' \rb^2 \rangle} \overline{\langle \lb u' u' \rb^2 \rangle}}},
\ee
and analogously for the cross correlation for the $y$-component of the baroclinic velocity field and the baroclinic vorticity, $C(v',\zeta')$.
We note that this definition leads to $C=0.5$ for perfect correlation between one component of the baroclinic velocity vector and the baroclinic vorticity, since the statistics are isotropic in the horizontal plane when sufficiently time-averaged.
The coefficients were computed over the entire investigated range of $\Rat$ for the case $\Pr=1$. Figure \ref{fig:corr} shows the average value 
\be
C(\textbf{u}',\zeta') = \frac{C(u',\zeta') + C(v',\zeta')}{2}.
\ee  
We observe that $C(\textbf{u}',\zeta')$ decays as  $\Rat$ is increased from $\Rat>20$, suggesting one possible reason for the reduced rate of growth of the barotropic kinetic energy with increasing Rayleigh number.

\begin{figure*}
\centering
\subfloat[][]
{\includegraphics[width=0.45\textwidth]{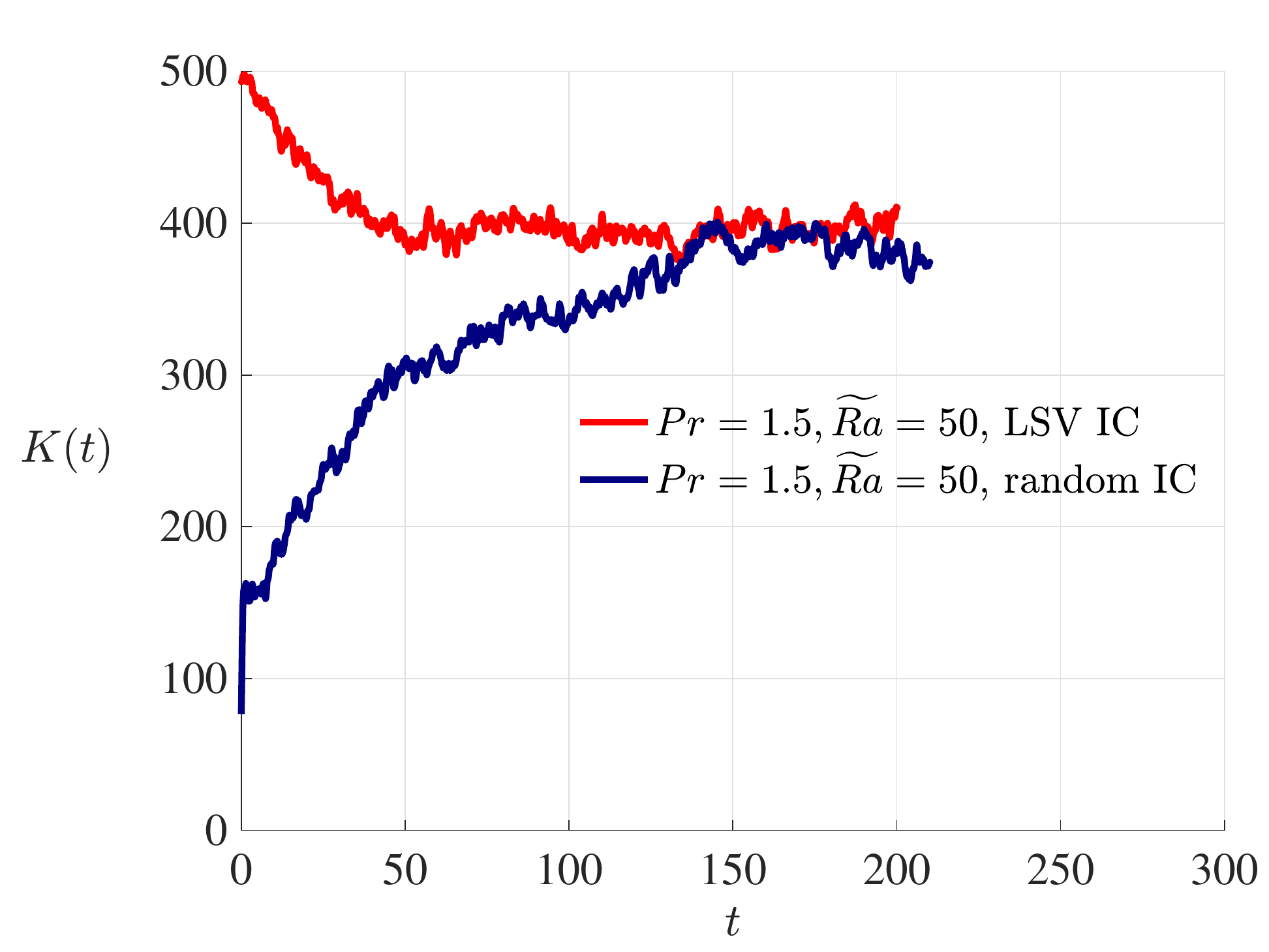}}\quad
\subfloat[][]
{\includegraphics[width=0.45\textwidth]{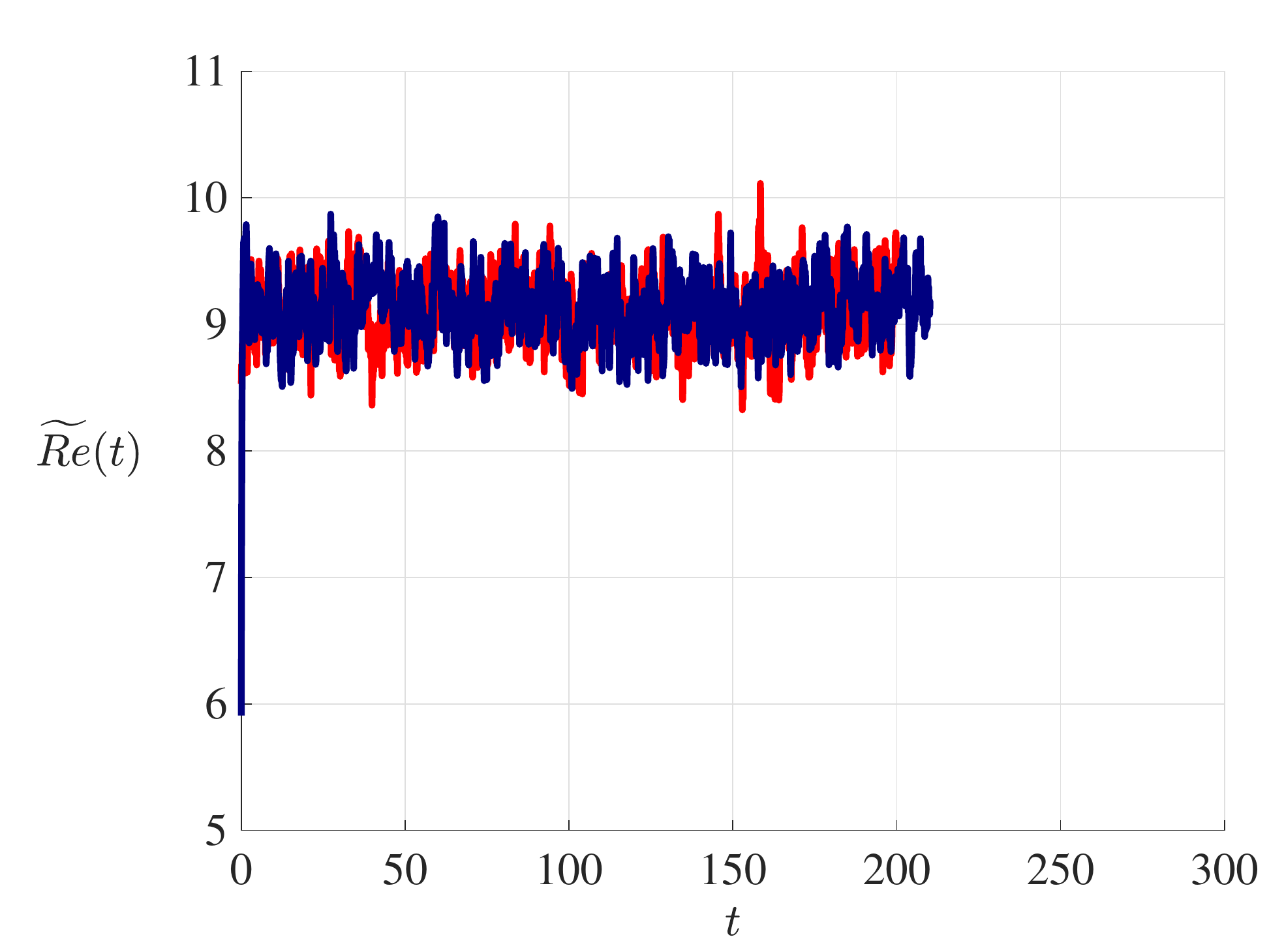}}
\caption{Influence of initial conditions on large-scale vortex (LSV) formation. Kinetic energy (a) and reduced vertical Reynolds number (b) for the parameters $\Pr=1.5$ and $\Rat=50$ from two different initial conditions: the case marked as ``random IC" has a random initial condition with no initial LSV present; ``LSV IC" marks an initial condition with a well-developed LSV in the system.}
\label{fig:Pr1point5Ra50ICs}
\end{figure*}

\subsection{The influence of initial conditions}\label{s:IC}

For the cases indicated by the superscript $*$ in table \ref{tab:results}, additional simulations were carried out to test the influence of initial conditions on the occurrence of LSVs. In particular, for cases capable of forming an LSV ($\Retm>5.812$), we checked that the kinetic energy of the saturated state is independent of the presence of an LSV in the initial condition. 
Our results indicate that both baroclinic (or convective) amplitude (measured by $\Ret$) and the barotropic kinetic energy in the saturated state do not depend on the initial condition, but only depend on $\Pr$ and $\Rat$. As an example, in figure \ref{fig:Pr1point5Ra50ICs} we show this by illustrating the time evolution of the kinetic energy per unit volume, $K(t)$, and the vertical Reynolds number, $\Ret(t)$, for the case $(\Rat=50, \Pr=1.5)$. Two simulations were run for this case: one started from a random initial condition that does not contain a pre-existing LSV (labeled as ``random IC'' in figure \ref{fig:Pr1point5Ra50ICs}), and one started from an initial condition with an LSV already present in the domain, given from the saturated state of the $(\Rat=40, \Pr=1)$  case (labeled as ``LSV IC''). In the former case, the initial growth of kinetic energy is due to the formation of the LSV due to the imbalance $T_k + F_k>|D_k|$ for $k=1$. In the latter case, the LSV initially present in the system was formed at a higher $\Ret$ and due to a stronger inverse cascade than the one developed for $(\Rat=50, \Pr=1.5)$. Therefore, initially the imbalance $T_k+F_k<|D_k|$ leads to a kinetic energy decay as the LSV cannot be energetically sustained. For both cases, a new state is eventually reached for which, statistically speaking $|D_k|=T_k + F_k$. 

Similarly, we also found that for cases in which $\Retm<5.812$, an LSV would eventually decay if it was present in the initial conditions. This result is in contrast with DNS calculations where large-intensity, domain-scale cyclonic vortices appear to be long lived when injected in a convective system in which large-scale structure would not spontaneously form \citep{bF19}.

\subsection{The influence of box dimensions}

The horizontal dimensions of the simulation domain are represented in terms of integer multiples of the critical wavelength $\lambda_c$. We indicate the horizontal size of the computational domain by $n_c \lambda_c \times n_c\lambda_c$, with $n_c$ being an integer. Most of the simulations were carried out with horizontal dimensions of $10\lambda_c \times 10 \lambda_c$ (i.e. $n_c=10$), which represents a trade-off between using a box size that is large enough to allow for computing converged statistics, and using a horizontal spatial resolution that is computationally feasible for an extensive exploration of the parameter space. Previous work has used values up to $n_c = 20$, but, to our knowledge, no systematic investigations of the box size on key quantities such as the Nusselt number and Reynolds number have been reported for rotating convection. For non-rotating convection, however, \citet{Rs18} showed that surprisingly large box dimensions are needed to obtain convergence in all statistical quantities; in contrast, the same authors found that globally averaged quantities such as the Nusselt number converged with relatively small box dimensions. For the present work we have carried out simulations for fixed Rayleigh number and Prandtl number of $\Rat=40$ and  $\Pr=1$. A robust LSV is present with this parameter combination. Time-series of these simulations are available in the supplementary material (see supplementary figure 1).

Figure \ref{fig:boxsize_Kbt_Re_Nu} shows the convective Reynolds number and Nusselt number, and the barotropic and baroclinic kinetic energy for a range of box sizes. The solid lines in figure \ref{fig:boxsize_Kbt_Re_Nu}(a) show the Nusselt and Reynolds number for a simulation in which the barotropic mode was set to zero at each timestep. We observe a nearly 23\% increase in the heat transport when the barotropic mode is not present. This result might be interpreted in terms of the horizontal mixing that is induced by the barotropic mode; the vertical transport of heat is reduced when horizontal motions sweep heat laterally. In addition, we find that the Reynolds number is reduced by $\approx 4.4 \%$ with respect to the $n_c=20$ case. This observation suggests that the inverse cascade plays a relatively small role in influencing the amplitude of the convective flow speeds. 

\begin{figure*}
\centering
\subfloat[][]
{\includegraphics[width=0.45\textwidth]{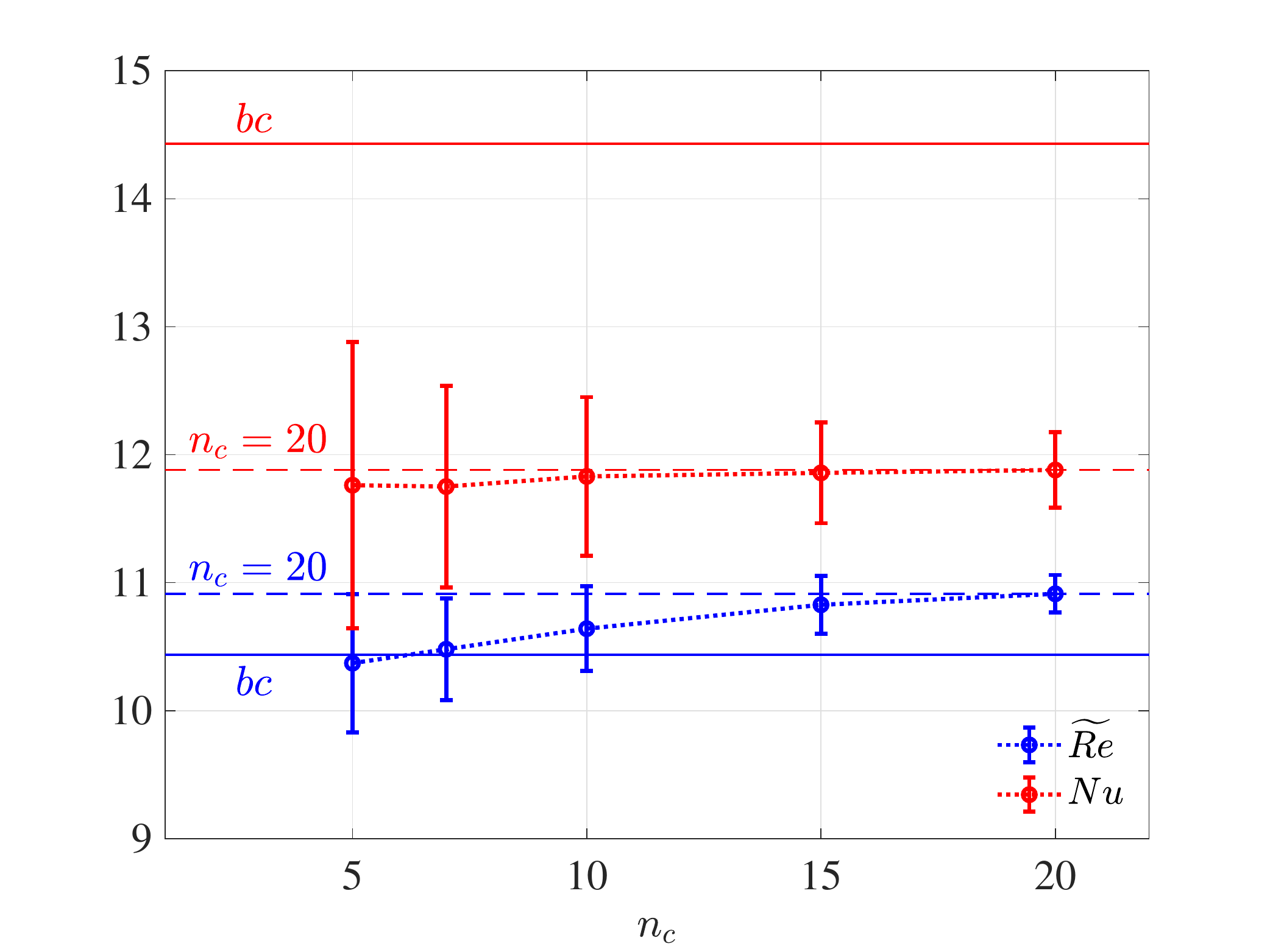}}
\quad
\subfloat[][]
{\includegraphics[width=0.49\textwidth]{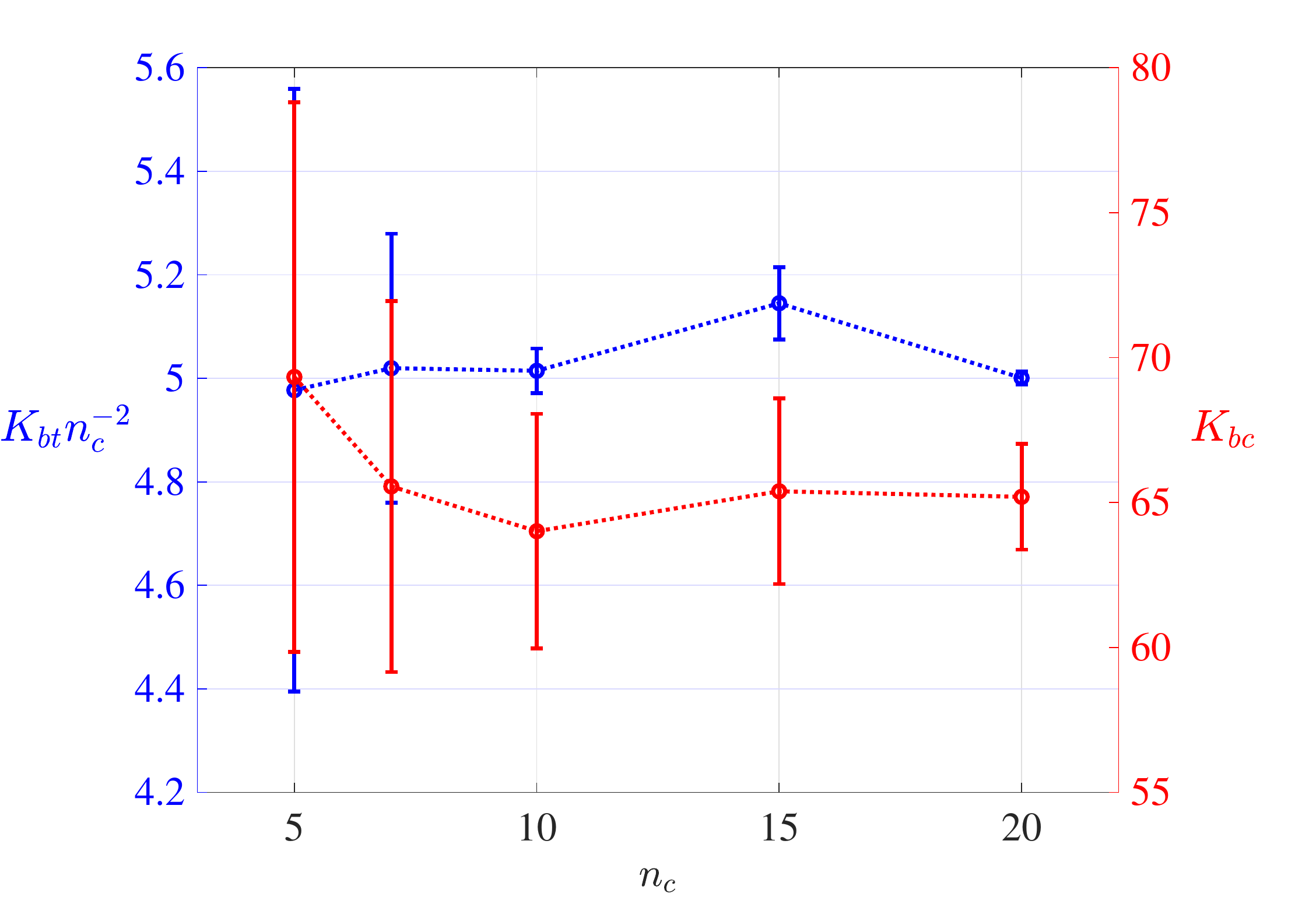}\label{subfig:boxsize_effect}}
\caption{Influence of the horizontal dimensions of the simulation domain on various quantities. Results for simulations with different horizontal dimensions, as characterized by $n_c\lambda_c\times n_c\lambda_c$, where $n_c$ is an integer and $\lambda_c$ is the critical horizontal wavelength. (a) Time-averaged Reynolds number $\Retm$ and Nusselt number $\Num$ versus $n_c$; (b) Normalized barotropic kinetic energy $ \Kbtm n_c^{-2}$ and baroclinic kinetic energy $ \Kbcm$. For all simulations shown here $\Pr=1$ and $\Rat=40$. The horizontal solid blue and red lines labeled '$bc$' represent the average values of $\Retm$ and $\Num$, respectively, calculated for a simulation with $n_c=10$ in which the barotropic flow is set to zero. The horizontal dashed lines indicate the $n_c =20$ values for comparison with the baroclinic case.}
\label{fig:boxsize_Kbt_Re_Nu}
\end{figure*}

An estimate for the intensity of the LSV based on the domain size can be made from the following simple argument. When a well developed LSV is present, the dominant component of the kinetic energy spectra ($k\lesssim 5$) scales approximately as $K_{bt}(k^*)\sim {k^*}^{-3}$ \citep{kraichnan1967inertial,smith1999transfer,rubio_upscale_2014}, where $k^* = k \widetilde{k}_{box} $, with $\widetilde{k}_{box} = 2\pi L_{box}^{-1}$ and $L_{box}=n_c\lambda_c$, is the dimensional box-scale wavenumber. Calculating the total kinetic energy we obtain:
\begin{equation}
K_{bt} = \int K_{bt}(k^*)dk^*\sim {k^*}^{-2} \sim L_{box}^2
\label{eqn:Kbt_L}
\end{equation}
where $K_{bt}\simeq K_{bt}(k^*=\widetilde{k}_{box})$ when a robust LSV is observed in the system. In particular, by doubling the linear size of the (squared) domain the kinetic energy of the LSV is allowed to quadruple in magnitude. The DNS study of \citet{bF14} also observed an increase in the barotropic kinetic energy with increasing box size. Our QG data shown in figure \ref{fig:boxsize_Kbt_Re_Nu}b is supportive of this quadratic dependence on box size.

\subsection{Scaling laws for the baroclinic dynamics}
\label{s:Scaling}

\begin{table}
  \begin{center}
\def~{\hphantom{0}}
\begin{tabular}{ccccc}
	$\Pr $ & $ \Rat $ & $ \alpha_r $ &  $\beta_r $ & $\gamma_r $  \\
	\hline
		[1,10] & [20,200] &   \alphaRaNum  &   \betaRaNum  &  \gammaRaNum \\[0pt]
		[1,7] & [20,200] &0.1899  &  1.1502  & -1.2376 \\[0pt]
		1 &  [20,200] & 0.1354  &  1.2198 &  -  \\[0pt]
		1.5 &  [20,160] &    0.1567  &  1.0758  &   -\\[0pt]
		2 &  [20,200] &    0.1330  &  1.0399  & - \\[0pt]
		2.5 &  [20,200] &    0.1375  &  0.9864 &   -\\[0pt]
		3 &  [20,120] &    0.1638  &  0.9007 &  -  \\[0pt]
		7 &  [20,160] &   0.0522 &   0.9912 &  -  \\[0pt]
		10 &  [20,120] &   0.0245 &   1.0792 &  - \\[0pt]
\end{tabular}
\caption{Least-squares fits to the Reynolds number, $\Retm = \alpha_r (\Rat - \Rat_c)^{\beta_r} \Pr^{\gamma_r}$ (for data encompassing multiple $\Pr$) or $\Retm = \alpha_r (\Rat - \Rat_c)^{\beta_r}$ (when a single $\Pr$ is considered). }
\label{tab:fits}
  \end{center}
\end{table}

Here we discuss least-squares fits to the baroclinic quantities $\Retm$ and $\Num$. Power law scalings were computed from all data collected in this study (see table \ref{tab:results} and figures \ref{fig:ReRa}) for various subsets of $\Rat$ and $\Pr$. 
For $\Retm$ with varying $Pr$, we used power law fits of the form 
\begin{equation}
\Retm = \alpha_r (\Rat - \Rat_c)^{\beta_r} \Pr^{\gamma_r} ,
\label{eqn:Ref1}
\end{equation} 
where $\alpha_r$, $\beta_r$ and $\gamma_r$ are all numerically computed constants. For constant $Pr$, we used 
\begin{equation}
\Retm= \alpha_r (\Rat - \Rat_c)^{\beta_r}.
\label{eqn:Ref2}
\end{equation} 
The numerically computed constants are denoted by $\alpha_r$, $\beta_r$ and $\gamma_r$ and given in table \ref{tab:fits}.

\begin{figure*}
\centering
\subfloat[][]
{\includegraphics[width=0.48\textwidth]{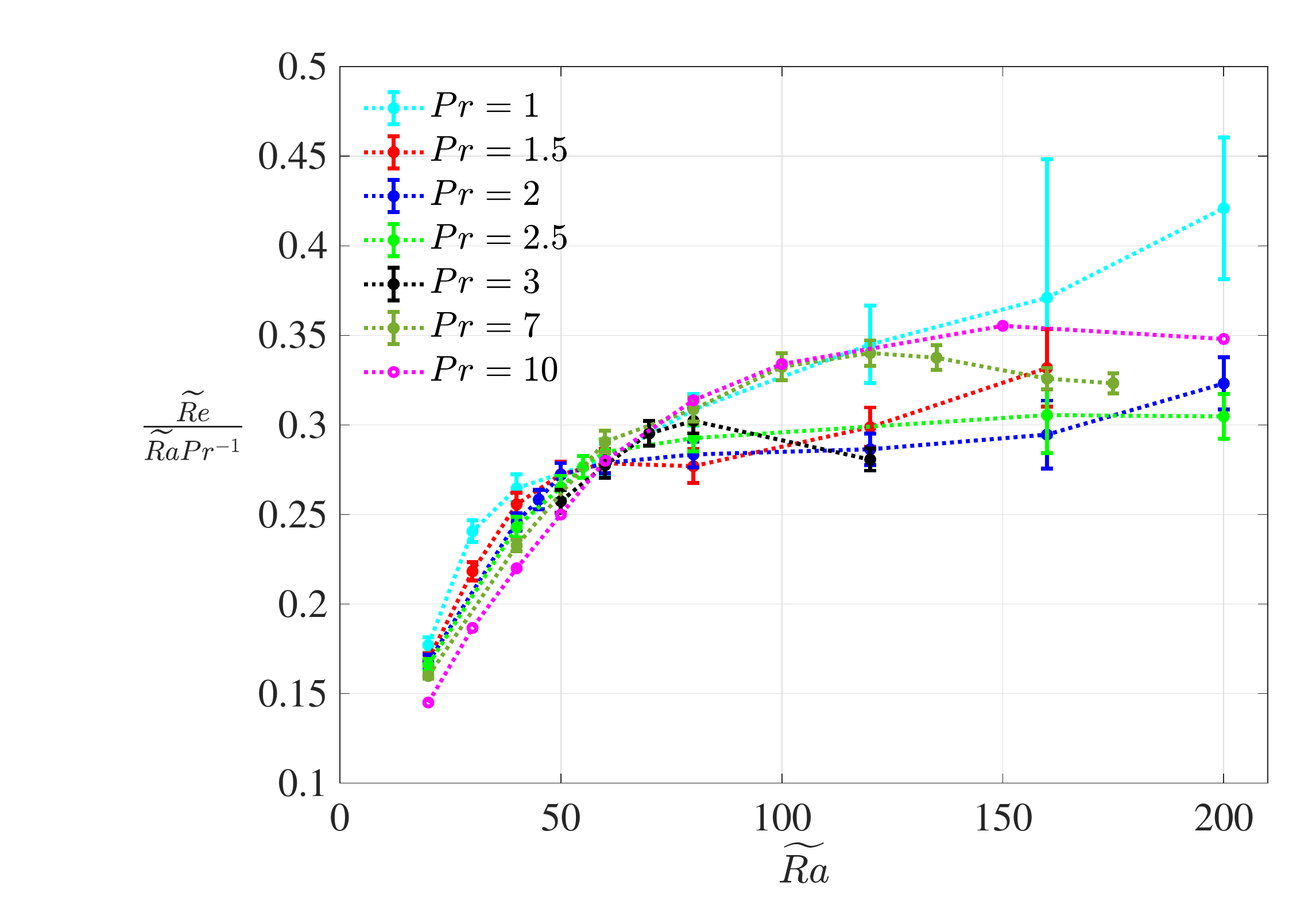}\label{subfig:ReducedRe_v2}}
\quad
\subfloat[][]
{\includegraphics[width=0.48\textwidth]{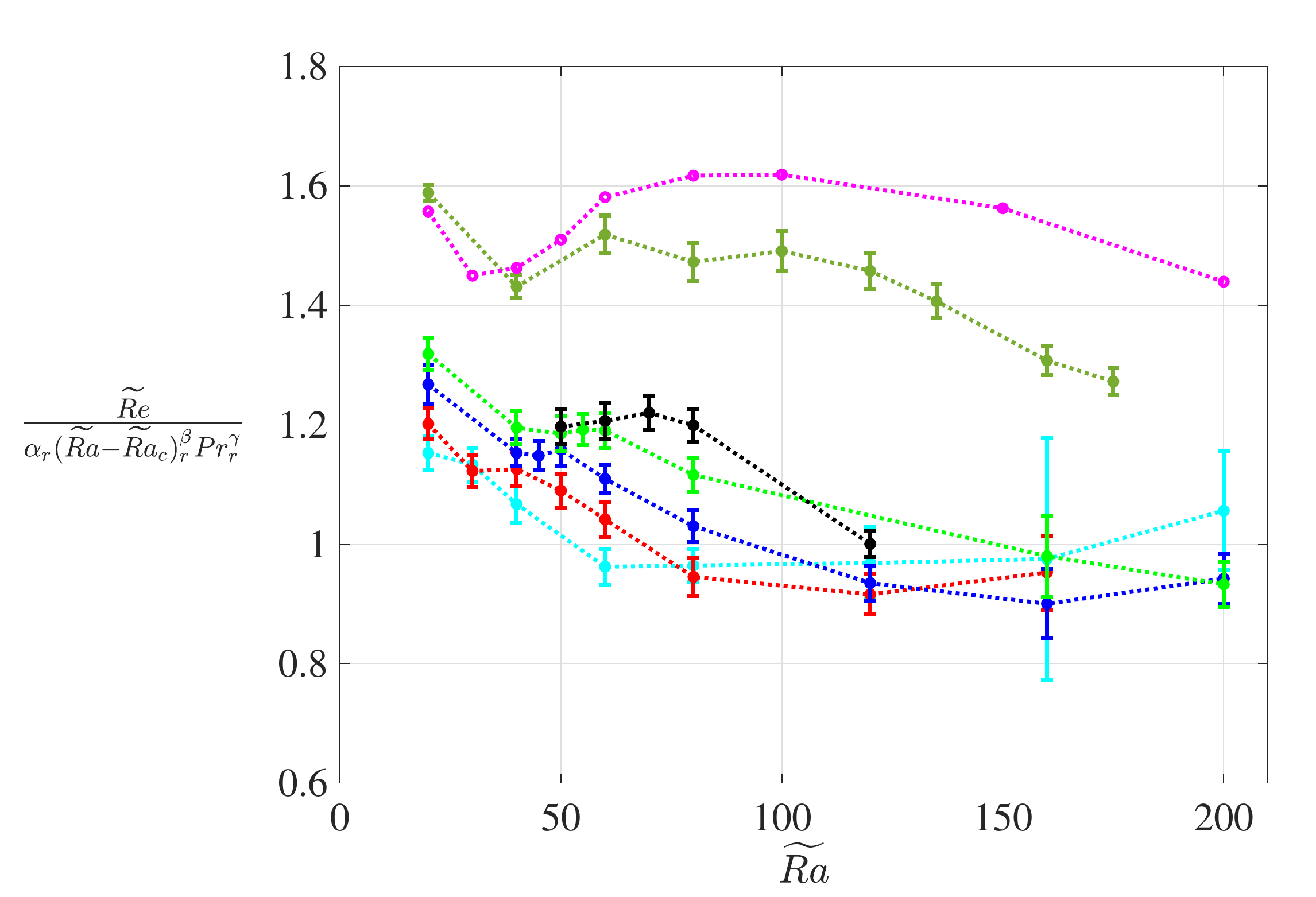}\label{subfig:ReducedRe}}
\caption{Scaling of the Reynolds number with $\Rat$. (a) Compensated $\Retm$ calculated according to $\Retm\sim \Rat \Pr^{-1}$. (b) Compensated $\Retm$ calculated according to the law \eqref{eqn:Ref1} and with values of $\alpha_r$, $\beta_r$ and $\gamma_r$ reported in table \ref{tab:fits} for $\Pr\in[1,10]$ and $\Rat\in[20,200]$ (i.e., all available $\Retm$ data). }
\label{fig:reducedReRa}
\end{figure*}

Fitting to all available data reported in this study (including the $\Pr=10$ dataset from  \citet{mC16}) we obtain $(\alpha_r,\beta_r,\gamma_r) =  ( \alphaRaNum, \betaRaNum , \gammaRaNum)$. We notice that these values are not too different from a linear scaling of the form $\Retm\sim \Rat \Pr^{-1}$, again suggesting that the reduced Grashof number plays a key role in controlling the dynamics. 
For many of the cases we find that $\beta_r$ is closer to unity when a single value of $Pr$ is used. 
Figure \ref{subfig:ReducedRe_v2} shows the compensated Reynolds number $\Retm Pr/\Rat$, where we see that there is a range of $\Rat$ values over which this scaling provides a reasonably good fit. However, significant departure from this linear Grashof number scaling is observed for the lower values of $Pr$, i.e.~those simulations that are characterized by the largest values of $\Retm$. Interestingly, this departure seems to be correlated with the behavior of the kinetic energy ratio $\Gamma$; the largest departures from the linear Grashof number scaling are observed for cases that possess the peak $\Gamma_{max}$, i.e.~those cases in which $\Retm\gtrsim24$.

We note that because the QG model employed here is asymptotically reduced, the Ekman number does not appear explicitly in the governing equations. However, we can relate our small-scale Reynolds number to the large-scale Reynolds number typically employed in DNS studies by noting that the convective length scale and fluid depth are related by $\ell = H \Ek^{1/3}$. Thus,
\be
\Ret = \frac{\left<W_{rms}\right> \ell}{\nu} = \lb \frac{\left<W_{rms}\right> H}{\nu} \rb \lb \frac{\ell}{H} \rb = \Re \Ek^{1/3} .
\label{E:asyRe}
\ee
Substituting the definition of the reduced Rayleigh number into the linear scaling $\Retm \sim \Rat /Pr$ we have
\be
\Retm \sim \frac{\Rat}{Pr} = \frac{Ra \Ek^{4/3}}{Pr}. 
\ee
Upon dividing through by $\Ek^{1/3}$ and using equation \eqref{E:asyRe} we have the relationship
\be
Re \sim \frac{Ra \Ek }{Pr} .
\ee
The above scaling is consistent with the recent spherical convection study of \citet{cG19}. However, we note that although the above large-scale Reynolds number scaling is diffusion-free, the corresponding small-scale scaling is not. 

\begin{figure*}
\centering
\subfloat[][]
{\includegraphics[width=0.48\textwidth]{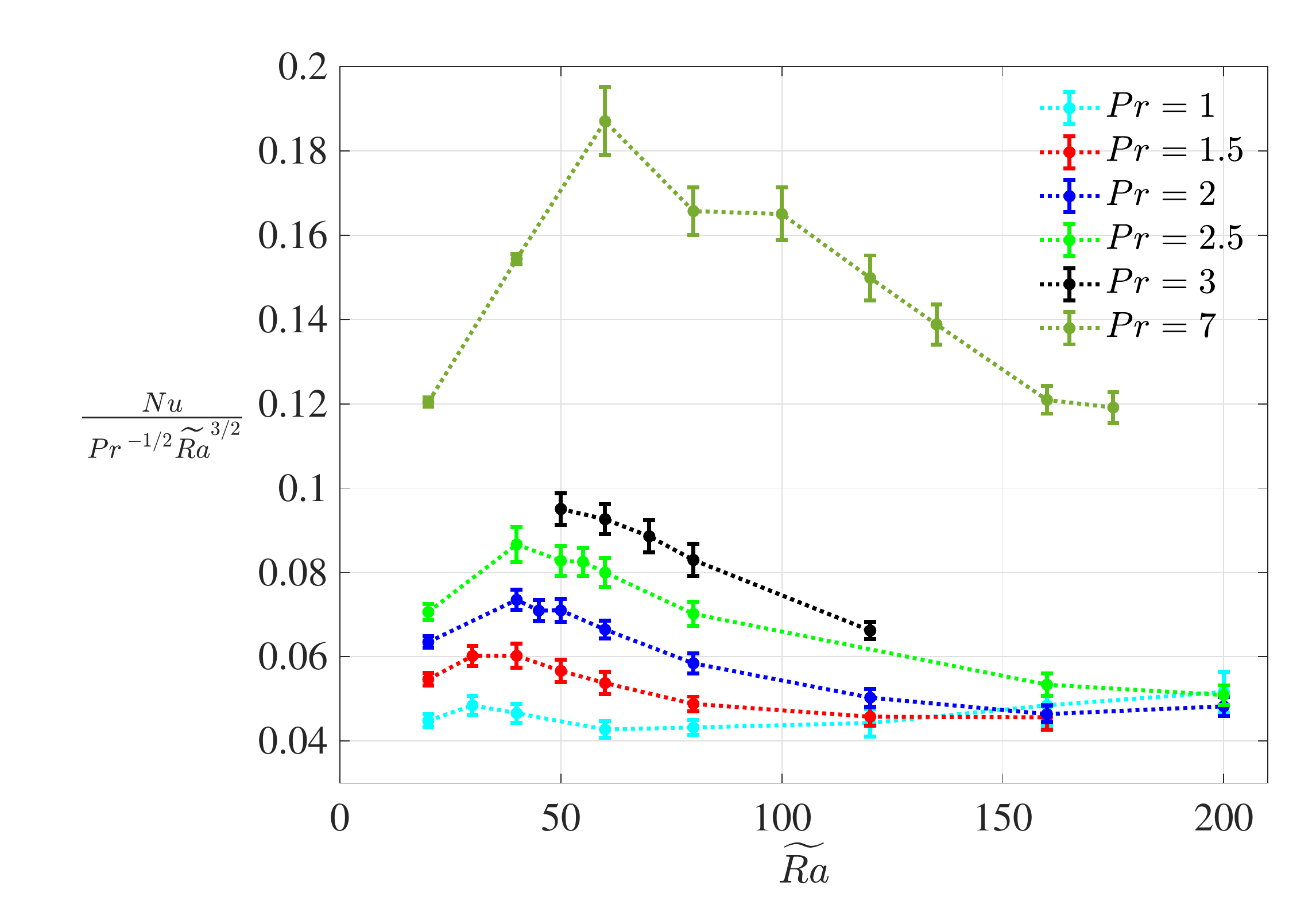}\label{subfig:ReducedNuRa_ultimate}}
\quad
\subfloat[][]
{\includegraphics[width=0.48\textwidth]{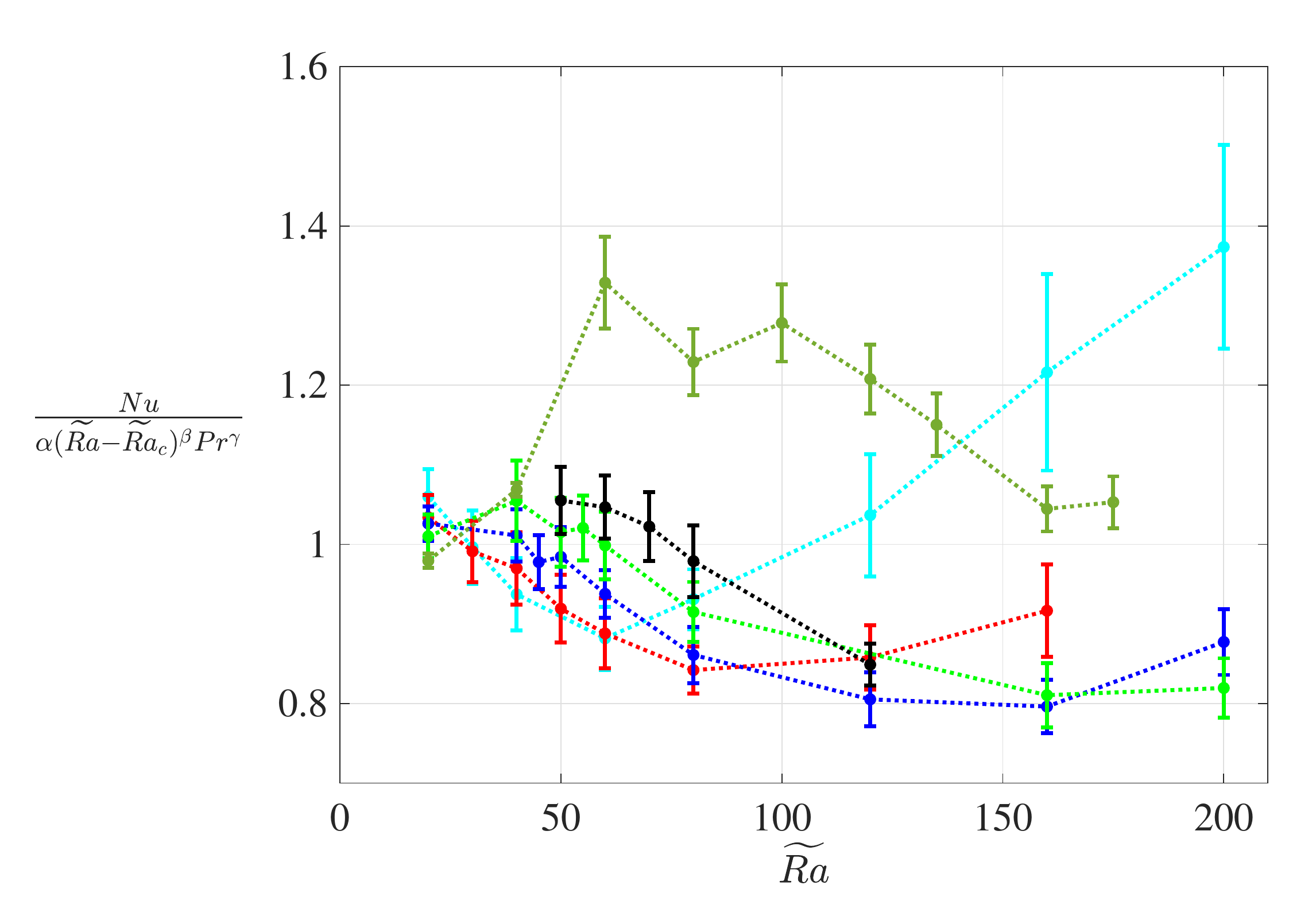}\label{subfig:ReducedNuRa}}
\caption{Scaling behavior of the heat transport with $\Rat$. (a) Compensated $Nu$ calculated according to $\Num\sim \Rat^{3/2} \Pr^{-1/2}$. (b) Compensated $\Num$ calculated according to the law \eqref{eqn:Ref1}, and with values of $\alpha_n$, $\beta_n$ and $\gamma_n$ reported in table \ref{tab:fitsNu} for $\Pr\in[1,7]$ and $\Rat\in[20,200]$ (i.e., all available $\Num$ data).}
\label{fig:reducedNuRa}
\end{figure*}

Analogous least-squares fits to the Nusselt number ($\Num$) are given by \begin{equation}
\Num = \alpha_n (\Rat - \Rat_c)^{\beta_n} \Pr^{\gamma_n} ,
\label{eqn:Nuf1}
\end{equation} 
or
\begin{equation}
\Num = \alpha_n (\Rat - \Rat_c)^{\beta_n}
\label{eqn:Nuf2}
\end{equation} 
for fixed values of Prandtl number. Results for various subsets of the explored parameters space are given in table \ref{tab:fitsNu}. Figure \ref{subfig:ReducedNuRa} shows the compensated $\Num$ according to \eqref{eqn:Nuf1} using all available data from the present study. 
From table \ref{tab:fitsNu} we see that the same fit using only the $1\le\Pr\le2$ cases suggests a fit that is roughly consistent with the ultimate scaling 
\be
\Num\sim\Rat^{3/2}\Pr^{-1/2},
\label{eqn:ultimateNu}
\ee
in agreement with \citet{kJ12}. Cases characterised by a lower $\Ret$ ( e.g.~$\Pr\ge3$) lead to a lower value for the exponent $\beta_n$ while using only the highest $\Ret$ cases ($\Pr=1$, $\Rat \ge120$) leads to a higher value of $\beta_n$. Compensated $\Num$ values, based on the ultimate scaling \eqref{eqn:ultimateNu}, are illustrated in figure \ref{subfig:ReducedNuRa_ultimate}. This plot suggests that cases that reach the ultimate $\Num$ regime are the same as those that are characterised by an increase in reduced $\Retm$ values in figure \ref{fig:reducedReRa}, and a corresponding drop in $\Gamma$ in figure \ref{fig:Kz_over_K_Re}.

\begin{table}
  \begin{center}
\def~{\hphantom{0}}
\begin{tabular}{ccccc}
	$\Pr $ & $ \Rat $ & $ \alpha_n $ &  $\beta_n $ & $\gamma_n $  \\
	\hline
		[1,7] & [20,200] &  \alphaNuNum  &  \betaNuNum  &  \gammaNuNum \\[0pt]
		[1,3] & [20,200] &  0.1969 &   1.2356 &  -0.3767 \\[0pt]
		[1,2] & [20,200] &  0.0993  &  1.3776 &  -0.5276 \\[0pt]
		[1,2] & [40,200] &  0.0933   & 1.3899  & -0.5275 \\[0pt]
		1 &  [20,200] & 0.0372  &  1.5725  &  -  \\[0pt]
		1 &  [120,200] &  0.0194  &  1.6989  &  -  \\[0pt]
		1.5 &  [20,160] &   0.1781  &  1.2011  &   -\\[0pt]
		2 &  [20,200] &    0.2196  &  1.1504  & - \\[0pt]
		2.5 &  [20,200] &    0.3798 &   1.0402 &   -\\[0pt]
		3 &  [20,120] &   0.6799  &  0.9147 &  -  \\[0pt]
		7 &  [20,160] &   0.6495  &  0.9957 &  -  \\[0pt]
\end{tabular}
\caption{Least-squares fits to $\Num = \alpha_n (\Rat - \Rat_c)^{\beta_n} \Pr^{\gamma_n}$ (for data encompassing multiple $\Pr$) or $\Num = \alpha_n (\Rat - \Rat_c)^{\beta_n}$ (when a single value of $\Pr$ is considered).}
\label{tab:fitsNu}
  \end{center}
\end{table}

\subsection{Balances}

\begin{figure*}
\centering
\subfloat[][]
{\includegraphics[width=0.3\textwidth]{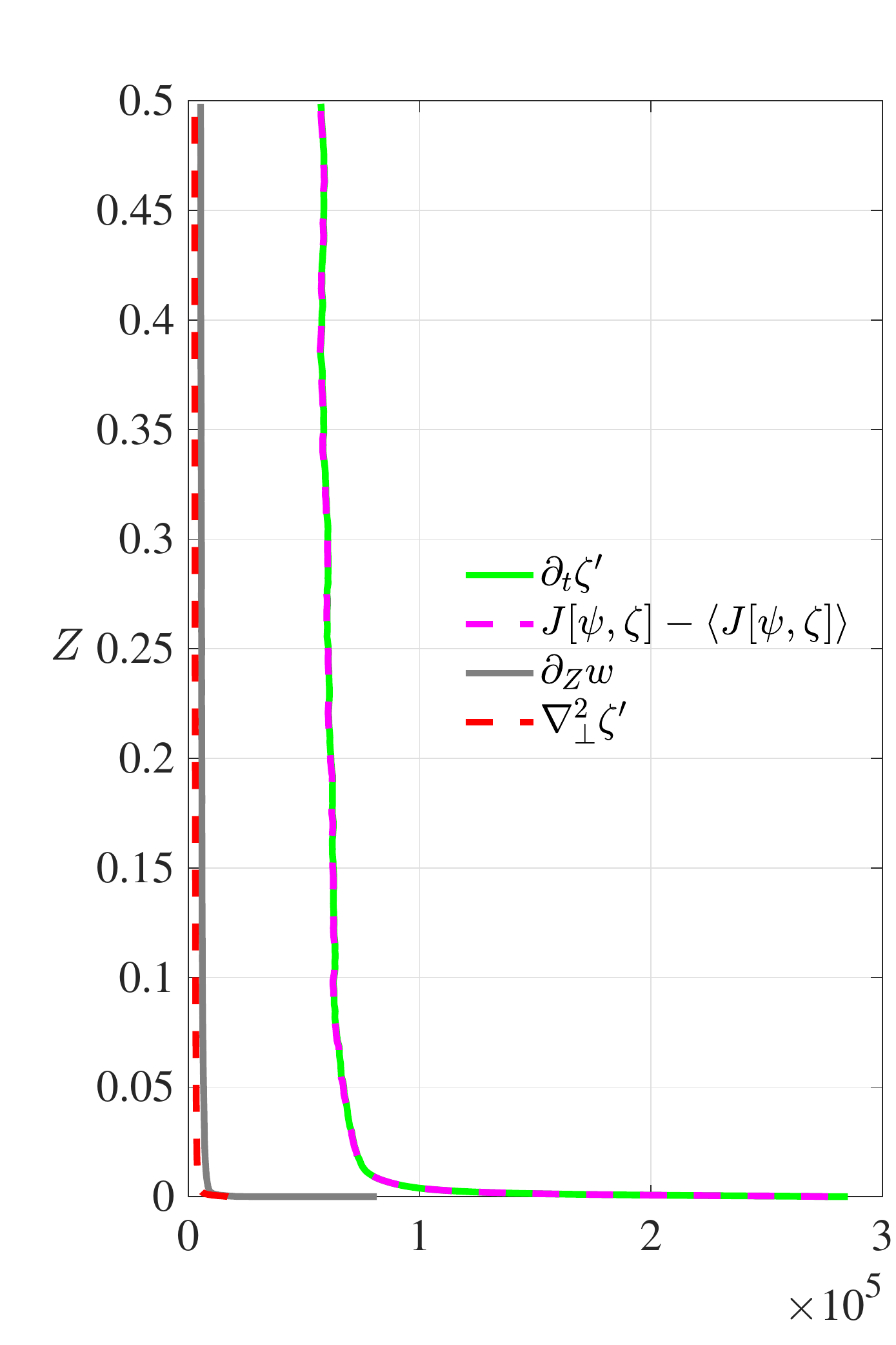}\label{subfig:AV_forces_bc_profile}}
\quad
\subfloat[][]
{\includegraphics[width=0.3\textwidth]{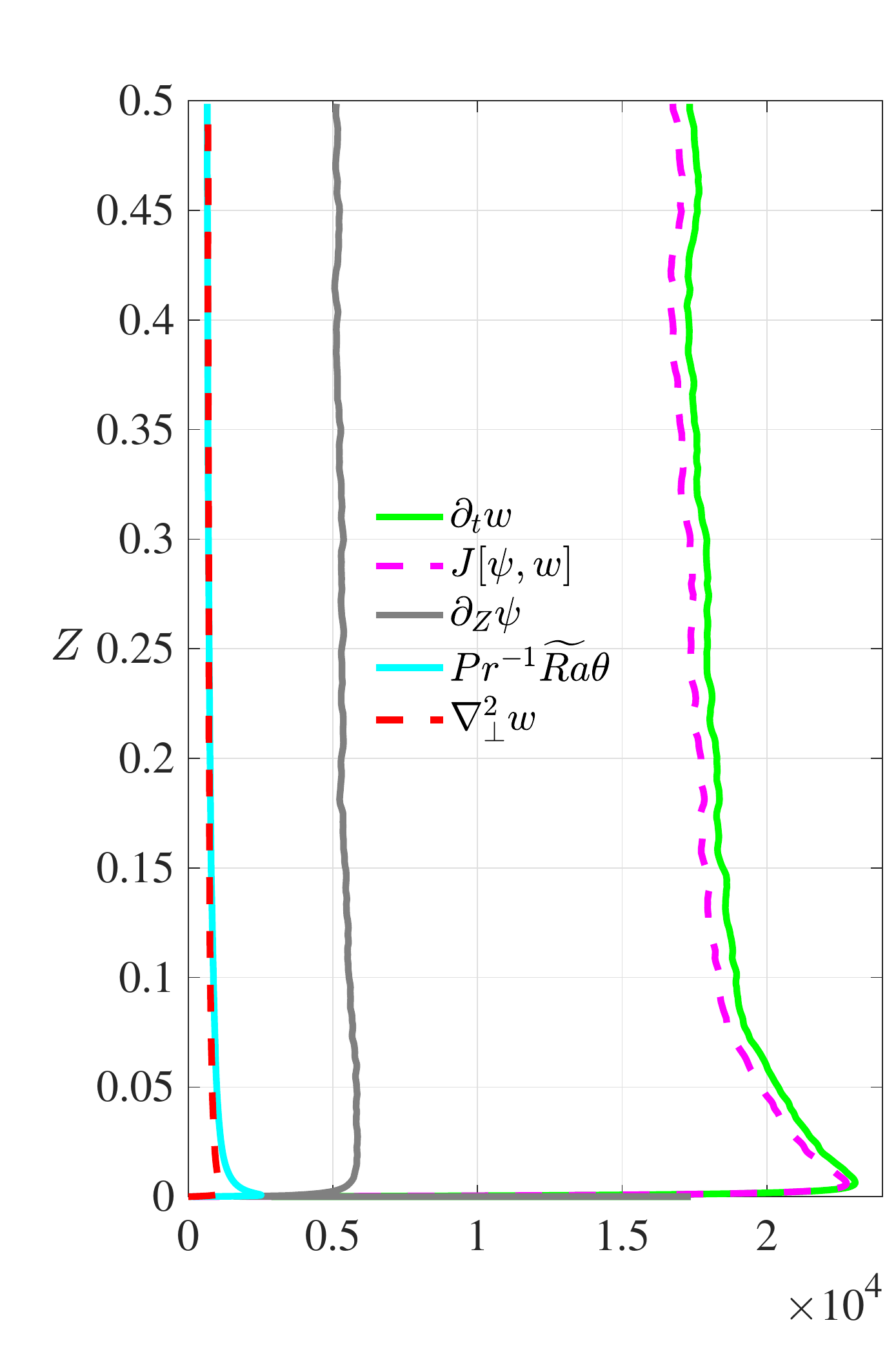}\label{subfig:z_forces_bc_profile}}
\quad
\subfloat[][]
{\includegraphics[width=0.3\textwidth]{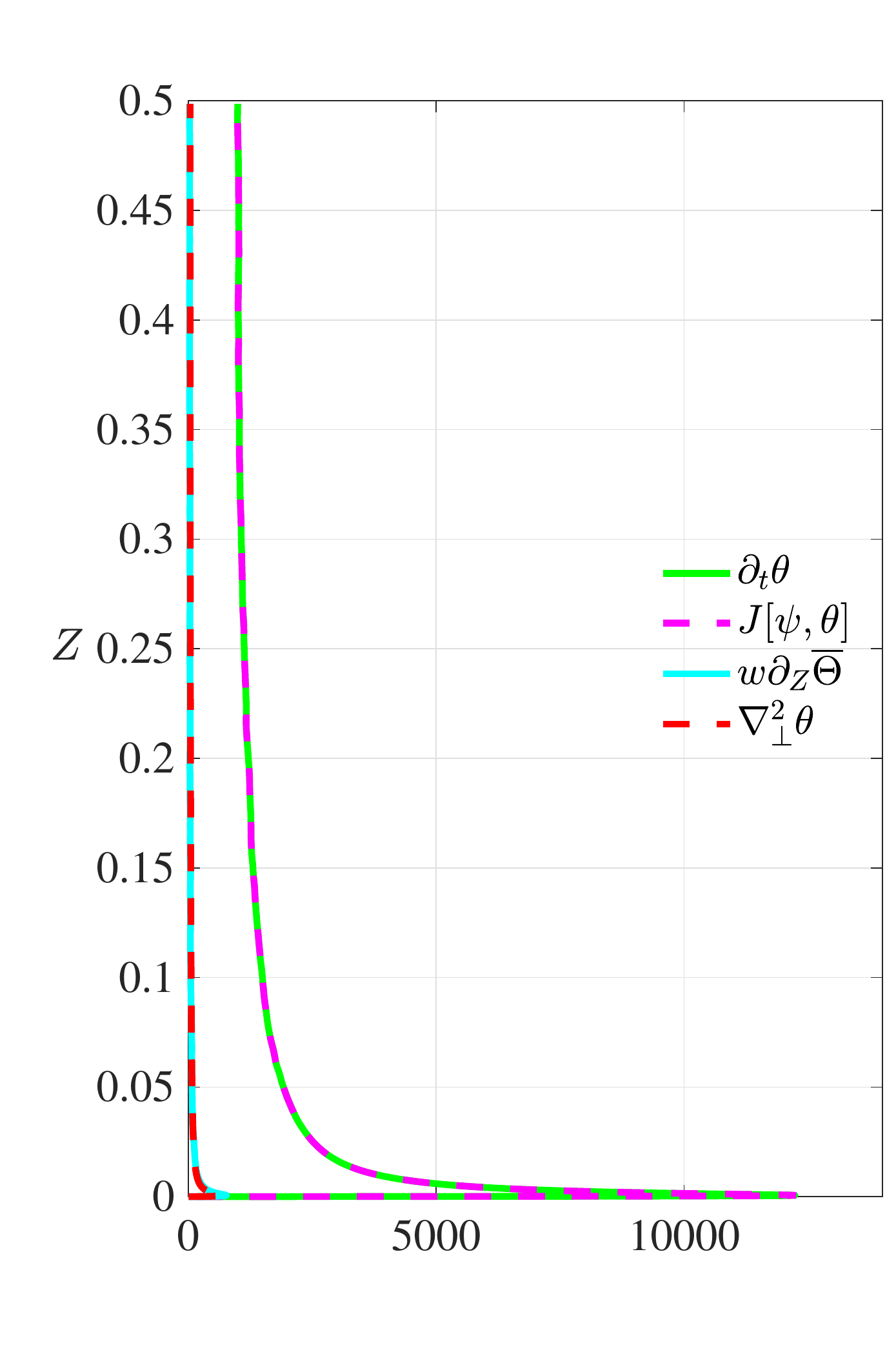}\label{subfig:heat_eq}}
\quad
\caption{Vertical profiles of rms terms in: (a) the baroclinic vorticity equation (obtained by subtracting \eqref{E:baro} from \eqref{E:vort0}); (b) the vertical momentum equation \eqref{E:mom0}; and (c) the fluctuating heat equation \eqref{E:theat0}. Profiles have been calculated as temporal averages for the case $\Rat=200$, $\Pr=1$ and are characteristic of all cases in the geostrophic turbulence regime, where LSV formation is robust.}
\label{fig:forces}
\end{figure*}

Vertical profiles of the horizontal rms of each term present in the baroclinic vertical vorticity, vertical momentum and fluctuating heat equations are shown in figures \ref{subfig:AV_forces_bc_profile}, \ref{subfig:z_forces_bc_profile} and \ref{subfig:heat_eq}, respectively, for the most extreme calculation of $\Rat=200$ and $Pr=1$ ($\Retm \approx 84 $). All of the quantities shown have been time-averaged. As shown previously \citep{kJ12}, within this high-$\Rat$ regime, the dominant terms in the governing equations are given by
\begin{gather}
\partial_t \zeta' + J[\psi,\zeta] - \langle J[\psi,\zeta] \rangle \approx 0,\\
\partial_t w + J[\psi,w] \approx 0,\\
\partial_t \theta + J[\psi,\theta] \approx 0,
\label{eqn:balance_dt}
\end{gather}
which shows that horizontal advection of all these quantities is a key characteristic of this regime. Close inspection of the first of these balances reveals that, as $\Ret$ grows, the advection of vorticity is increasingly dominated by the advection due to the barotropic flow, i.e. $J[\langle\psi\rangle,\zeta] \gg J[\psi',\zeta]$ for $\Ret\gg 0$.

On their own, the `balances' given above reveal little about the resulting dynamics. Higher-order, or subdominant, effects are necessary in the dynamics, especially with regard to heat transport. Figure \ref{subfig:z_forces_bc_profile} suggests that small differences between the rms values of $\partial_t w$ and $J[\psi,w]$ are necessary to balance the vertical pressure gradient, $\dz \psi$. This perturbative effect repeats again at even higher order, as figure \ref{subfig:z_forces_bc_profile} shows that the buoyancy force and vertical viscous force are approximately balanced, i.e.
\be
\frac{\Rat}{\Pr}  \pth  \approx \lp w .
\ee
Moreover, we find a subdominant balance in the fluctuating heat equation between the advection of the mean temperature and horizontal thermal diffusion, 
\be
w \dz \mth \approx \lp \theta.
\ee

\begin{figure*}
\centering
\subfloat[][]
{\includegraphics[width=0.45\textwidth]{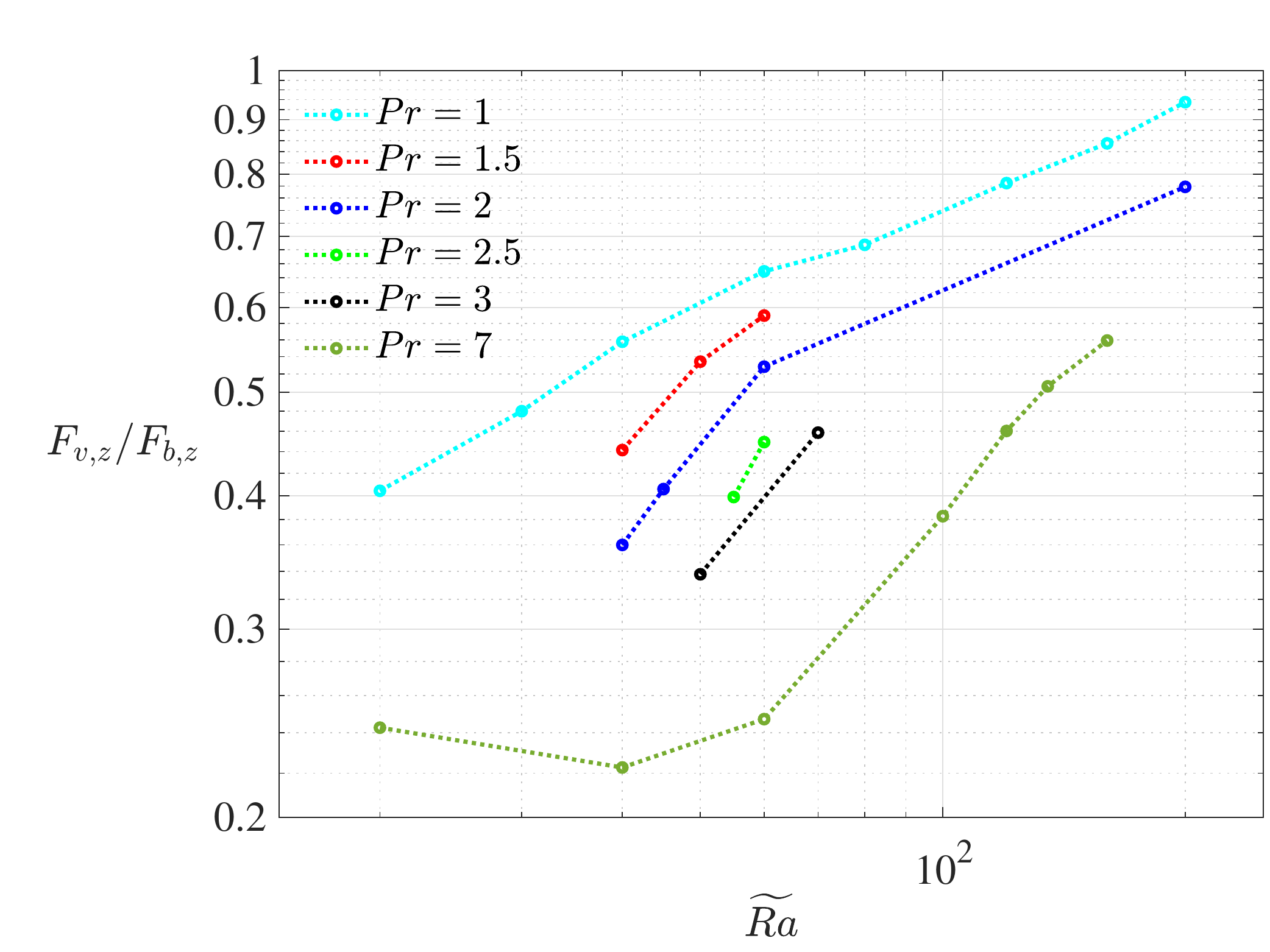}} \quad
\subfloat[][]
{\includegraphics[width=0.45\textwidth]{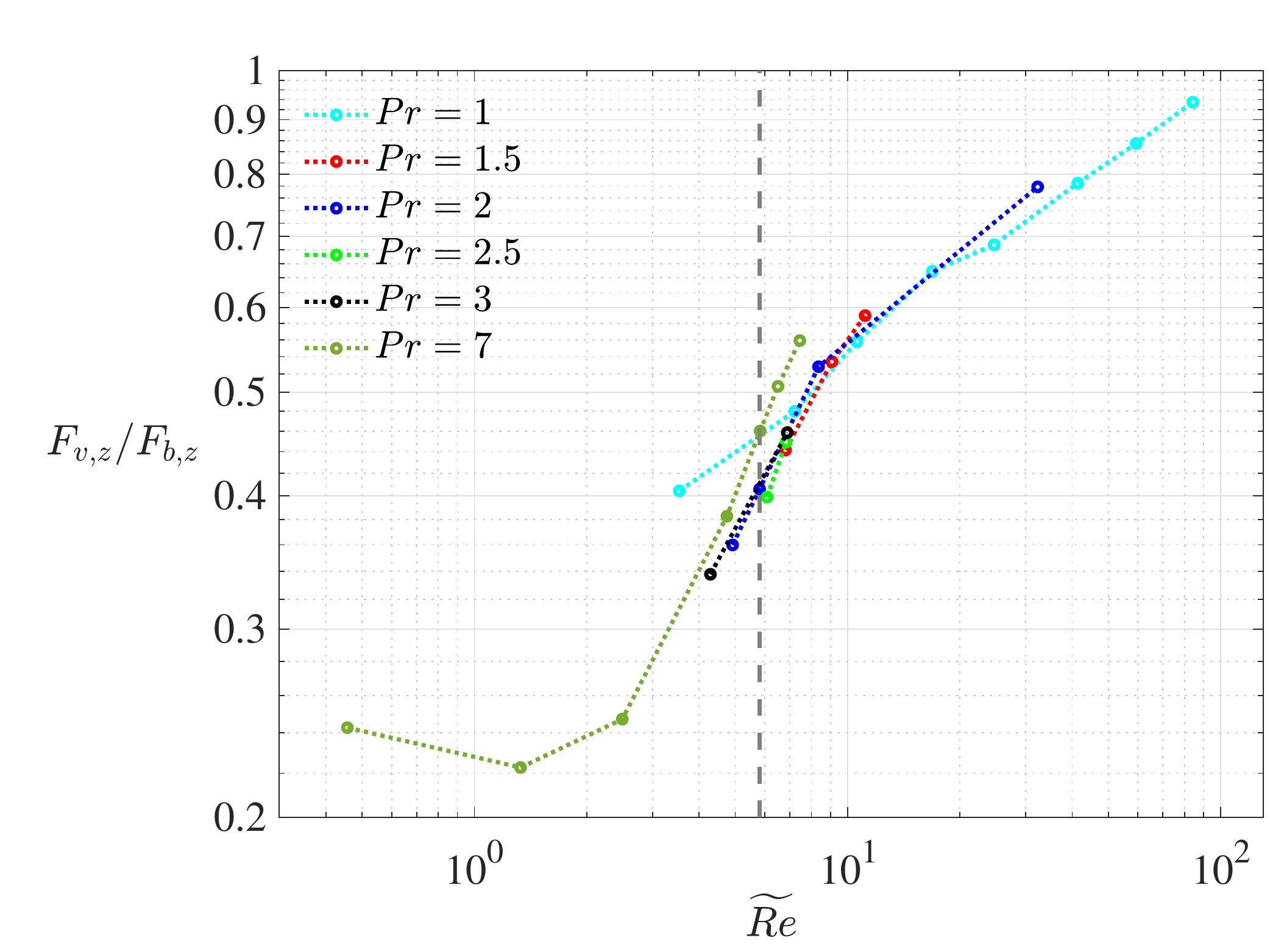}}
\caption{Ratio of the vertical components of the rms viscous force, $F_{v,z} =  \lp w$, to the rms buoyancy force, $F_{b,z}=\Rat \Pr^{-1} \pth$: (a) force ratio as a function of $\Rat$; (b) force ratio as a function of $\Retm$.}
\label{fig:FRatio}
\end{figure*}

To better understand the role of the subdominant balance between viscosity and buoyancy, we show in figure \ref{fig:FRatio} the ratio of of the vertical components of the rms viscous force, $F_{v,z} =  \lp w$, to the rms buoyancy force, $F_{b,z}=\Rat \Pr^{-1} \pth$, both as a function of $\Rat$ and $\Retm$. Surprisingly, this ratio is an increasing function of $\Rat$ (and therefore also of $\Retm$). This result is in stark contrast to non-rotating convection in which viscous forces become ever smaller (relative to other forces) with increasing Rayleigh number. Indeed, the so-called `free-fall' scaling for convective flow speeds, characterized by a balance between buoyancy and inertia, relies on the influence of viscosity being weak \citep[e.g.~see][]{mY19}. All of the different Prandtl number cases appear to show qualitatively similar behavior, and, as figure \ref{fig:FRatio}(b) shows, a reasonable collapse of the data can be obtained when  the force ratio is plotted versus the Reynolds number.

\section{Discussion and conclusions}
\label{S:Discussion}

A systematic investigation of rapidly rotating convection was carried out to determine  the necessary conditions under which large-scale vortices (LSVs) form, and how the amplitude of such vortices and associated convective flow speeds scale with the input parameters. To achieve the extreme parameter regimes that are thought to be representative of natural systems such as planetary and stellar interiors, we have made use of an asymptotic description of the governing equations that rely on the assumption of a leading-order geostrophic balance. Varying the thermal Prandtl number has allowed us to determine the influence of fluid properties on the convective dynamics, and has also allowed for a more detailed control of the convective Reynolds numbers over our investigated range of Rayleigh numbers. 


The LSVs form as a consequence of an inverse cascade that transports kinetic energy from small scale, convective motions up to the system-wide scale, characterized by a box-normalized wavenumber of $k=1$. These LSVs grow in time until the energy input from the convection is balanced by large-scale viscous dissipation. All of the simulations presented show evidence of this equilibration process, regardless of the particular values of $\Pr$ and $\Rat$. We find that LSV-dominant convection can be characterized by a critical convective Reynolds number $\Retm \approx 6$ across the range of investigated Prandtl numbers, in agreement with low-$\Ek$ DNS simulations performed at $\Pr=1$ \citep{bF14,cG14}. Although an increase in  $\Pr$ leads to a concomitant increase in viscous dissipation for a fixed value of $\Rat$, we find, for the first time to our knowledge, evidence of LSV-dominant convection in the ``plume'' regime. In particular, we observed the formation of barotropic vortices with a Prandtl number as high as $\Pr=7$, a value that is representative of water at typical laboratory conditions. This finding suggests that LSVs may be detectable in laboratory experiments that use water as the working fluid.
 From the data reported in this study we can estimate a threshold value of $\Rat_t\simeq 120$ for LSV to form at $\Pr=7$ which can be translated into large-scale $Ra_{t}$ for a given $\Ek$ via \eqref{eqn:RatRaH}. State-of-the-art laboratory experiments can reach  $\Ek=10^{-8}$ \citep{jC15,jC19} giving $Ra_{t}\simeq 5.6\cdot 10^{12} $, a value for which heat transfer data suggest convection to be in a transitional regime between rotationally-dominated and non-rotating dynamics. We note that the presence of no-slip boundaries (not used in the present study) has been shown to suppress the formation of LSVs \citep{mP16}, so additional studies are needed to determine the threshold for LSV-dominant convection with no-slip boundary conditions.

Several properties of LSVs have been studied. In agreement with the DNS study of \citet{cG14}, we find that the relative size of the kinetic energy of the barotropic flow to that of the convection reaches a maximum value at a particular value of the Rayleigh number. However, with DNS studies, there is a corresponding increase in the Rossby number with increasing Rayleigh number. In contrast, the asymptotic model used here only captures the asymptotically small Rossby number limit, showing that this change in the growth of the LSV must be present in the rapidly rotating regime. When data from the entire range of  $\Pr$ is plotted as a function of Reynolds number, the peak in the barotropic kinetic energy occurs near $\Retm \approx 24$. Therefore, the growth of the barotropic kinetic energy slows as the Rayleigh number is increased, suggesting that there is an optimum forcing level. We find that this change in behavior is related to a decrease in the velocity and vorticity correlations that are necessary to drive the inverse cascade. Our findings suggest that additional regimes, beyond the accessible limits of the present investigation, may be present in the convective dynamics as the Rayleigh number is increased further. 

The horizontal dimensions of the simulation domain are shown to have a direct influence on the energy present in the LSV. It is found that the LSV energy grows quadratically with the horizontal dimension of the simulation domain (assuming domains of square cross-section), in agreement with DNS calculations \citep{bF14}. This finding is likely linked to the total available convective kinetic energy, which also grows quadratically with the horizontal dimensions of the simulation domain. Although a detailed investigation of the dynamical effect of this scaling was beyond the scope of the present investigation, this geometry-dependent effect may nevertheless have implications on the resulting dynamics.

The simulations suggest that there is no obvious scaling regime in the convective flow speeds with increasing Rayleigh number. A linear scaling of the form $\Retm \sim \Rat/Pr$ appears to collapse the data over a limited range in $\Rat$, but the highest $\Retm$ cases diverge from this scaling at the highest accessible values of $\Rat$. We note that this linear scaling can be translated to an equivalent large-scale Reynolds number scaling of the form $\Re \sim \Ek \Ra/\Pr$, which has been noted in previous studies of rotating convection \citep{cG19}. Although this scaling is independent of the diffusion coefficients $\nu$ and $\kappa$ when viewed on the largest scales of the system, the small-scales remain influenced by viscosity. Indeed, the simulations have revealed that the ratio of the rms viscous force to the rms buoyancy force in the vertical component of the momentum equation is an \textit{increasing} function of $\Rat$ (or equivalenty $\Ra$). This observation may simply be a result of the energetics of the Boussinesq system that requires the net heat transport to be balanced by viscous dissipation. In this regard, it might be argued that viscosity is fundamental to rotating convective dynamics.

\section*{Acknowledgements} 
This work was supported by the National Science Foundation under grant EAR \#1620649 (SM, MAC and KJ). M.~Krouss was partially supported by the Undergraduate Research Opportunities Program at the University of Colorado, Boulder. This work utilized the RMACC Summit supercomputer, which is supported by the National Science Foundation (awards ACI-1532235 and ACI-1532236), the University of Colorado Boulder, and Colorado State University. The Summit supercomputer is a joint effort of the University of Colorado Boulder and Colorado State University. The authors acknowledge the Texas Advanced Computing Center (TACC) at The University of Texas at Austin for providing high performance computing and database resources that have contributed to the research results reported herein. Volumetric renderings were produced with the visualization software VAPOR, a product of the Computational Informations Systems Laboratory at the National Center for Atmospheric Research.

\bibliographystyle{jfm}
\bibliography{jfm-bib,References}

\end{document}